\documentclass[cits,hyper]{JINST}

\makeatletter
\let\set@color\relax
\makeatother

\usepackage{color}

\usepackage[figuresright]{rotating} 
\usepackage[caption=false]{subfig}
\usepackage{textcomp}
\usepackage{ulem}

\usepackage{url}

\usepackage{tikz}
\usetikzlibrary{positioning}
\usetikzlibrary{shapes.misc} 
\usetikzlibrary{shapes.arrows}
\usetikzlibrary{arrows,chains}
\usetikzlibrary{scopes,fit,calc}

\newcommand{\Eqref}[1]{Eq.~\ref{#1}}
\newcommand\Sectref[1]{Sect.~\ref{#1}}

\newcommand{\ac}[1]{#1}
\newcommand{\acp}[1]{#1}
\usepackage{xspace}
\usepackage{amsmath}

\newcommand\Figref[1]{Fig.~\ref{#1}\xspace}
\newcommand\Figsref[1]{Figs.~\ref{#1}\xspace}

\newcommand\Tabref[1]{Tab.~\ref{#1}\xspace}
\newcommand\Tabsref[1]{Tabs.~\ref{#1}\xspace}

\title{Any Light Particle Search II \\ Technical Design Report}

\author{Robin B\"ahre$^a$,
Babette D\"obrich$^b$,
Jan Dreyling-Eschweiler$^b$,
Samvel Ghazaryan$^b$,
Reza Hodajerdi$^b$,
Dieter Horns$^c$,
Friederike Januschek$^b$,
Ernst-Axel Knabbe$^b$,
Axel Lindner$^b$\thanks{Corresponding
author.},
Dieter Notz$^b$,
Andreas Ringwald$^b$,
Jan Eike von Seggern$^b$,
Richard Stromhagen$^b$,
Dieter Trines$^b$,
Benno Willke$^a$\\
\llap{$^a$}Albert Einstein Institute,\\
  Callinstr. 38, 30167 Hannover, Germany\\
\llap{$^b$}Deutsches Elektronen-Synchrotron (DESY),\\
  Notkestr. 85, 22607 Hamburg, Germany\\
  \llap{$^c$}University of Hamburg,\\
  Institut f\"ur Experimentalphysik, Luruper Chaussee 149,
  22761 Hamburg, Germany

  E-mail: \email{axel.lindner@desy.de}}

\abstract{This document constitutes an \textcolor{blue}{excerpt} of the Technical Design Report for the second 
stage of the ``Any Light Particle
Search'' (ALPS-II) at DESY as submitted to the DESY PRC in August 2012 and reviewed in November 2012.
ALPS-II is a ``Light Shining through a Wall'' experiment which searches for photon oscillations into
weakly interacting sub-eV particles. These are often predicted by extensions of the Standard Model and 
motivated by astrophysical phenomena. 
The first phases of the ALPS-II project were approved by the DESY management on February 21st, 2013.
}


\begin{document}

\section{Introduction}
\label{chap:tdr:introduction}
Understanding the fundamental constituents and forces of nature is one of the strongest motivations for society to support experimental and theoretical physics.
This has been again proven for example  by the enormous public resonance on the likely discovery of the Higgs boson at the LHC and shows up continuously in the 
interest in astronomical 
and cosmological questions.
Indeed, the progress is impressive. 
Mainly accelerator based research has led to the development of the so-called Standard Model
of particle physics, which is able to describe experimental 
results up to
\textperthousand-level accuracy.
However, while such experiments have been extremely successful to fill the gaps in the Standard Model as predicted a few decades ago, 
evidence for physics beyond this theoretical framework has mounted from astrophysical and cosmological observations.
Most prominent are Dark Energy and Dark Matter.
So far accelerator based experiments at the energy or intensity frontiers have 
not been able to identify explanations towards understanding these phenomena.

Hence it appears to be natural to ask whether physics beyond the Standard Model may hide at another frontier.
And in fact such a possible frontier is known since the early 1980ties, when the apparent CP-conservation in QCD was 
traced back to a hypothetical new particle, the axion.
It was quickly shown that the axion, if it exists, should be extremely light and weakly interacting to escape experimental bounds. 
Such an axion was named ``invisible''. 
It poses a new ``low-energy frontier'' to the Standard Model hinting at uncharted territory beyond the reach 
of accelerator based experiments.
In recent years it became clear that the axion might be just one example of many different 
``weakly interacting slim particles'' (WISPs) as we will describe in \Sectref{chap:tdr:goals}. \\

Experiments searching for WISPs usually rely on photon-WISP interactions.
As will be shown in the subsequent chapter, one can distinguish between WISPs with the same quantum numbers as photons (allowing for kinetic mixing with photons), 
and pseudo-scalar or scalar bosons which require two-photon interactions. 
The proposed second stage of the ALPS experiment will be sensitive to both kinds of WISPs and even other predicted WISPy particles.

Typically in such an experiment one of the two photons is provided by a laser beam, while the other is supplied by a strong magnetic field.
In such environments the interactions between WISPs and photons take place in a coherent fashion. 
The interaction probability is many orders of magnitude larger than in accelerator based beam dump experiments for example.
A virtual or real production of WISPs could give rise to spectacular experimental observations ranging 
from polarization effects to the sudden appearance of light 
in a seemingly perfectly shielded environment.
The ALPS-II proposal concentrates on the latter idea by trying to shine light through a wall.

\begin{figure}[htb]
\centering
\includegraphics[angle=0,width=100mm]{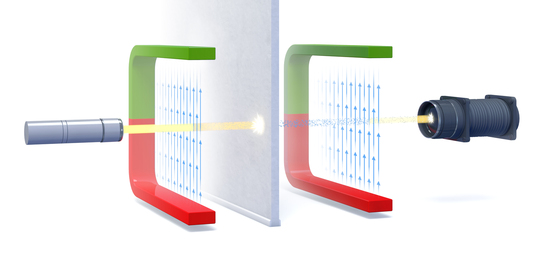}
\caption{The principle of a light-shining-through a wall experiment:
light, typically from a strong laser, is shone into a magnetic field. WISPs could be created by interaction of the 
laser light with the magnetic field or just by kinetic mixing. 
A barrier blocks the light, but WISPs easily traverse it. In the second part of the experiment behind the wall,
some WISPs convert back to photons giving rise to the impression of light shining through a wall. 
The WISPs have to move over macroscopic distances, 
so that only the production of on-shell particles is relevant here.
 \label{fig:LSW}} 
\end{figure} 

A clear observation of light generated in a very well shielded environment could only be explained by yet unknown WISPs which pass 
any barrier.
The principle of ``light-shining-through-a-wall'' (LSW) experiments is shown in Figure\,\ref{fig:LSW}.

WISPs produced by laser light as well as reconverted photons originating from these WISPs have laser-like properties.
This allows to
\begin{itemize}
\item
guide them through long and narrow tubes inside accelerator dipole magnets and
\item
 to exploit resonance effects by setting up optical resonators (cavities).
\end{itemize}
The ALPS-I experiment at DESY has improved previous results by a factor of 10 in 2010.
Future increases in sensitivity rely basically on three corner stones:
\begin{itemize}
 \item An optical resonator in front of the wall is the best means to ``store'' the injected 
 laser light and thus to amplify the available laser light power. The resonator houses a standing light
 wave created by the superposition of the light waves reflected by the two mirrors defining the resonator. 
 In the resonator the field strength of the light wave moving towards the ``wall'' is much larger than the field
 strength of the injected laser light. 
 The ratio of the square of both field strengths is called the power built-up factor in the following.
  \item With a second resonator behind the wall the back 
  conversion probability of WISPs into photons can be greatly enhanced. 
  To understand this, one has to keep in mind that an optical resonator also works 
  for very faint light in exactly the same way as sketched above. Indeed it has been verified 
  experimentally in the microwave regime that even a light wave with an expectation value for the number of photons inside 
  the resonator much below one is amplified as anticipated. Now the wave function of the WISPs propagating 
  in the resonator behind the wall can be described as consisting out of a large ``sterile'' part and a tiny 
  photonic component. Consequently, the photonic component is amplified in the resonator in exactly the same 
  fashion as ``ordinary'' light and the conversion probability into photons increased accordingly. However, 
  this approach requires that the resonators in front and behind the wall are tuned to the same frequency and 
  spatial mode (as the laser photons and the photons from regenerated WISPs have exactly the same properties) 
  and have a negligible relative phase jitter only (so that the WISP beam passing the wall can inject
  photons into the second resonator in the same manner as laser light is injected into a resonator).
  \item The magnetic length is increased by using a string of dipole magnets. As the production of WISPs 
  and the regeneration of photons takes place in a coherent fashion, a 20-fold increase in the 
  magnetic length would result in the same increase in the sensitivity to the WISP-photon coupling strength.
\end{itemize}

The ALPS-II proposal describes how to combine existing infrastructure at
DESY (HERA dipole magnets, long straight sections in the HERA tunnel,
cryogenics) with world-leading expertise in laser technology (derived from
experience at gravitational wave interferometers) and new optical detectors
(superconducting single-photon counters) to achieve sensitivities in WISP searches more than
three orders of magnitude better than at existing laboratory experiments.\\

The interest in the low-energy frontier has changed in the last five to ten
years. This shows up in the growing participation in the PATRAS workshop
series (see 
{\url {http://axion-wimp.desy.de}}), 
recent workshops
 in the US 
and the
document submitted by international authors for the discussion of the update
of the European strategy for particle physics
({\url{http://indico.cern.ch/contributionDisplay.py?contribId=105\&confId=175067}}).
Cornerstones were the CAST helioscope at CERN, which searches for axion
emission from the sun compatible with existing solar models, new theoretical
insights into extensions of the Standard Model, which very often predict
numerous new particles at the low-energy frontier, and recently observed
astrophysical puzzles which could be explained by such new particles.
Consequently new experimental activities started world-wide. Very often they
made use of the infrastructure of large accelerator laboratories. One
example was the first stage of the
ALPS (``Any Light Particle Search'')
 experiment at DESY.
It utilized a HERA dipole magnet combined with a strong cw laser driving an optical resonator inside the magnet.
By combining the know-how of particle physicists and experts of gravitational wave interferometers, ALPS-I reached the world-wide best sensitivity
on the photon coupling for new weakly-interacting,
low mass particles amongst
laboratory searches.\\
ALPS-I as well as competing experiments at CERN, FNAL, JLAB and elsewhere can be described as quick ``shots into the dark''. 
They aimed for fast returns using existing equipments (in the ALPS-I case a HERA dipole magnet accidentally still installed at a test bench) 
given strong budget constraints.
None of these experiments have shown evidence for physics beyond the Standard Model.\\
However, the realizability of such experiments based on new collaborations of different branches of physics has been 
proven successfully in these prototype approaches.
A second generation ALPS-II experiment promises to increase the experimental sensitivity in axion-like particle searches by 
more than three orders of magnitude (!).
It will allow to access the parameter space of new hypothetical particles predicted by the abovementioned astrophysical 
phenomena and predictions of low-energy incarnations of string theory in the laboratory.
It will surpass the strongest constraints on axion-like particles provided presently by CAST in the low-mass range whilst offering full 
laboratory control over all experimental parameters.
ALPS-II will use existing HERA dipoles to be installed in a straight section of the HERA tunnel.
This will be achieved at comparably modest costs while the impact of any discovery could hardly be overestimated. 
However, there are competing activities mainly in the US, which pose time demands on the realization of ALPS-II.
For these reasons, a number of theorists and experimentalists have proposed in the recent 
submission to the European Strategy Group to realize ALPS-II.\\

\section{Motivation and science goals for ALPS-II}
\label{chap:tdr:goals}

In the following, we first outline in detail the motivation arising  from theory and astrophysics  for the search for new physics at the 
low-energy frontier,
and give a first sketch of the potential impact of ALPS-II in this field, see Sect.~\ref{sec:reasons}.
Secondly, we detail on the prospected sensitivity of ALPS-II  derived from the experimental parameters, 
cf. Sect.~\ref{sec:discovery_pot}.
Details on the experiment itself will follow in Sect.~\ref{chap:tdr:experiment}.

\subsection{Fundamental physics with low-mass, weakly coupled particles \label{sec:reasons}}

\subsubsection{The hunt for WISPs and their connection to fundamental physics questions}
Whilst at the moment the LHC strongly supports our current view of fundamental 
particle interactions, the Standard Model, 
it is disturbing that our standard cosmological model still requires at 
least two more components whose nature is
at the moment completely unknown: these are Dark Matter (DM) and 
Dark Energy. In addition, recent cosmological data suggests even a 
third component, Dark Radiation, a weakly interacting relativistic
particle contributing to the cosmological radiation density like a neutrino. 
Besides the missing explanation for this ``dark sector'', a
second shortcoming of the Standard Model is its omittance of a quantization of gravity.
A worthwhile and enormous effort is being made to gain an 
understanding of these tantalizing mysteries of nature, most prominently at the high 
energy frontier. However, only a {\it comprehensive} search for physics beyond the
Standard Model can make sure that we do not overlook the answers 
to the questions we have to nature.

In particular, new particles may have eluded our experimental searches
so far not only if they are very massive,
but also easily if they are rather light but weakly coupled \cite{Jaeckel:2010ni,Ringwald:2012hr}.

For this reason, in this document, we propose an unprecedented 
search for weakly-interacting slim particles (WISPs) in a 
LSW setup,  using
high laser intensities as a lever arm. 
LSW makes use of the fact that WISPs coupled to photons
can be produced inside an optical
resonator (``generation region''). Subsequently
WISPs can traverse a barrier (``wall'') which is opaque to
photons due to the WISPs' feeble interaction with the barrier. 
Thereupon some WISPs can be reconverted into photons behind the
barrier (``regeneration region''). Thus, via WISPs,
light could seemingly ``shine through a wall''\footnote{Note
that the ``Standard Model background for LSW'' due to neutrinos or gravitons
is negligible for the proposed setup \cite{Ahlers:2008jt}.}.

We herein build on the pioneering
experience with the ALPS-I experiment 
\cite{Ringwald:2003nsa,Ehret:2007cm,Ehret:2010mh,Ehret:2009sq},
which still remains the currently most sensitive LSW-setup  \cite{Redondo:2010dp}
worldwide, see \cite{Ehret:2010mh}. Other optical LSW setups included
LIPSS \cite{Afanasev:2006cv,Afanasev:2008jt} at Jefferson Lab, GammeV \cite{Chou:2007zzc,Steffen:2009sc} at Fermilab, OSQAR
\cite{Pugnat:2007nu,Schott:2011fm} at CERN as well as a LSW setups with BMV \cite{Robilliard:2007bq,Fouche:2008jk} and BFRT \cite{Cameron:1993mr}
 (see also Tab.~\ref{tab:magnets2} for the respective magnetic lengths).
Note that for OSQAR, the evaluation of data is not yet completed \cite{schottprivate}. 
Also, a proposal for a consecutive experiment at Fermilab, ``REAPR'' has been made \cite{Mueller:2010zzc}.

In a nutshell, the enhanced sensitivity of ALPS-II (three orders of magnitude in comparison to ALPS-I) will
be due to the increased magnetic length, an optical resonator on the regeneration side and 
the introduction of advanced detector technology.

As argued below, the proposed ALPS-II experiment might thus contribute to our answering the above 
mentioned
important fundamental physics questions.
ALPS-II will be sensitive to uncharted parameter ranges of several WISPs, 
in particular to ``axion-like particles'' (ALPs), 
hidden photons (HPs), minicharged particles (MCPs) and certain scalar fields
of massive gravity theories.
The theory case for these particles and the corresponding ALPS-II sensitivity is outlined 
below.

Pertaining to experiments which exploit the WISPs' coupling to photons, three major 
search strategies are implemented today  \cite{Sikivie:1983ip}:
Firstly, so-called Haloscopes work under the assumption that axions, ALPs or HPs constitute a large 
fraction of the Dark Matter halo also present in our 
solar system. If this holds true,
a sizable number of Dark Matter  axions, ALPs or HPs are present within state-of-the-art 
haloscope detectors, making WISPs detectable 
through a WISP-photon conversion process.
Secondly, so-called Helioscopes make use of the WISP particle flux which should reach us 
from the sun. The most prominent example
for such a setup is the ``Cern Axion  Solar Telescope'' (CAST).
Thirdly, light-shining-through-walls setups, as also proposed in this document, 
have become an indispensable tool in the search 
for WISPs.
Although they are less sensitive than Haloscopes and Helioscopes at first sight, 
they have at least two different advantages 
which make them indispensable:
They are particularly flexible in the sense that they maintain full access and 
control over the WISP production and 
regeneration regions.
For example, controlling and altering the laser polarization as well as 
the magnetic field strength
often provides access to more particles species. Also, control
over the production side eventually enables the locking of a second optical cavity 
on the regeneration side. This will 
eventually 
make the proposed LSW setup ALPS-II more sensitive than today's best existing 
helioscopes in the lower mass region.
In addition, note that LSW setups are sensitive whether or not WISPs are only a 
fraction of Dark Matter or no Dark Matter 
at all.

\subsubsection{The case for axions and axion-like particles}
\begin{figure}
\centering
\includegraphics[width=1\textwidth]{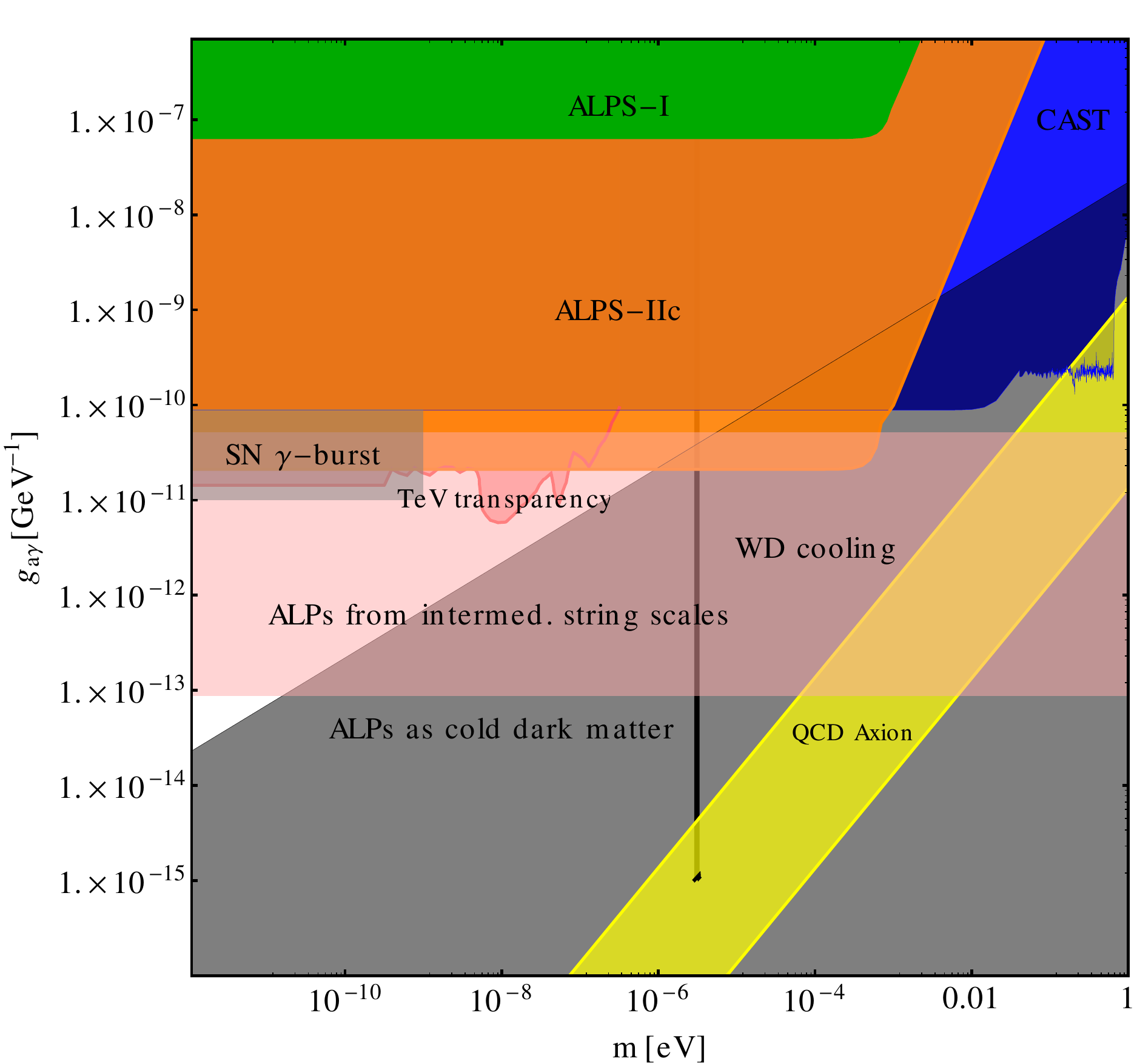}
\caption{Schematic overview of the sensitivity reach of the final stage of ALPS-II, ALPS-IIc depicting its scientific 
impact (The detailed sensitivity
range of ALPS-IIc is given in 
Fig.~\protect\ref{fig:ALPsvac}.). 
As a yellow band, generic QCD axion models are indicated.
For comparison, the ALPS-I results (in green, max. sensitivity $g_{a\gamma}\gtrsim 7 \times 10^{-8} \mathrm{GeV}^{-1}$) and the 
currently most stringent bounds 
on ALPs in this mass region from the CAST helioscope \cite{Arik:2011rx} (in blue) are shown.
As visible, ALPS-IIc (in orange, max. sensitivity $g_{a\gamma}\gtrsim 2 \times 10^{-11}\mathrm{GeV}^{-1}$) will surpass the CAST bounds in the 
lower mass region and 
tackle parameter regions in which ALPs are motivated by fundamental particle physics questions (cold Dark Matter candidates, in gray, and predictions
from string theory around $g_{a\gamma}\sim 10^{-12}$~GeV$^{-1}$, not colored) as well as astrophysical hints: TeV transparency (red, 
fiducial region, cf. \cite{Meyer:2013pny})
and WD cooling (light red band)).
In addition the parameter region excluded from
the 1987a super-nova burst is indicated. The predictions from theory and astrophysical considerations
are generally certain within about an order of magnitude. Note that the parameter range between $\mu {\rm eV} \lesssim m \lesssim 1$meV 
is particularly interesting in the context of axion Dark Matter, and currently
most successfully probed by ADMX  \cite{vanBibber:2013ssa} (cf. black line).
ALPS, on the other hand, can be Dark Matter in a much wider mass range. For a recent comprehensive overview of the ALPs parameter space,
see \cite{Hewett:2012ns,Arias:2012mb}.
}
\label{fig:ALPshints}
\end{figure}

Besides the Higgs particle, the most established proposal for a fundamental spin-0 particle 
in nature is the axion
\cite{Wilczek:1977pj,Weinberg:1977ma}. As a consequence
of the spontaneous breakdown of the so-called Peccei-Quinn (PQ) \cite{Peccei:1977hh} symmetry, 
it provides the most viable solution to the strong CP problem.
In a nutshell, evidence of the axion particle would explain why the observed CP-violation 
in the sector of strong interactions
is off by at least 10 orders of magnitude from expectation.
The mechanism which gives rise to the axion can be generalized to generic (pseudo-) 
scalars $a$ coupled weakly to two-photons
$\sim g_{a \gamma}$, so-called axion-like particles (ALPs).
For axions, the relation between their mass and coupling is given through the color anomaly of 
the PQ symmetry,
cf. the yellow band in Fig.~\ref{fig:ALPshints}. Thus, intrinsically,
for typical axion models this mass-coupling relation is such that only a small part of its parameter 
space could be searched yet.

Generic ALPs, however, do not fall under the mass-coupling relation of axions, and thus 
can be realized in a much wider parameter space, see Fig.~\ref{fig:ALPshints}.
Like the axion, ALPs can be realized as pseudo-Nambu-Goldstone bosons of symmetries broken 
at very high energies.  
Very strong motivation for their existence indeed comes from string theory
which tries to tackle the Standard Model's shortcoming of omitting gravity. 
Generically
in these theories,
a rich spectrum of light (pseudo-)scalars 
with weak couplings arises.  
Of particular relevance to the physics case of ALPS-II 
are intermediate string scales $M_{\mathrm{s}}\sim 10^{9}$ to $10^{12}$GeV, which can contribute to the 
natural explanation of several hierarchy problems in the SM. 
In these theories, ALPs generically exhibit a coupling to photons which lies in a 
parameter range that is accessible to the final stage of the ALPS-II experiment, ALPS-IIc:
The ALP coupling to two-photons is $g_{a \gamma} \sim \frac{\alpha}{2\pi} \frac{c}{f_a}$, 
where $f_a$ is the symmetry
breaking scale, $\alpha$ the fine structure constant, $c$ is a constant of $\mathcal{O}(1)$ and $f_a\sim M_{\mathrm{s}}$. Thus
ALPs from intermediate string scales can very well occur in the 
$g_{a \gamma} \sim10^{-12}$ GeV$^{-1}$ regime \cite{Cicoli:2012sz},
as sketched in Fig.~\ref{fig:ALPshints}.
Unlike for axions, for axion-{\it like} relation the ALP-photon
coupling to couplings to other particle species is
not straightforward\footnote{While it is true that for Goldstone bosons
the coupling to the respective particle types $x$ will obey
$g_{ax} \sim C_{ax}/f$, with $f$ the symmetry breaking scale,
however, the coefficient $C_{ax}$ is model dependent and could even
be 0 for some $x$.}.
It is important
to follow therefore complementary search strategies. While ALPS-II is only
sensitive to the ALP coupling to photons, other experiments
and astrophysical arguments probe also
(pseudo-) scalar couplings, e.g., to nucleons \cite{Raffelt:2012sp} or neutrons \cite{Dubbers:2011ns}.

However, axions and ALPs can not only be a low-energy window to new physics in the ultraviolet, 
but they are also well-motivated Dark Matter candidates.
Generally, one of the most prominent candidates for Dark Matter are ``weakly interacting massive particles'' (WIMPs) whose drawing power
is twofold: WIMPs at the weak scale provide the right relic abundance and new particles at 
these scales are favored in particular
by supersymmetry.
A lot of effort has thus be put into detecting WIMPs as Dark Matter directly
as well
as indirectly at, e.g., the LHC. 
However, although no final conclusive statements can yet be made: The fact that the LHC
as well as the currently
most sensitive direct search Xenon100
have found no such particle, posing heavy restrictions 
on some popular models \cite{Buchmueller:2012hv}, should remind us that other 
well-motivated Dark Matter 
candidates do exist.
In particular, axions and other WISPs constitute viable Dark Matter candidates. 

For axions, the Dark Matter option has been mainly explored since
the 80s particularly by the ``axion dark matter eXperiment'' (ADMX). With their ongoing effort and the contribution
of setups with complementary mass range \cite{Baker:2011na,Irastorza:2012jq,Horns:2012jf},
well-established methods are available to explore
the cold Dark Matter axion window. For other WISPs, the viable Dark Matter
parameter space is largely unexplored as recently 
demonstrated \cite{Arias:2012mb}.
Note that if produced non-thermally through the misalignment mechanism, even very light WISPs can be cold
Dark Matter.
Again, it is noteworthy that ALPS-II can explore
in part ALP Dark Matter options as indicated in Fig.~\ref{fig:ALPshints},
and might therefore complement upcoming experiments dedicated to WISPy Dark Matter detection.

In addition to the motivations for ALPs that arise from UV-completions of the Standard
Model and their possible connection to Dark Matter,
two hints from astro-particle physics strengthen the physics case for 
searching axion-like particles ALPS-II even further.

The first hint pertains to the propagation of cosmic gamma rays 
with energies above\footnote{Note
that renormalization group studies formally justify a comparison of ALPs constraints arising 
from different momentum regimes \cite{Eichhorn:2012uv}. However,
in any case there is strong motivation for purely laboratory
measurements, due to some inevitable uncertainties inherent to bounds from astrophysical sources \cite{Jaeckel:2006xm}.}
$\mathcal{O}(100)$GeV.
Even if no absorbing matter blocks the way of these high energy photons,
absorption must be expected as the gamma rays deplete through
electron-positron pair production through interaction with extragalactic background light. 
However, the observed energy spectra do not seem to match the absorption feature inferred from this argument~\cite{Horns:2012fx}.
Axion-like particles could provide a resolution to this puzzle. 
Here, the anomalous transparency can be explained if photons convert into ALPs in astrophysical magnetic fields. The ALPs then
travel unhindered due to their weak coupling to
normal matter. Close to the solar neighborhood, ALPs could then be reconverted to high-energy photons.
A sizable number of authors is intensely studying this resolution to the transparency problem,
see, e.g., \cite{DeAngelis:2007dy,SanchezConde:2009wu,Horns:2012kw} to name only a few.
Typically,
ALPs related to the transparency hint are predicted to lie in the $g_{a \gamma} \sim10^{-11}$ to $10^{-12}$ GeV$^{-1}$ coupling-region. 
In  Fig.~\ref{fig:ALPshints} it can be seen that ALPS-II
can explore a sizable parameter space related to this hint.
Note that the favored parameter regions related to this hint can be covered with ALPS-II almost entirely \cite{Meyer:2013pny}.

The second hint pertains to the longstanding puzzle of the anomalously large cooling rate of white dwarfs (WD)
which receives additional support by the most recent 
studies of the WD
luminosity function~\cite{Isern:2012ef} and in some cases to the observed decrease in pulsation period of these objects~\cite{Corsico:2012ki}. 
The cooling excess can be attributed to axions with $f_a\sim 10^{8}$ to $10^{9}$GeV if they couple to electrons. In a generic case the axion 
(or ALP) will also have a photon coupling. In Fig.~\ref{fig:ALPshints}, the corresponding target region\footnote{The
broadness of the target region is due to the model-dependent factors which translate the ALP-electron to an ALP-photon
coupling. With further studies, a sharpening of the target region should occur.} is depicted, 
labeled ``WD cooling''.

A number of additional astrophysical observations not indicated
in Fig.~\ref{fig:ALPshints} have been considered
to search for signatures of ALPs or -- in the absence
of a signal -- constrain the parameters. 
Observations of linearly polarized emission from 
magnetized WDs \cite{2011PhRvD..84h5001G} and changes of the linear polarization
from radio galaxies (e.g., \cite{2012PhRvD..85h5021H})
 provides complementary approaches to search for signatures of ALPs. 
The current limits are close to $g_{a\gamma}\sim 10^{-11}$~GeV$^{-1}$ albeit
with uncertainties related to the underlying assumptions.

Note that of course upcoming data from Cherenkov or space-bound telescopes might very well 
further strengthen or -- 
on the contrary --
weaken the case for astrophysical ALPs. However, it is in any case mandatory to 
find definitive proof or disproof 
for these theories in the 
laboratory as only there uncertainties 
in experimental parameters of WISP production and identification can be optimally controlled.

To put this into the right context we emphasize that as
a rare exception to astrophysical claims in general, testing scenarios of
the ALP solution to the transparency anomaly and the
anomalous WD cooling is (astonishingly) possible in
a small laboratory setup.

\subsubsection{Motivation for hidden photons and minicharged particles}

Hidden sectors, i.e., particles characterized by very weak interactions with ordinary matter, 
are a very generic feature of field- and string-theory 
completions of the Standard Model. 
For example, a hidden sector is a viable instrument to realize supersymmetry breaking.
Hidden photons, i.e., gauge bosons of an extra U(1) gauge group, are natural ingredients 
of these hidden sectors.
Most prominently, hidden photons couple to the visible sector via kinetic 
mixing~\cite{Holdom:1985ag,Abel:2008ai,Cicoli:2011yh}, 
parameterized in the following by $\chi$. 

Even if in a fundamental theory $\chi=0$, we know from prominent 
examples
within the 
Standard Model, 
that effectively $\chi\neq0$ can be easily induced
through integrating out degrees of freedom at higher 
energy scales\footnote{One such example is the renowned Heisenberg-Euler effective action \cite{Heisenberg:1935qt}
leading to effective photon self-interaction through integrating out fermionic degrees of freedom.},
e.g., degrees of freedom arising from string extensions \cite{Abel:2008ai,Goodsell:2009xc,Cicoli:2011yh}. 
The $\chi\sim 10^{-8}$ to $10^{-9}$ region, accessible to the first stages of ALPS-II 
is naturally realized in string compactifications with intermediate string scales \cite{Goodsell:2009xc}.
Also, like ALPs, HPs are a viable cold Dark Matter candidate, see \cite{Nelson:2011sf,Arias:2012mb}.

In addition to hidden photons, the hidden sector naturally can also contain matter (scalars or Dirac fermions)
with fractional electric charge, called 
minicharged particles \cite{Gies:2006ca,Abel:2006qt}. 
Most prominently, they can emerge in theories which contain a hidden photon. 
Search for these particles is crucial, as MCP searches provide an alternative observational window 
to hidden sectors and in particular can provide insight 
if the hidden photon turns out to be massless
 and thus not directly traceable.
Here, ALPS-II will chart MCP parameter space which is so-far only indirectly accessible through cosmological arguments.

\subsubsection{Further scientific impact of ALPS-II}

Interestingly, although the new physics models listed above are the most well-reviewed and thus discussed great in detail here, ALPS-II would be
sensitive
to other, very topical models, which aim at resolving the puzzle of Dark Energy. Only recently, it has been shown that theories of ghost-free massive
gravity can indeed be constructed, leading to a broad phenomenology for new scalar degrees of freedom  \cite{deRham:2012az}.
In particular scalar fields with very light mass are a consequence of using massive gravity to explain 
the acceleration of the expansion of the universe and thus Dark Energy.
For some of these models, it has been pointed out that ALPS-I already provides the best constraints \cite{Brax:2012ie} which would be even further
improved at ALPS-II or, even better lead to the discovery of Dark Energy particle candidates in a laboratory.
However, as this is still a young field, where the model-building roads are not as firmly paved as for WISPs,
we will not consider these scenarios here in more detail. 

We still want to strongly emphasize that due to these 
developments, the future scientific impact of ALPS-II could very well be even wider than described in the previous sections.

\subsection{Discovery potential at the three stages of ALPS-II \label{sec:discovery_pot}}

ALPS-II is subdivided into three stages, see Sect.~\ref{subsec:tdr:overview-stages} for details.
In the following we focus on a theoretical sensitivity analysis for these stages.
The installation of the magnet string, required to detect ALPs and MCPs will be realized only in the final stage, ALPS-IIc.
In preparation for this final stage, two stages without magnets at different optical cavity lengths will be implemented.
ALPS-IIa ($L=10$m) and ALPS-IIb ($L=100$m) will thus be designed to search for hidden photons.

\subsubsection{Experimental aims in the search for ALPs at ALPS-IIc \label{sec:alps_osci}}

ALP-photon oscillations \cite{Raffelt:1987im} occur through a pseudo-scalar coupling 
term $\mathcal{L}_{\mathrm{int, P}}
= g_{a \gamma} a \vec{E} \vec{B}$, here enabling the 
conversion of laser photons characterized by $\vec{E}$ into 
ALPs $a$. At ALPS-IIc, this happens 
in an external magnetic field $\vec{B}$ of a HERA dipole magnet string, where the laser photons are polarized in parallel 
to the magnetic field\footnote{For scalar ALPs, accordingly 
$\mathcal{L}_{\mathrm{int, S}} = g_{a \gamma} a \ (\vec{E}^2-\vec{B}^2)$ and only
the orthogonal mode of the laser
light couples to the external field. For ALPS-IIc, measurements in both polarization modes are foreseen. 
We emphasize that in the following all statements about the 
prospective sensitivity for pseudo-scalars apply equally well for the case of scalars.}. The theory of
these oscillations for light with frequency $\omega$ has been
discussed widely in the literature,
see \cite{Redondo:2010dp} for a recent overview with a focus on LSW setups.

For the envisaged measurements at ALPS-IIc, it is crucial to 
realize that for a string of $N$ HERA magnets,
field-free regions are present at the magnet junctions.
Thus the magnetic field is segmented across the entire magnet string.
This slightly modifies the formula for the oscillation probability valid for
ALPS-I  \cite{Ehret:2010mh}. 
The corresponding implications for the situation with periodic field-free intersections
has been discussed in detail in \cite{Arias:2010bh},
and one finds that the oscillation probability for photon $\leftrightarrow$ ALP 
oscillations becomes:
\begin{equation}
P_{\gamma\leftrightarrow a}=\frac{\omega}{\sqrt{\omega^2-m^2}}
\frac{4 \ g_{a \gamma}^2 \omega^2 B^2}{M^{4}} \sin\left(\frac{M^2 L}{4 \omega N}\right)^2
\frac{\sin\left(\frac{M^2 N (L/N+\Delta)}{4 \ \omega}\right)^2}
{\sin\left(\frac{M^2 (L/N+\Delta)}{4 \ \omega}\right)^2} \ ,
 \label{eq:ALPs_osci_prob}
\end{equation}
which reduces to the well-known expression for single magnet setups for $N=1$.
Also, we have set $\hbar=c=1$.

In a notation analogous to \cite{Ehret:2010mh}, 
we have introduced the quantity $M^2=(m^2+2\omega^2 (n-1))$, where $n$ is the index of refraction.
In addition, $\Delta$ defines the 
field-free gap-length,
where each single HERA dipole has a magnetic length of $l=8.83$m and 
the total magnetic length is $L=Nl$.
For ALPS-IIc, the gap-length is $\Delta=0.936$m and we envisage the usage 
of $N=10$ magnets on each side of the light-blocking barrier
at field strengths
of $B=5.3$T. Details on the setup of the magnet string will follow in Sect.~\ref{sec:tdr:magnets}.

Note that whilst increasing $\omega$ formally increases the conversion 
probability for WISPs,
an appropriate compromise has to be found regarding the photon number flux in front of the wall and the aperture requirements.
Hence photon energies beyond the optical regime are disfavored mainly  due the lack of appropriate high finesse
resonators. For example LSW with X-rays has been performed already
\cite{Battesti:2010dm,Inada:2013tx}  but the sensitivity is much smaller than the experiment proposed here, 
except in the high mass range.
In the optical photon energy range, experience at ALPS-I (operated at 532\,nm) 
has shown that the mirrors could stand high power intensities only for some 10 hours.
We suspect that the single-photon energy of 2.33\,eV is sufficient to break chemical bounds which 
in turn damage the mirror coatings.
Such effects have not been observed for 1064\,nm photons, where much expertise and optical component qualification 
procedures exist in the gravitational wave interferometer community. 
For this reason, we have switched to $\omega=1.16$~eV for ALPS-II.
At even lower energies, LSW (with resonant enhancement) is also 
possible, see ongoing experiments at CERN \cite{Betz:2012tp,Jaeckel:2007ch}.
However, this option is incompatible with the aperture available 
in accelerator dipole magnets limiting the sensitivity in ALP searches.

\begin{table}
\begin{tabular}{|l|c|c|c|c|c|}
\hline
Parameter & Scaling & ALPS-I & ALPS-IIc & Sens. gain \\ [1pt] \hline
Effective laser power $P_{\rm laser}$ & $g_{a\gamma} \propto P_{\rm laser}^{-1/4}$ & 1\,kW & 150\,kW & 3.5\\[1pt] \hline
Rel. photon number flux $n_\gamma$ & $g_{a\gamma} \propto n_\gamma^{-1/4}$ & 1 (532\,nm) & 2 (1064\,nm) & 1.2\\[1pt] \hline
Power built up in RC $P_{\rm RC}$ & $g_{a\gamma} \propto P_{reg}^{-1/4}$ & 1 &  40,000 &  14\\[1pt] \hline
 $BL$ (before\& after the wall) & $g_{a\gamma} \propto (BL)^{-1}$ & 22\,Tm & 468\,Tm  & 21\\[1pt] \hline
Detector efficiency $QE$ & $g_{a\gamma} \propto QE^{-1/4}$ & 0.9 & 0.75 & 0.96\\[1pt] \hline
Detector noise $DC$ & $g_{a\gamma} \propto DC^{1/8}$ & 0.0018\,s$^{-1}$ & 0.000001\,s$^{-1}$ & 2.6\\[1pt] \hline
Combined improvements & &  &  & 3082\\[1pt] \hline
\end{tabular}
\caption{
Parameters of the ALPS-I experiment in comparison to the ALPS-II proposal. The second column shows 
the dependence of the reachable ALP-photon coupling on the experimental parameters. The last column lists the approximate 
sensitivity gain for axion-like particle searches compared to ALPS-I. The main sensitivity gain is due to the enhanced magnetic length arising from the
2$\times$10 HERA-dipole
magnet string, cf.~\protect\Tabref{tab:magnets1} for details.
A sizable additional gain arises from the 
installation of a regeneration cavity (RC), cf.~\protect\Sectref{sec:tdr:laser} for details.
A further upgrade comes also from the detection side. The numbers given in
this table are for a transition edge sensor detector, as this
is the detector which is expected to be used in the final version of all
three stages of ALPS-II, 
see \protect\Sectref{chap:tdr:detector} 
and \protect\Tabref{tab:qfactor} 
For hidden photons, there is no gain from the magnetic field. 
Thus the sensitivity gain follows as above except for the factor coming from $BL$
and amounts to 147 for hidden photons.
\label{tab:param}
}
\end{table}

\begin{figure}
\begin{center}
\includegraphics[width=1\textwidth]{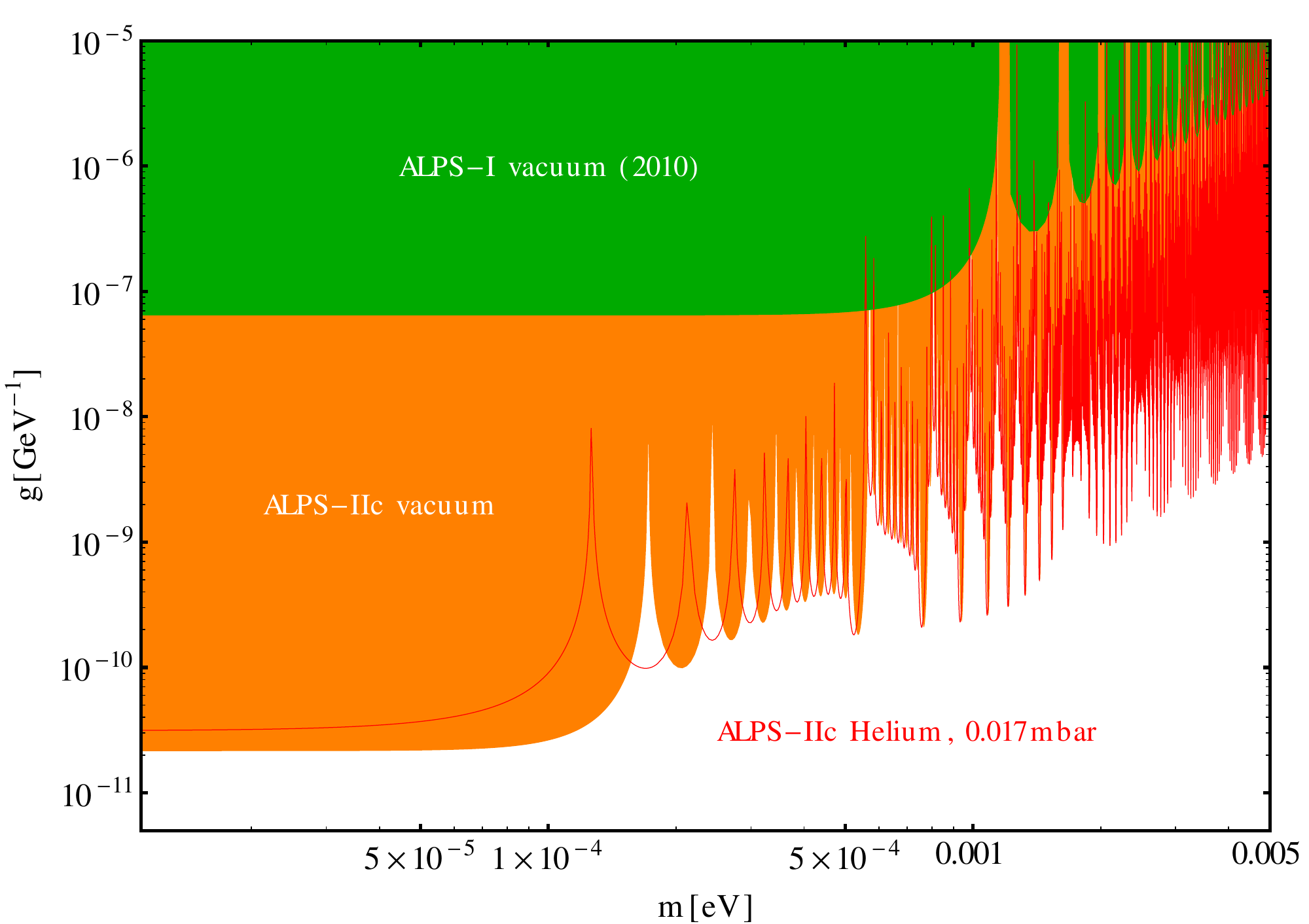}
\end{center}
\caption{Prospective sensitivity reach for axion-like particles in the final
stage of the ALPS-experiment, ALPS-IIc.
The uppermost, green shaded area gives the currently most sensitive laboratory bounds
on ALPs as reported in \cite{Ehret:2010mh}
for comparison. Below, in orange, we give the expected sensitivity reach of 
ALPS-IIc, surpassing the currently most sensitive
laboratory limits by more than three orders of magnitude, see main text for 
details. In addition we give exemplarily
the sensitivity reach with  Helium gas at a fixed pressure of 0.017mbar.
With varying pressures, the Helium insertion allows to cover the sensitivity
gaps of the vacuum measurements and thus scan the entire accessible parameter range.
For clarity, parameter regions favored by theory and astrophysics hints are shown
in \protect\Figref{fig:ALPshints} 
only.
}
\label{fig:ALPsvac}
\end{figure}

To arrive at the prospective ALPS-IIc exclusion bounds shown in Fig.~\ref{fig:ALPsvac}, 
we infer the expected performance based on
the results
of ALPS-I. Note, that for an LSW-configuration which relies on photon-ALP-photon conversions, 
it is the square of \Eqref{eq:ALPs_osci_prob} which determines the experimental sensitivity. 
In Tab.~\ref{tab:param}, 
the respective upgrades are listed and their impact on the testable photon-ALP 
coupling $g_{a\gamma}$ is shown to culminate in
a total sensitivity gain of about 3082. Thus, in the most sensitive low-mass region
$m\lesssim 10^{-4}$eV the explored coupling values at ALPS-I 
$g_{a\gamma}\gtrsim 7 \times 10^{-8}$ GeV$^{-1}$ will be extended to
 $g_{a\gamma}\gtrsim 2 \times 10^{-11}$ GeV$^{-1}$ at ALPS-IIc.
For a recent comprehensive overview
of the ALP parameter space also at larger masses, see, e.g., \cite{Hewett:2012ns,Arias:2012mb,Ringwald:2012hr}.

As visible in Fig.~\ref{fig:ALPsvac}, for masses above $m\gtrsim 10^{-4}$eV, the sensitivity 
curve adopts an oscillatory behavior.
Physically this corresponds to phase-mismatches in the ALP production. This is of course unwanted, 
as ALPs could very well just 
hide in these regions where the sensitivity is decreased.
In order to fill these gaps in sensitivity, one has the possibility to shift the effective
photon mass by the insertion of gas
into the beam tube. This method has already proven to be very successful at ALPS-I using Argon \cite{Ehret:2010mh}.

For a single magnet configuration $N=1$, it is easy (and analytically possible) to spot the
minima of the transition probability in vacuum.
The gas index of refraction which optimally closes the gaps
can then be computed by minimizing the expression in \Eqref{eq:ALPs_osci_prob} with respect 
to $(n-1)$, see \Eqref{eq:tan_formula}
below. For $N>1$, 
all gaps will be covered ideally through a parameter scan 
with different gas pressures through the insertion of Helium, as detailed in Sect.~\ref{sec:tdr:vacuum}.

In Fig.~\ref{fig:ALPsvac}, we plot the transition probability for vacuum  
and exemplarily for one single Helium pressure corresponding to 0.017mbar 
(see paragraph including \Eqref{eq:pressure_vac} for details on the index of refraction to pressure relation).

To conclude, ALPS-IIc charts thus-far unexplored parameter space, surpassing even current helioscope limits at low masses
and, for the first time, simultaneously tackles 
regions which are favored by theory
(Dark Matter, ALPs from intermediate scale string theory) 
as well as astrophysical hints (WD cooling, TeV transparency), cf. Fig.~\ref{fig:ALPshints}.

\subsubsection{Discovery potential for HPs and MCPs at ALPS-II}

\subsubsection*{Projected Sensitivity in the Search for Hidden Photons \label{sec:th_hps}}

The theory of photon-hidden photon oscillations differs from the case of photon-ALP-oscillations most prominently by the fact
that former
also occur without an external magnetic field (both particles have equal spin). 
If HPs have a mass, kinetic mixing
behaves as mass mixing and thus photon-HP 
oscillations occur (in analogy to oscillations among different neutrino flavors)~\cite{Ahlers:2007rd}. 
This leads to the disappearance and regeneration of photons as they propagate in vacuum and makes their discovery particularly
easy in light-shining-through a wall setups.

Photon-HP oscillations in LSW-experiments are reviewed in \cite{Redondo:2010dp} and  
the oscillation probability for one oscillation (in either direction) $\gamma\leftrightarrow \gamma'$ is given as \cite{Ahlers:2007qf}
\begin{equation}
 P_{\gamma\leftrightarrow \gamma'}\simeq 4 \chi^2 \frac{m_{\gamma'}^4}{M_{\gamma'}^4} \sin^2 \left(
 \frac{M_{\gamma'}^2 L}{4\omega}\right)\label{eq:hp_osci}  \ ,
\end{equation}
where, in analogy to the definition for ALPs $M_{\gamma'}^2=(m_{\gamma'}^2+2\omega^2 (n-1))$.
Again, for the LSW-setup of ALPS-II, the square of \Eqref{eq:hp_osci} is the relevant quantity.

\begin{figure}
\begin{center}
 \includegraphics[width=1\textwidth]{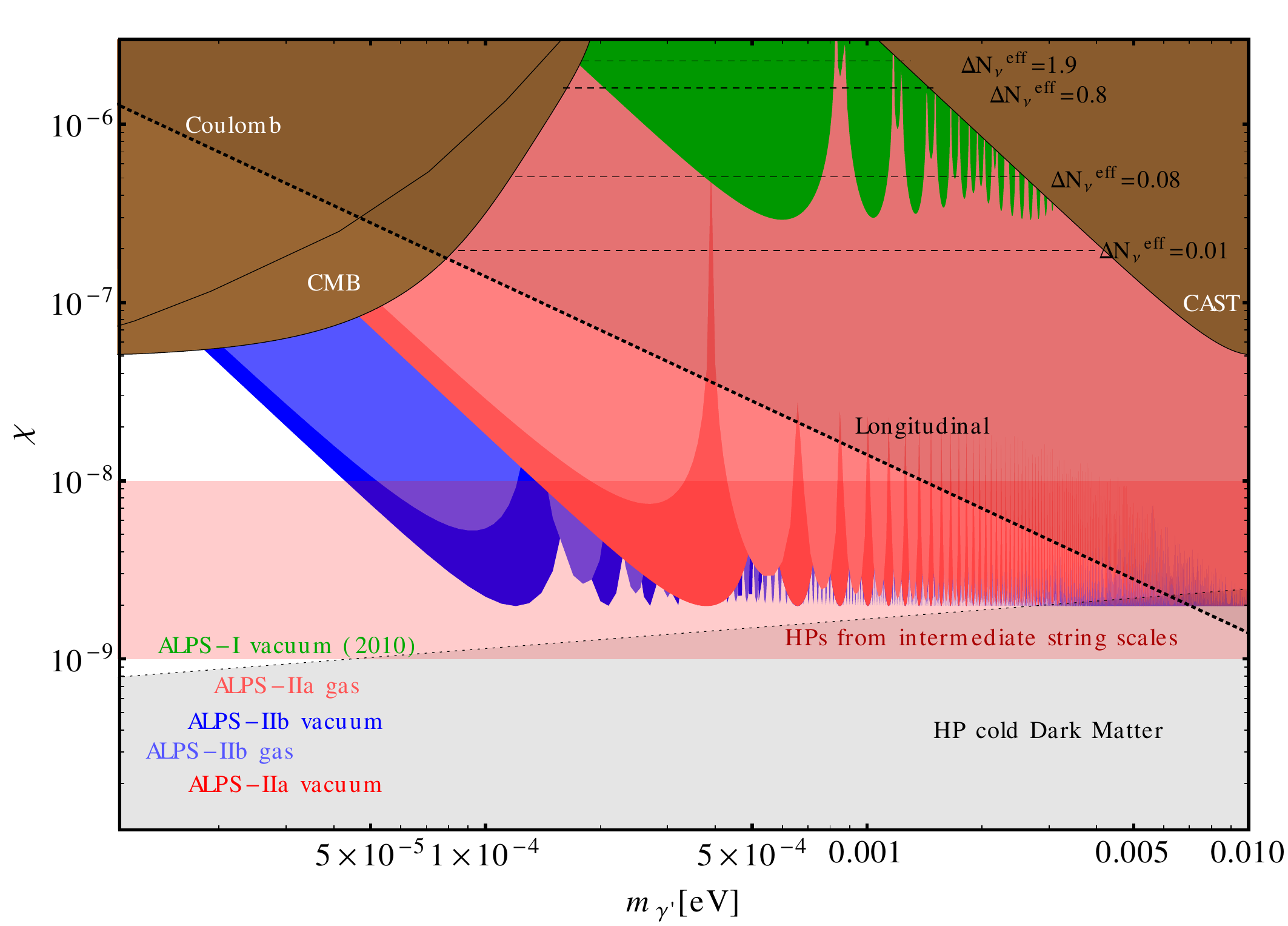}
\end{center}
\caption{Coupling-mass plane for hidden photons. The shaded areas indicate the prospective 
 parameter reach for HPs at ALPS-IIa with regeneration cavity and ALPS-IIb in setups with vacuum and gas-filled tubes, respectively. 
 The uppermost, green shaded oscillatory curve denote the results
 obtained by ALPS-I in the vacuum case for comparison. 
 Below, in light and dark red, the parameter reach for ALPS-IIa with regeneration cavity and gas/no-gas is indicated. From there,
 the light and dark blue areas to the left denotes the reach of ALPS-IIb, with and without gas, respectively.
 For the parameter choices with gas, see main text.
The black dashed lines enclose the HP Dark Radiation hint \cite{Jaeckel:2008fi}, however
recent analyses \cite{An:2013yfc,Redondo:2013lna} exclude the parameter range above the line labeled
``Longitudinal, thus disfavoring this interpretation.
The gray shaded lower area indicates where HPs could be cold DM \cite{Arias:2012mb}.
Note that the $\chi\sim 10^{-8}$ to $10^{-9}$ region is naturally realized in string compactifications 
with intermediate string scales \cite{Goodsell:2009xc} (beam-like red-shaded area).
The brown areas in the upper right and left corners denote exclusion bounds from other experiments, see text for details.}
\label{fig:hpgas}
\end{figure}

Figure \ref{fig:hpgas} shows the prospective exclusion limits obtained in the various ALPS-II stages.
Note that at each single stage,
ALPS-II will chart thus-far unexplored HP parameter space, implying that at each stage a particle discovery is indeed possible.
The black vertical dashed lines in Fig.~\ref{fig:hpgas} refer to the Dark Radiation hint  \cite{Jaeckel:2008fi}
derived from CMB data
\cite{Komatsu:2010fb,Dunkley:2010ge}. Although the newest Planck results \cite{Ade:2013zuv} still
contain indication for an extra
relativistic degree of freedom in the CMB (although reduced to $N_{\rm eff} = 3.36 \pm 0.34$ at 68 \% confidence), the hidden photon is 
unlikely to be under the viable candidates to account for this phenomenon.
It has been shown that, the total production of hidden photons
in the sun has been underestimated,
since the production of longitudinal modes
has not been taken correctly into account.
Following \cite{An:2013yfc,Redondo:2013lna}, the parameter space
above the line labeled ``Longitudinal'' in Fig.~\ref{fig:hpgas} is already excluded
by the bounds on the energy loss due to hidden photon production in the sun's plasma.
However, one should note that for some models
the bounds labeled ``Longitudinal'' might not apply,
e.g. in models, where the hidden photon decays.
In a situation where, for example a sizable amount of hidden 
photons from the sun 
cannot reach earth due to decay, a laboratory search is indispensable.
This holds true also for limits on HPs from direct Dark Matter
searches \cite{An:2013yua}.

Our measurements will nicely complement the exclusion limits derived from CAST helioscope measurements \cite{Arik:2011rx} as well as regions covered by 
measurements of Coulomb's law \cite{Jaeckel:2010xx} and cosmological implications \cite{Mirizzi:2009iz}. For a recent comprehensive overview
of the hidden photon parameter space, see, e.g., \cite{Hewett:2012ns,Arias:2012mb}.

The prospective exclusion limits for the different vacuum stages of ALPS-II are estimated as follows.
In contrast to the ALPS-I setup\cite{Ehret:2010mh}, infrared lasers with $\omega=1.16$eV will be used and the length of the vacuum tubes $L$ is 
increased\footnote{Note that for ALPS-I, the length of the vacuum tubes was slightly different in the production- and regeneration-regions,
leading to a ``non-saturation'' of the exclusion limits for at $m_{\gamma'}\gtrsim 10^{-3}$eV. 
E.g., this led to a minimum of an oscillatory envelope function at about $m_{\gamma'}\approx 2\times 10^{-3}$eV,
cf. the green shaded area in \ref{fig:hpgas}.
For ALPS-II however, the length of the vacuum tubes on the generation and regeneration side will be equal
and such unwanted features are avoided.}.
Thus, quite generally, as the argument of the sine depends on 
$L m^2/\omega$, the set-in of the oscillatory behavior within \Eqref{eq:hp_osci} is shifted to lower mass-values for
enhanced lengths.
The frequency of our laser sets also the mass scale up to which our experiment is sensitive to hidden photons.
At higher masses, HPs are well probed through helioscopes and their production in the 
sun and horizontal branch stars, see, e.g. Fig. 6-3 in \cite{Hewett:2012ns}.
The HP coupling to electrons is particularly well accessible at accelerators, see, e.g. Fig 6-2
of \cite{Hewett:2012ns}
for an overview.

Quantitatively, the expected increase in the sensitivity to $\chi$ can be also estimated from Tab.~\ref{tab:param}. 
The crucial difference to the situation with ALPs is that the ``magnetic length'' $BL$ will not give an enhancement factor
as photon-HP oscillations do not rely on external fields.
Rather, the main sensitivity gain will arise from the use of a regeneration cavity.
Note that the resonant regeneration principle has been tested with microwaves in the sub-quantum
regime \cite{Hartnett:2011zd}. An early optical setup is described in \cite{Fukuda}.

For ALPS-IIa with production and regeneration cavity as well as the transition edge sensor, the enhancement factor with respect to $\chi$ is 147.
ALPS-IIb will further enlarge the covered parameter space by the length increase $L=100$m and a corresponding 
shift in the covered mass range to lower masses. These statements are summarized within the Fig.~\ref{fig:hpgas}, showing the ALPS-II sensitivity
in the hidden photon parameter space. 
In addition, ALPS-IIb will begin to tackle the HP cold DM region.
Note that recent \cite{Mizumoto:2013jy}
and upcoming  helioscope searches (SHIPS \cite{Schwarz:2011gu})  are
sensitive in particular for
$m_{\gamma'}>10^{-2}$eV.

As visible from Fig.~\ref{fig:hpgas}, cusps appear in the sensitivity range for HP masses $m_{\gamma'}\gtrsim 10^{-3}$eV,
corresponding to phase-mismatches in the HP production in the vacuum measurements.
These minima lie at points where $m_{\gamma'}^2 L/4\omega= s \pi$, with $s\in \mathbb{N}^{+}$.
A viable method to close these gaps in the sensitivity reach is the variation of the effective photon mass through insertion of gas into the tubes.
This slightly decreases the amplitude of the conversion probability.

To compute index of refraction which optimally closes the gaps, one has to minimize the expression in \Eqref{eq:hp_osci} with respect to $(n-1)$. 
Thus, defining $y =M_{\gamma'}^2 L/(4\omega)$, the transition maxima lie at\footnote{Note that in \cite{Ehret:2010mh}
the theoretical derivation had a minor fault which however has no influence on presented results therein.
} solutions to a condition with infinite but numerable solutions
\begin{equation}
 \tan \left(y \right) =  y \\ \Leftrightarrow y= \{0, 4.49, 7.73... \} \ . \label{eq:tan_formula}
\end{equation}
To cover the lowermost gap in the sensitivity for ALPS-IIa ($L=10$m) and ALPS-IIb ($L=100$m), indices of refraction
of $(n-1)=4.57 \times 10^{-8}$ and  $(n-1)=3.52 \times 10^{-9}$ are optimal, respectively. As $s\rightarrow \infty$, the best indices of 
refraction
adequate to cover the gaps optimally shift to  $(n-1)=5.31 \times 10^{-8}$ and  $(n-1)=4.43 \times 10^{-9}$, respectively.
Thus for ideal coverage of the gaps, for ALPS-IIa we should have  $(n-1) \simeq 5 \times 10^{-8}$ and for ALPS-IIb  $(n-1)\simeq4 \times 10^{-9}$.
This choice is depicted in Fig.~\ref{fig:hpgas}. As can be seen, the gas insertion technique, also pioneered by ALPS-I  \cite{Ehret:2010mh}
will enable to cover the higher mass reach entirely.

To conclude, the HP search at ALPS-II will not only cover a parameter space not accessible by astrophysical considerations,
but also tackle a kinetic mixing region naturally predicted by string compactifications with intermediate string scales
as well as a fraction of the HP DM parameter space.

\subsubsection*{Projected Sensitivity in the Search for Minicharged Particles}

Light-shining-through a wall for minicharged particles at ALPS-IIc comes about as follows:
Photons can be converted into (massless) hidden photons through a virtual MCP loop within an external magnetic field.
Subsequently, the hidden photons can
traverse the barrier as in Sect.~\ref{sec:th_hps} and
can thereafter be reconverted into photons through the intermediate MCP loop \cite{Ahlers:2007rd}.

The corresponding conversion probability for this process reads~\cite{Burrage:2009yz}
\begin{equation}
\label{hpprob1}
P(\gamma \leftrightarrow \gamma^\prime  )\simeq
\left| \frac{m_{\tilde{\gamma}'}^2}{M^2}\right| \left|e^{i k_+ L}-e^{i k_- L}\right|^2 , 
\end{equation}
\begin{equation}
k_\pm = \frac{1}{4\omega}\left(2\omega^2 (n-1)-m_{\tilde{\gamma}'}^2 \pm M^2 \right)(1\pm \chi^2 m_{\tilde{\gamma}'}^2/M^2) , 
\end{equation}
where now $m_{\tilde{\gamma}'}$ is the hidden photon effective mass (accounting for the MCP-induced change in the dispersion relation).
It can be written as $m_{\tilde{\gamma}'}^2=-2\omega^2 \Delta N (\epsilon,B,m_{_{\rm MCP}})$ with
$\Delta N (\epsilon,B,m_{_{\rm MCP}})$ being the complex index of refraction due to minicharged particles of charge $\epsilon$ and mass $m_{_{\rm MCP}}$.
The minicharge $\epsilon$ is related to the kinetic mixing parameter $\chi$ of the hidden photons through
$\chi e_{\rm H} = \epsilon e$, where $e_{\rm H}$ is the  scalar or Dirac spinor coupling in the hidden sector and $e$ the ordinary electromagnetic coupling.
The explicit expression for $\Delta N$  for scalar or Dirac spinor MCPs is given in Ref.~\cite{Ahlers:2007rd}. 

In a scenario where $e_{\rm H}$ matches the coupling in the visible 
sector\footnote{Although probing MCPs in an LSW setup with direct access to the MCP coupling is possible,
this is not an immediate option for ALPS-II, as in this situation a solenoidal
magnetic field configuration is preferred \cite{Dobrich:2012sw,Dobrich:2012jd}.},
ALPS-II can improve the best limits provided by ALPS-I by around two orders of magnitude.
This is due to the fact that the transition probability scales as $B^{2/3}$ in the low-mass limit\footnote{This is a consequence of the
strong-field limit of the polarization tensor, applicable for minicharged particles \cite{Ahlers:2006iz,Karbstein:2011ja}.},
and the sensitivity enhancement factor presented for ALPs (cf. Tab.~\ref{tab:param}) is reduced accordingly.

As a consequence, ALPS-II will chart MCP parameter space which is so-far only 
indirectly accessible through cosmological or astrophysical arguments \cite{Melchiorri:2007sq}.
For a recent comprehensive overview
of the MCP parameter space, see, e.g., \cite{Hewett:2012ns}.

\section{The experiment}
\label{chap:tdr:experiment}

Here we describe the experimental setup starting with a summary of its main
characteristics. For a short summary see also \cite{Dobrich:2012du}.

\subsection{Overview}
\label{sec:tdr:overview}

This section gives a brief overview on the different stages of the
ALPS-II experiment and sketches its operation modes which will allow us
to verify the achieved sensitivity and, eventually,  to pin down the nature
of the WISP responsible for light shining through the wall.

\subsubsection{Stages of the experiment}
\label{subsec:tdr:overview-stages}

It is planned to realize the ALPS-II experiment in three main stages.
This will allow for a thorough development of the required new techniques
without the necessity of large infrastructure efforts (i.e., long magnet
strings) right from the beginning. As we plan to enter new territory in
parameter space as well as in optical setups, the staged approach is also
advantageous because it allows to adapt to unforeseen new developments. We
will show below, that all three stages will provide sensitivities for
different kinds of WISPs beyond present day laboratory experiments
allowing for interesting physics results, see also
\Sectref{sec:discovery_pot}.

It should be stressed that the ALPS collaboration considers the
potential technological risks as manageable as will be shown in this
document, but the staged approach described here is also very well
compatible with DESY's commitments in other larger projects.

\begin{figure}[!tbh]
  \centering
  \subfloat[][ALPS-I]{%
    \label{subfig:alpsI:schematic}%
    \includegraphics{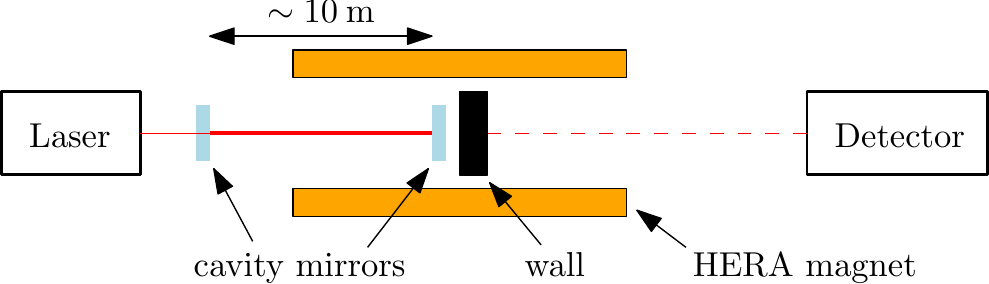}%
  }
  \\[0.5cm]
  \subfloat[][ALPS-IIa]{%
    \label{subfig:alpsII:phases:schematic:a}%
    \includegraphics{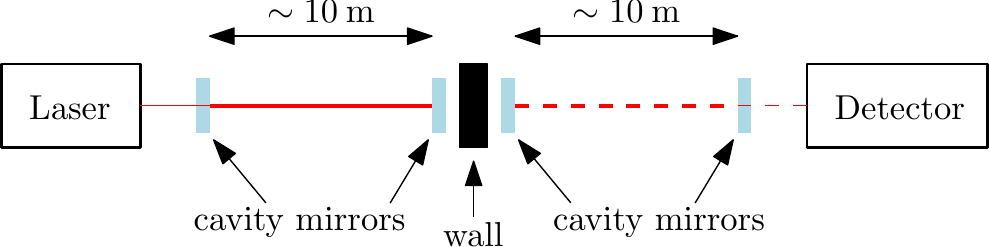}%
  }
  \\[0.5cm]
  \subfloat[][ALPS-IIb]{%
    \label{subfig:alpsII:phases:schematic:b}%
    \includegraphics{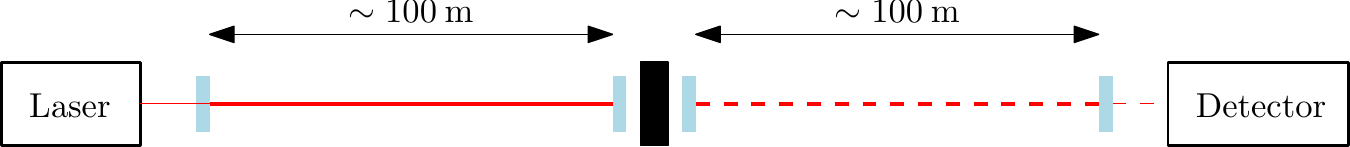}%
  }
  \\[0.5cm]
  \subfloat[][ALPS-IIc]{%
    \label{subfig:alpsII:phases:schematic:c}%
    \includegraphics{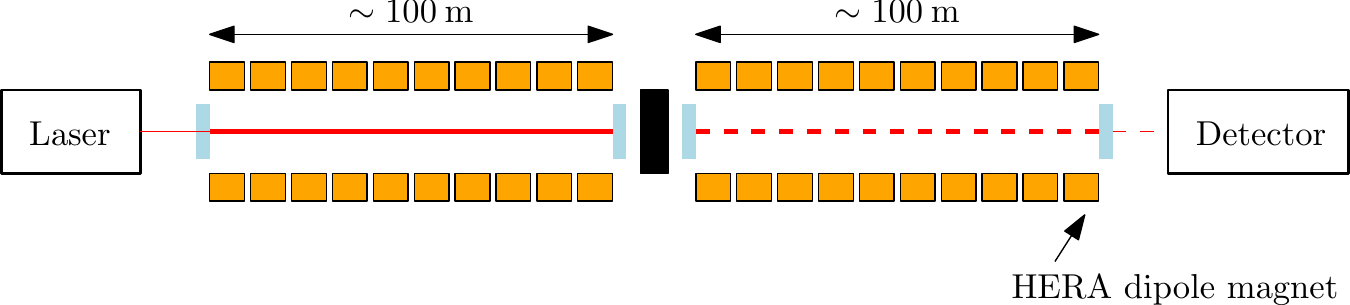}%
  }
  \caption[Different phases of ALPS-II]{%
    Schematic view of the different phases of
    ALPS-II and of ALPS-I.
    \subref{subfig:alpsI:schematic}
    ALPS-I with a $\sim 10$~m long production cavity and cavity
    end mirror and wall inside of a HERA superconducting dipole magnet.
    \subref{subfig:alpsII:phases:schematic:a}
    ALPS-IIa: prototype with two 10~m long cavities.
    \subref{subfig:alpsII:phases:schematic:b}
    ALPS-IIb: prototype with two 100~m long cavities.
    \subref{subfig:alpsII:phases:schematic:c}
   ALPS-IIc: prototype with two 100~m long cavities using the HERA
    superconducting dipole magnets.
  }
  \label{fig:alpsII:phases:schematic}
\end{figure}

\begin{figure}[htb]
\centering
\includegraphics[angle=0,width=130mm]{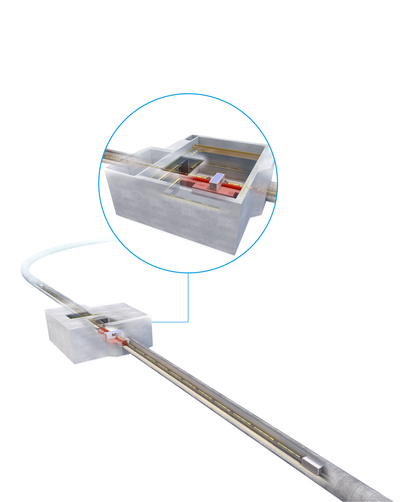}
\caption{Artist's view of the ALPS-IIc setup. The picture shows a straight section of the HERA tunnel
equipped with 20 HERA dipoles. The middle part, accommodating the central breadboard including the ``wall'' is highlighted.
 \label{fig:IIc}} 
\end{figure}

The basic three stages of ALPS-II are shown in
\Figref{fig:alpsII:phases:schematic}.
\begin{itemize}
  \item ALPS-IIa:\\
  This stage aims for demonstrating all features of the optical system
  and the superconducting 
  transition-edge sensor (TES) detector in a dedicated laser
  laboratory with 10~m long production and regeneration cavities.
  ALPS-IIa itself will be subdivided into several steps. A 1~m tabletop experiment in 
  air with reduced optical 
  power and cavity finesses is already being conducted, 
  which will allow to prove the consistency of the optical design, develop and test the control systems
  and yield first performance
  data of the overall system. 
  Subsequently, the high power production cavity will be realized allowing for a first data run using 
  the 
  CCD camera and perhaps 
  already the 
  TES detector to search for hidden photons. 
  This sub-step is expected to be concluded in early 2013. Afterwards, the
  regeneration cavity including all features of the final ALPS-IIc
  stage will be set up. In 2014, we plan to search for
  hidden photons with full sensitivity. Besides interesting physics
  results, this experiment corresponds in all aspects to ALPS-IIc,
  having only a shorter length and lacking magnets. By the middle of 2014, we
  also envisage to have sufficient experience with the straightening of HERA
  dipoles so that all major ingredients of the final ALPS-II stage are
  known.
 
  It should be mentioned that most costs of ALPS-IIa have been already covered. The new
  laser laboratory has been constructed in 2011. We are
  reusing the laser of ALPS-I.
 Nearly all optical components as well
  as the vacuum system have been purchased already.

\item ALPS-IIb:\\
  In principle this stage corresponds to ALPS-IIa, but with about
 100~m
  long cavities. It will be installed in one straight section of HERA
  (around the experimental hall HERA-West) using the long straight vacuum
  pipe of the HERA proton ring. This second stage will demonstrate the
  functionality of the system on the length scale foreseen for the final
  setup with HERA dipole magnets as well as the ability to cope with the
  situation in the HERA tunnel, e.g., seismic vibrations (which are expected
  to be smaller than in the laser laboratory of ALPS-IIa), cleanroom
  installation, and operation given the tight spatial constraints. The
  aperture boundaries are a little relaxed compared to ALPS-IIc, so that
  setting up and operating the experiment will allow us to gain valuable
  experience for the final ALPS-IIc setup.

  The experience with ALPS-IIb should allow for a fast realization of
  ALPS-IIc. As ALPS-IIb will reuse many components of ALPS-IIa the
  additional costs will be moderate. Mainly new infrastructure, new
  cleanrooms and new mirrors for the long cavities have to be provided.
  Construction for ALPS-IIb could start in the second half of 2013 with
  preparation of the site and installation of the cleanrooms. This should
  allow to install the optics and the detector starting late 2014 being
  ready for data taking in the second half of 2015. 

\item ALPS-IIc:\\
  This is the final version of ALPS-II with full length cavities and
  straightened HERA dipoles. If the approval of this stage is given in the
  middle of the year 2014, site preparations of a straight section in the
  HERA tunnel (most likely at HERA North) could start in the beginning of
  2015, when the construction of the linear accelerator of the European XFEL
  project is completed. Note that much more effort is required here compared
  to ALPS-IIb as the site is to be prepared to accommodate 20 straightened
  HERA dipole magnets, see \Figref{fig:IIc}. The time schedule for the installations strongly
  depends on the available manpower of the DESY infrastructure groups
  involved. At present, it is estimated that both the site preparation and
  the installation of the magnets (which are to be straightened in 2014)
  will require about one year so that ALPS-IIc will be ready for
  commissioning and data taking in the year 2017.  At the end of the same
  year, physics results will be available.  
\end{itemize}

In \Tabref{tab:param} the main parameters and the sensitivity of ALPS-I
and ALPS-IIc are compared. We aim for surpassing the present laboratory
sensitivity in the coupling constant of axion-like particles to photons by
more than three orders of magnitude. In the low mass regime we aim to even
surpass bounds provided by helioscope measurements,
cf.~\Figref{fig:ALPshints}. DESY, coincidentally, seems to be an ideal place
for such an enterprise. First, the HERA dipole magnets will possibly allow
for a relatively easy and cheap way of straightening to increase their
aperture. This aperture allows to set up ALPS-IIc with up to 24 magnets
given the lower bound of tolerable clipping losses. Second, this setup just
fits into a straight section of the HERA tunnel, so that ALPS-IIc does
not require any new building. Third, and to finalize the number of lucky
coincidences, 24 spare HERA dipole magnets are available at DESY so that the
construction of ALPS-IIc does not demand demounting of the arcs of the
HERA accelerator (besides using the cryoboxes as discussed later). 

\subsubsection{ALPS-II operation modes and signal interpretation}
\label{subsec:tdr:overview-modes}
While operating the experiment the following main issues have to be
guaranteed:
\begin{itemize}
\item
the alignment of production and regeneration cavity,
\item
the power build-up for 1064 nm light in the regeneration cavity,
\item
and the pointing of regenerated light from WISPs onto the detector.  
\end{itemize}

All these critical items can be probed by opening a shutter in the wall
separating the production and regeneration part of the experiment. The
shutter is shown on the central breadboard in \Figref{fig:optical:layout},
allowing a tiny fraction of the infrared light of the production cavity to
enter the regeneration cavity. If both cavities are aligned and the
resonance condition of the regeneration cavity is met, the 
quadrant photodiodes
at the end of the regeneration cavity (see
\Figref{fig:optical:layout}) will notice the expected intensity of
1064~nm radiation. At the same time, the detector will sense
the infrared light in the regeneration cavity. As this light moves
exactly along the same path as photons from reconverted 
WISPs, the pointing
can be checked easily. However, when the shutter is opened, about
1.7~mW of 1064~nm  light ($10^{16}$ photons per
second) will reach the detector. This intensity is to be damped by dedicated
filters to allow the detectors to cope with the flux.

The analysis of the data will rely on a thorough understanding of any
backgrounds to detect a potentially very small photon flux from reconverted
WISPs. A signal will be identified by comparing data taken under
conditions where WISPs can be expected (so-called ``data frames'') with
data which by construction cannot contain reconverted WISPs (``dark
frames''). To minimize changes in the experimental set-up for data and dark
frames the regeneration cavity (for the ALPS phases IIb and IIc) will
be detuned to not amplify 1064~nm  radiation while staying in
lock for 532~nm light. This can be achieved with the help of
the 
AOM on the central breadboard (see \Figref{fig:optical:layout} in the optics section), again
checked by opening the central shutter. With the regeneration cavity being
ineffective for photons from WISPs, their flux at the detector will be
damped by the power build-up factor (about 40000) of the regeneration
cavity. Hence, these dark frames correspond to background runs without any
WISPs, while the ambient conditions (i.e., stray light, fluorescence
effects) remain unchanged. Of course, additional test data, where, for example,
the main laser is shut off, the frequency doubled green light suppressed or
the detector totally blocked, will be taken to understand the origin of any
spurious background photons. Experience at ALPS-I showed that
continuous improvements and investigative searches resulted finally in well
understood remaining backgrounds (essentially only from detector effects).

Surely, a discovery of WISPs by a photon signal at ALPS-II is to
be proven by systematic studies even if the statistical significance is
beyond doubts. In addition, such studies will allow to pin down some
properties of the detected WISP. The following list gives an overview
on possibilities to test the WISPy nature of a photon excess:
\begin{itemize}
\item For any WISP the signal is expected to rise with the power
  build-up in the regeneration cavity. This prediction can be easily probed
  by changing the lock condition for the regeneration cavity.

\item A comparison of runs with magnets on and off allows to discriminate
  between WISPs which mix kinetically with photons (hidden photons)
  and WISPs which couple to more than one photon (like ALPs). 

\item The polarization of the laser light can be changed with respect to the
  direction of the magnetic field. Hence, the parity of a WISP, which
  shows only up in data with magnets on, can be determined. Scalars are
  produced with the polarization perpendicular to the magnetic field,
  pseudo-scalars with a polarization parallel to the field.

  A production not strongly depending on these polarization directions might
  indicate the existence of minicharged particles and massless hidden photons.

\item In \Figsref{fig:ALPsvac} and \ref{fig:hpgas} the dependence of the
  WISP production and reconversion on the gas pressure within the cavities
  is shown. This effect would allow to estimate the mass of the discovered
  WISP, ranging from relatively accurate measurements, if the mass is
  sufficiently large, to upper limits for very lightweight WISPs. 
\end{itemize}

In summary, the ALPS-II setup will allow for many different systematic
checks to prove the origin of a photon excess and to even determine
properties of a WISP particle. The sketched run procedure also enables
sufficient tests to monitor the performance of the experiment, so that a
WISP signal (in the accessible sensitivity regime) cannot escape
detection.

\subsection{Laser and optics}
\label{sec:tdr:laser}
The main goal of the optical design of ALPS-II is to make the
electro-magnetic field provided by the laser beam on one side of the wall as
large as possible and to detect a possibly regenerated field on the other
side with a very high sensitivity. Both tasks can be supported with optical
Fabry-Perot type resonators. On the side in front of the wall, the ALPS-II
production cavity (PC) can increase the optical power of the light beam directed towards
the wall by a factor of 5000 compared to the power of the injected laser.
Behind the wall, the regeneration cavity (RC) increases the production probability with which
photons are created from the axion field
\cite{Hoogeveen:1990vq,Sikivie:2007qm} by a factor of 40000.
Please note that operating the cavities requires the usage of a continuous wave laser.

\subsubsection{Technical challenges}
These main goals translate into several requirements on the optical system
that have to be fulfilled. Two high finesse optical cavities have to be
operated within the small aperture provided by the magnet string with
identical optical axes. Both cavities have to be resonant for the same light
frequency. The mirrors have to be controlled to form stable eigenmodes
co-linear to the central axes of the magnets. The laser beam has to be
matched to these modes spatially as well as concerning its frequency. While
the control of the laser and the PC is state-of-the-art in several
fields of modern optics, e.g., the interferometric detection of gravitational
waves \cite{blair2012}, a new challenge arises in controlling the RC. This cavity has to be kept 
aligned and resonant for the regenerated light without
the use of any control light in the spatial and spectral acceptance range of the ALPS-II detector.

\subsubsection{Conceptual design}

\begin{figure}[tbh]
  \centering
  \includegraphics[width=0.7\textwidth]{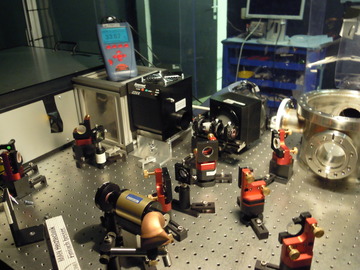}%
  \caption{This picture shows the injection stage of the production cavity in the laser laboratory in hall 
  50 at DESY.}
  \label{fig:cavity}
\end{figure}
The design goal of ALPS-II is to inject 30~W into the PC with a power buildup of 5000 and to operate the RC with an 
equivalent power buildup of 40000. The injection stage of the PC as of August 2012 is shown in \Figref{fig:cavity}.
The main limitations for the optical design are given by the aperture of the magnet strings, by the 
durability and losses of available dielectric mirror coatings and by the environmental disturbances,
namely the vibrational fluctuations and 
their coupling into the length degree of freedom of the optical cavities. To get a quantitative understanding 
of these design boundaries we use a 
staggered experimental approach: We first set up a phase ALPS-IIa prototype experiment with 10~m long cavities 
to understand the limitations 
from dielectric coatings and gain experiences in controlling the two cavities. Sensor and actuator range and
noise investigations as well as lock-acquisition 
studies are part of this phase. Furthermore, the analysis of vibrations and other environmental disturbances
will be an important outcome of this phase.\par
Based on the information gained, we will modify the setup to allow an operation of two 100~m long cavities in 
phase ALPS-IIb. This phase will not 
have the small aperture set by the magnets, which leads to somewhat relaxed requirements for the cavity alignment 
control. Once, both 100~m long cavities 
are operable, remaining fluctuations can be measured and optionally reduced. Furthermore, we can measure the
margin in control authority and bandwidth to estimate the 
acceptable level of additional vibrations related to magnet operation and environmental noise. These important 
results will feed into the final design of the ALPS-IIc 
phase, in which the two cavities need to be operated in the small aperture given by the magnets. The three 
different phases of the experiment are schematically shown 
in \Figref{fig:alpsII:phases:schematic}.\par
\begin{figure}[tb]
  \centering
  \includegraphics[scale=.5]{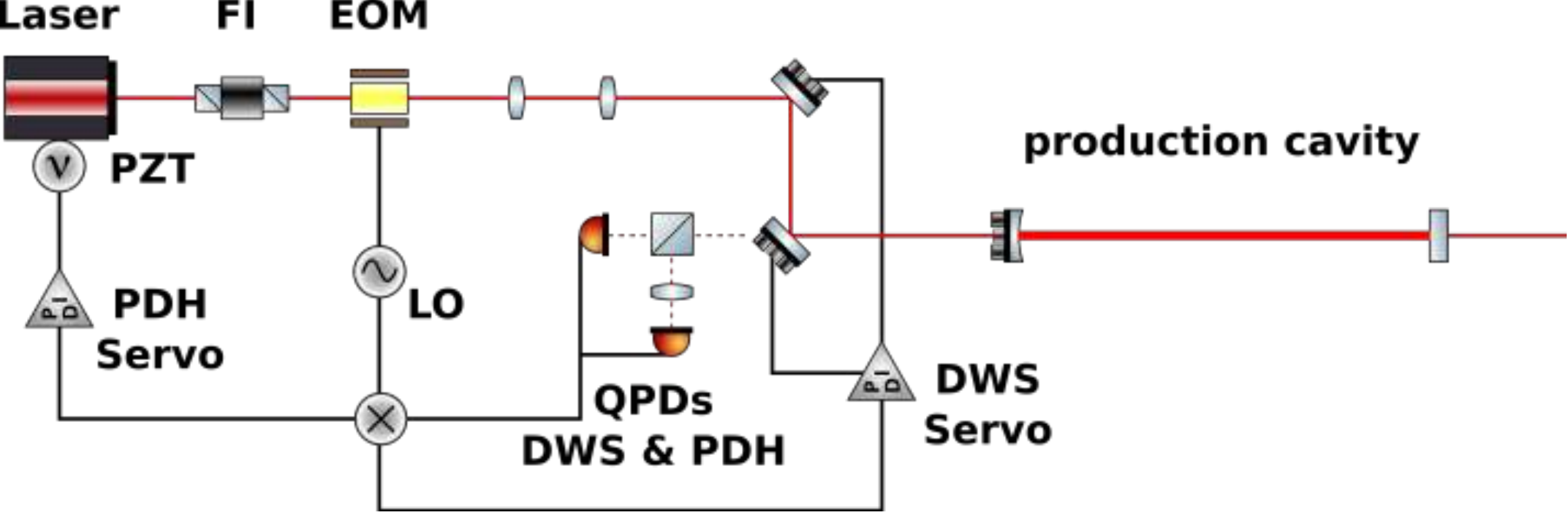}%
  \caption{Schematic of the ALPS-II injection stage including the production cavity.}
  \label{fig:injection:stage}
\end{figure}

A schematic sketch of the laser and the PC is shown in \Figref{fig:injection:stage}.  
We will use a 35~W high-power 
single-mode single-frequency laser system operating at 1064~nm. The master-oscillator 
power-amplifier design allows to amplify 
the power of the 2~W master laser (Innolight Mephisto) to 35~W while maintaining the 
frequency stability of the low noise master laser. 
Several of these lasers are currently in use in gravitational wave detectors and have proven to
be reliable with low maintenance efforts for many years. 
This laser was already successfully used in the ALPS-I experiment.
The laser light passes a Faraday isolator
(FI) to avoid back reflections
into the laser and an electro-optical modulator that imprints phase
modulation sidebands onto the light. These sidebands are used in a
heterodyne detection scheme (Pound-Drever-Hall (PDH) sensing,
\cite{black2001}) to gain information on the deviation of one of the
resonance frequencies of the PC from the laser frequency. Two lenses and
two steerable piezo-electric transducer
(PZT) mirrors are used to match the
Gaussian laser beam to the eigenmode of the PC. A fraction of the light
reflected by the cavity is transmitted by a turning mirror and detected with
two QPD that have a detection area split in four
quadrants. A lens is used in front of one quadrant photodiode
(QPD) to shift the Gouy phase of the
Gaussian beam by 90$^\circ$. Via the heterodyne differential wavefront sensing
(DWS) scheme \cite{morrison1994}, we will gain information on the parallel 
shift and tilt of the laser beam with respect to the cavity's optical axis, 
both in the horizontal and vertical plane. The sum of all quadrants of one
of the QPDs is used in the PDH sensing.\par
By appropriately feeding the alignment and length error signals back to the
laser frequency control actuator and to the PZT mirrors we can stabilize the 
high power laser beam to match the cavity eigenmode 
and to keep laser and cavity resonant. The goal is to keep the remaining spatial 
and frequency fluctuations small enough to limit the 
fluctuations of the power buildup in the cavity to less than 5\% (RMS).\par
The optical parameters of the PC and RC are designed to allow for high buildup factors.
To reduce the risk of mirror damage we limited 
the maximum intensity on the PC mirrors to 
$500 {\rm kW}/{\rm cm}^2$ 
(well below the damage threshold of 
about a few 
$1000 {\rm kW}/{\rm cm}^2$ for sputter and e-beam coatings), a level at which mirrors in the gravitational wave 
community have been operated for long durations without degradation. For a smallest beam radius of
5~mm at the central mirror this 
corresponds to approximately 
150~kW circulating power or a power buildup of the PC of $PB_{\rm PC}\approx5000$. 
The RC will not suffer from too high intensities on the mirrors such that its power buildup
will be limited by the optical losses. 
If we take a typical value of 8ppm losses per mirror and allow for an equal amount of
roundtrip clipping loss the power buildup in an impedance matched 
cavity will be $PB\approx40000$.
Based on the above arguments we have chosen $PB_{\rm PC}\approx5000$ and $PB_{\rm RC}\approx5000$ as the
ALPS-II cavity 
design parameters.\par
\begin{figure}[h]
  \centering
  \includegraphics[scale=.5]{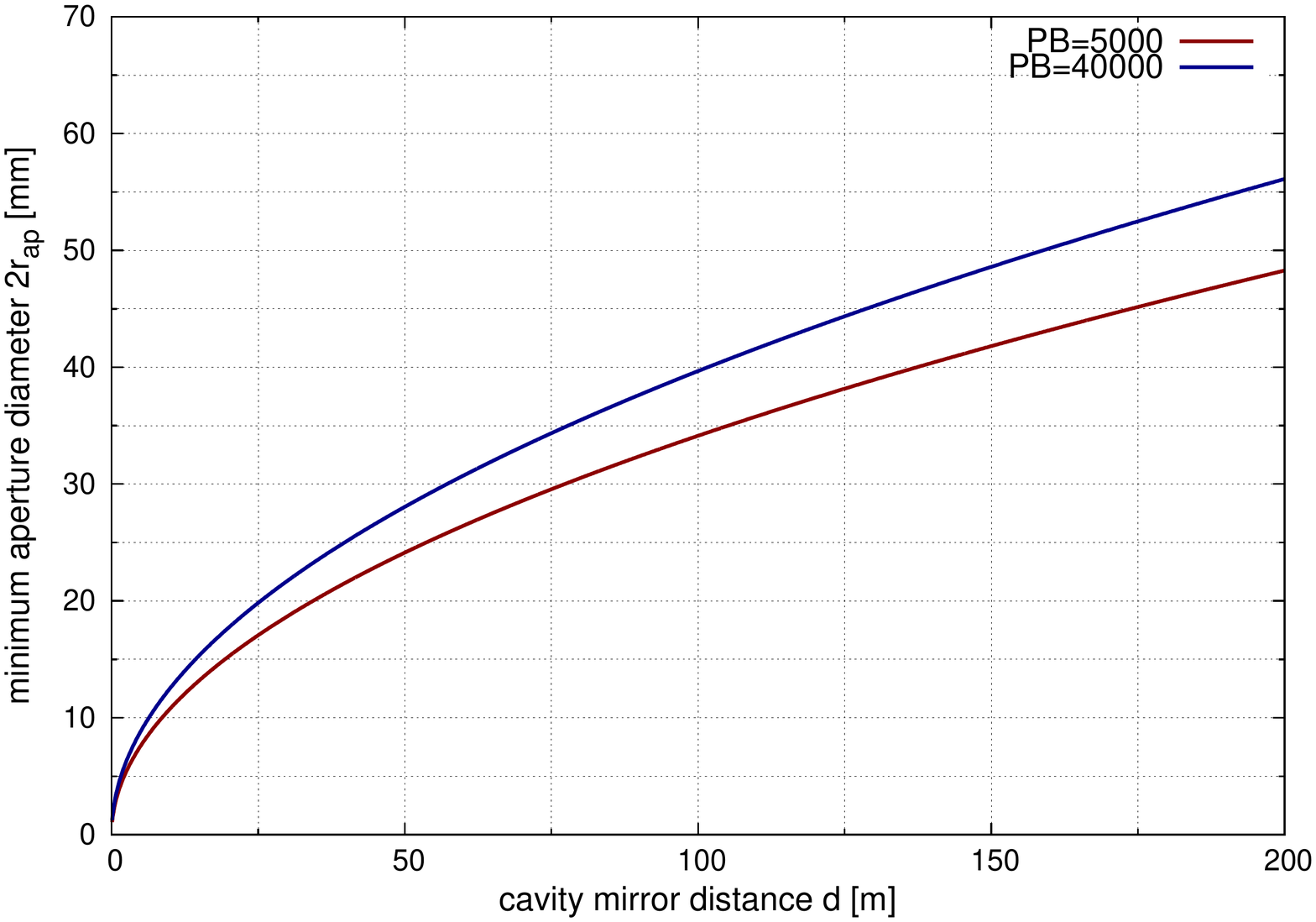}%
  \caption[Minimum aperture diameters for PC and RC power buildup 
  plotted over cavity mirror distance]{Minimum aperture diameters $2\cdot r_{ap}$ 
  required to allow for a power buildup of 5000 for the PC (red, lower line) 
  and 40000 for the RC (blue, upper line) plotted over the cavity mirror 
  distance $d$. Additional mirror losses of $8$~ppm are assumed for each cavity mirror.}
  \label{fig:minimum:aperture}
\end{figure}
The eigenmode in the RC has to be an extension of the PC eigenmode to match 
the Gaussian beam parameters of the regenerated field. 
It can be shown that the Gaussian beam with the smallest clipping losses for
a given constant magnet aperture and identical length of the PC and RC is 
the one with the beam waist in the center between the two cavities and a
Rayleigh range equal to the length of a single cavity. For such a beam the aperture 
radius required to keep the clipping losses small enough to achieve the power 
buildup goals is plotted over the cavity length \Figref{fig:minimum:aperture}.\par

Under the assumption that the HERA magnets can be straightened to give an aperture of 
$d=50$~mm -- with the free aperture taken to be 
10~mm
smaller, to account for alignment tolerances, the waviness of the vacuum pipe, and fluctuations
of the laser position -- 
a cavity length of 100~m
is possible.\par
In principle the PC could be longer as larger clipping losses are
acceptable in this cavity. Given that the total length is limited by the infrastructure, we have,
however, not included this option to
allow for mirrors with identical radius of curvature
(ROC) in both cavities.\par

\begin{figure}[tbh]
  \centering
  \includegraphics[width=0.7\textwidth]{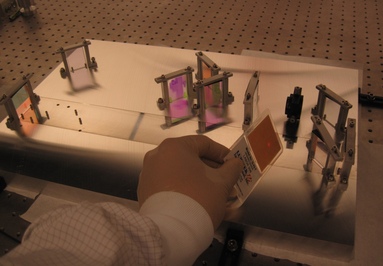}%
  \caption{Picture of a prototype of the ALPS-II central breadboard during setup of the 1~m
  tabletop experiment at AEI.}
  \label{fig:breadboard}
\end{figure}
A key component of the ALPS-II optical design is the central breadboard. This breadboard is 
marked with a beige-colored background in the optical layout 
shown in \Figref{fig:optical:layout}, a picture of the breadboard during setup of the 1~m
tabletop experiment can be seen in \Figref{fig:breadboard}. It provides a long term stable platform
for mounting the central cavity mirrors and the quadrant photodiodes 
used as alignment references.\par
The mirror labeled as CBS1a is the end mirror of the PC and is chosen to be
flat (ROC$=\infty$). This choice will force the optical axis of the
PC to be perpendicular to the surface of CBS1a. A fraction of 90~\% of the
beam transmitted by CBS1a is directed towards mirror M2. Mirror M2 transmits
1\% of the beam which is detected by the quadrant photodiode QPD3. A
control loop shown in \Figref{fig:optical:layout} is used to steer the input mirror PIC of the
PC
such that the beam on QPD3 hits a reference point. Once this control loop is
operating, all degrees of freedom of the PC axis are fixed with respect to the central breadboard.\par
\begin{figure}[tbh]
  \centering
  \includegraphics[scale=.5]{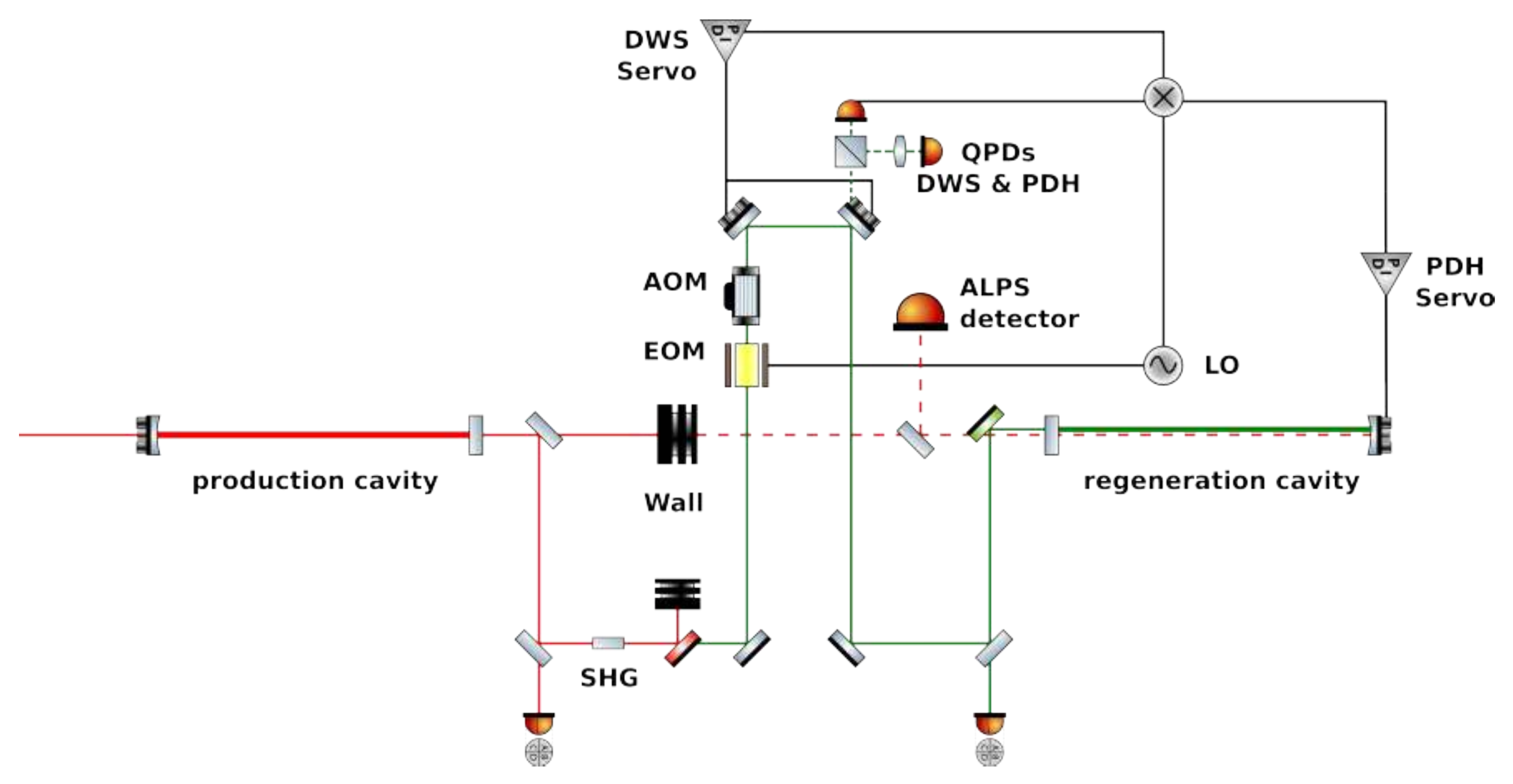}%
  \caption{Schematic of the ALPS-II regeneration cavity including control loops.}
  \label{fig:regeneration:stage}
\end{figure}
\begin{figure}[tbh]
  \includegraphics[scale=.7,angle=90]{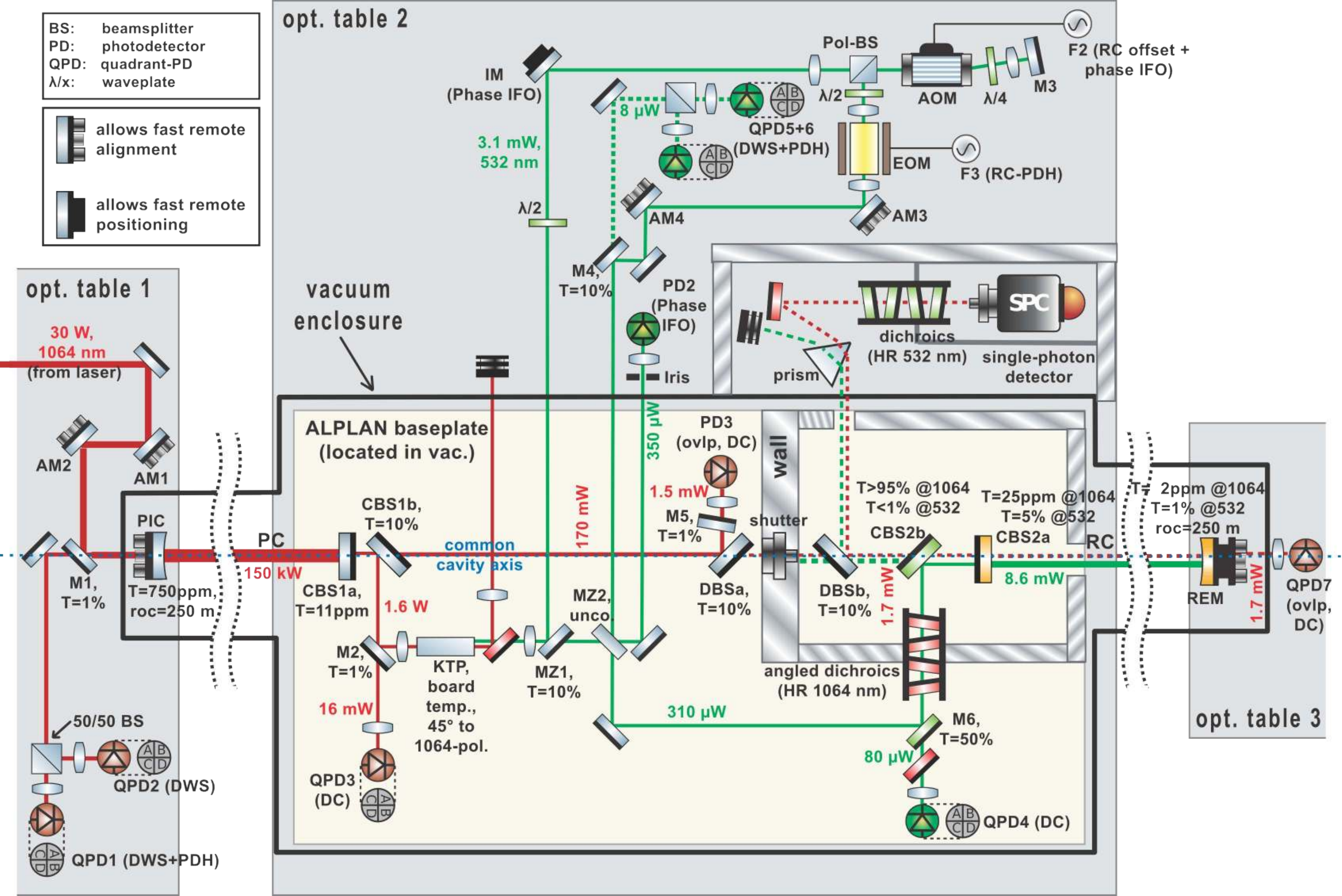}%
  \caption[Optical layout for ALPS-IIb and ALPS-IIc]{Layout of the ALPS-IIb and ALPS-IIc optical tables.}
  \label{fig:optical:layout}
\end{figure}
The main fraction of the beam is reflected by M2 and send into a KTP
non-linear crystal. A beam with a wavelength of 532~nm is
produced via the second harmonic generation
(SHG) process. A dichroitic
mirror separates the green and red beam and directs the red beam to a beam
dump. A fraction of 90~\% of the green beam will be directed towards an
optical setup not mounted on the central breadboard (and outside of the
vacuum tank which houses the central breadboard, see
\Figref{fig:optical:layout}). In this setup the green beam is frequency
shifted via an acousto-optic modulator
(AOM) and phase modulation sidebands are added to the beam
for the PDH locking of the RC. The conditioned green beam is now send
back to the central breadboard and passes the ``wall'' through a set of
dichroitic mirrors to make sure that no 1064~nm light enters
the RC and detection area. After the ``wall penetration'' the green beam
is injected into the RC. The RC cavity consist of the flat mirror CBS2a and 
the curved mirror REM which can be controlled in the direction of the cavity
axis for cavity length control and in rotation and tilt for cavity alignment
control. The mirrors of the RC are coated to form a resonant cavity for
1064~nm with a power buildup (PB) of 40000 and for 532~nm
with a PB of approximately 60. The green beam reflected by the RC traces back the path of the incoming
beam to the quadrant photodiodes
QPD 5 and QPD 6 which are used as sensors in differential wavefront sensing
(DWS) and PDH sensing schemes to generate error signals for the alignment control of the green 
beam with respect to the cavity eigenmode and the length of the PC with respect to the resonance of
the green beam. Control loops feeding 
back to the steering mirrors AM3 and AM4 will keep the green beam aligned to the cavity eigenmode and
feed-back to the cavity end mirror REM will 
change the cavity length to make it resonant for the green beam. A schematic of the control
loop layout can be seen in \Figref{fig:regeneration:stage}.

The remaining task is to make the eigenmode of the RC co-linear with the
eigenmode of the PC. For this we have to mount the two flat central mirrors
of these cavities in parallel and rigidly on the breadboard (see below for
required precision). This ensures that the cavity axis are parallel. Once
the red and green beams are resonant in the respective cavity we can open
the shutter in the wall to allow a 1064~nm beam to impinge on
the RC. By rotating and tilting mirror REM we can laterally shift the
RC
eigenmode until it matches the incoming 1064~nm beam. During
this process we might need to change the frequency offset between the green
and red beam via the AOM to meet the resonance condition for the green and red beam at the same time. 
Once the 1064~nm 
beam resonates in the RC we note the position of the green beam on QPD4 as the spatial reference for the
RC eigenmode. This reference 
position and the normal to the surface of mirror CBS2a fully define the RC axis. 
Hence a 
control loop feeding back to tilt and rotation of REM to keep the beam on
this reference position fixes the location of the RC eigenmode relative to
the central board and co-linear with the PC axis. If the positions of CBS1a,
QPD3, QPD4 and CBS2a do not move with respect to each other and all control
loops are closed the 1064~nm beam will be resonant in the
RC
as will be any light field regenerated from WISP particles. A typical
measurement sequence would involve an open central shutter at the beginning
and end of the measurement (and possibly in between) to ensure alignment and
resonance conditions are right and a closed central shutter for the WISP
search. During those searches possibly regenerated 1064~nm
photons will be transmitted by CBS2a and CBS2b and steered by DBSb to the
main ALPS detector (CCD or TES). It is essential that none of the 532~nm photons used to 
control the RC is accidentally 
detected and interpreted as regenerated 1064~nm 
photons. Hence the 532~nm photons have to be reflected,
absorbed or spatially split from the 1064~nm beam path after
leaving the RC.\par
Dedicated studies to determine the production probability of 1064~nm photons (for example by fluorescence effects) from the 532~nm 
light used to lock RC are under way. First results have shown that this probability is below $10^{-17}$. 
We plan to set up further experiments to mimic the 
optics of the regeneration cavity in more detail and to exploit polarization effects to discriminate between 
fluorescence photons and light from reconverted WISPs. 
For ALPS-II, a production probability for 1064~nm from 532~nm photons of less than $10^{-21}$ photons is to 
be achieved.\par
Once the shutter in the wall is closed, we rely on the stability of all components on the central breadboard. 
As the green beam used for the RC control 
leaves the breadboard and the vacuum system it might be subject to phase fluctuations 
caused by air turbulences and/or vibrations of the optical components outside 
of the vacuum system. Hence we will set up a Mach-Zehnder interferometer with the beam splitters
MZ1 and MZ2. The output port of this interferometer is sensed by PD2 
and will give us information about the phase fluctuations. If required, a control loop feeding
back to mirror IM will be used to correct for these fluctuations.

\subsubsection{Expected performance}
We expect that we can reach a light power level of 150~kW
traveling into the direction of the ``wall'' and to enhance the detection efficiency of regenerated
photons by an RC with a power buildup of 40000. 
Several requirements have to be fulfilled to achieve this performance:
\begin{enumerate}
\item The total lateral and angular beam shift introduced to the 
1064~nm beam by optical components between the two cavities on the central 
breadboard has 
to be smaller than 1~mm and 10$\mu$rad, 
respectively, because any beam shift is not seen by the particles transversing the wall
and hence reproduced light would not match the RC eigenmode.
\item To allow parallel alignment of the optical cavities, the
central mirrors have to be parallel to within 10$\mu$rad.  
\item Drifts (e.g., caused by thermal gradients) of the components on the central 
board have to be small enough to meet the pointing requirements given in 1. 
The mismatch between twice the red resonance frequency and the green resonance frequency
of the RC has to be stable to a fraction of 0.2 of the RC
linewidth (corresponding to approximately  95~\% of the maximum power buildup).
\item The actuator range of the different control loops has to be large enough to compensate the
free running peak-to-peak fluctuations of the relevant degrees-of-freedom.
\item The control-loop disturbance reduction has to be large enough to reduce the deviation of
the cavities from their operation points such 
that the power buildup factor can be kept at a value larger the 0.95 times the maximal buildup.
\end{enumerate}
Simulations of the optical setup and estimations of the control loop performance indicate that 
all the requirements can be met with state-of-the-art optics, 
electronics and fabrication processes. 

\subsection{Cleanroom design}
\label{sec:tdr:Cleanroom design}

\subsubsection{General considerations}
The setup of the optical components of the experiment has to fulfill the following requirements:
\begin{enumerate}
 \item The optical components 
 outside and inside the vacuum must be kept free of dust particles after being cleaned, 
 during the mounting process and during the experimental runs. Due to the high cavity power buildup aimed for, the optical components in the
 experiment, especially the cavity end mirrors, are most sensitive to dust particle
 contamination. Even a degradation of a few ppm in the reflectivity of an end mirror 
 would lower the performance of the regeneration cavity intolerably. 
 \item Mechanically and thermally 
 stable conditions have to be 
 maintained during the operation of the experiment. Only minimal position changes of optical components are tolerable
 due to the limited expansion range of the piezo-electrically driven mirror holders.
 A movement above a few $\mu {\rm m}$ in beam direction of an end mirror can not be compensated 
 by the feedback loop for cavity length stabilization. 
 \item For the operation of the infrared laser (continuous-wave output:
  35~W at 1064~nm) a safe 
 working environment has to be created and maintained at all times, with special attention to 
 the setup phase of the experiment. 	
\end{enumerate}

\subsubsection{The cleanrooms}
\begin{figure}
\includegraphics[width=0.95\textwidth]{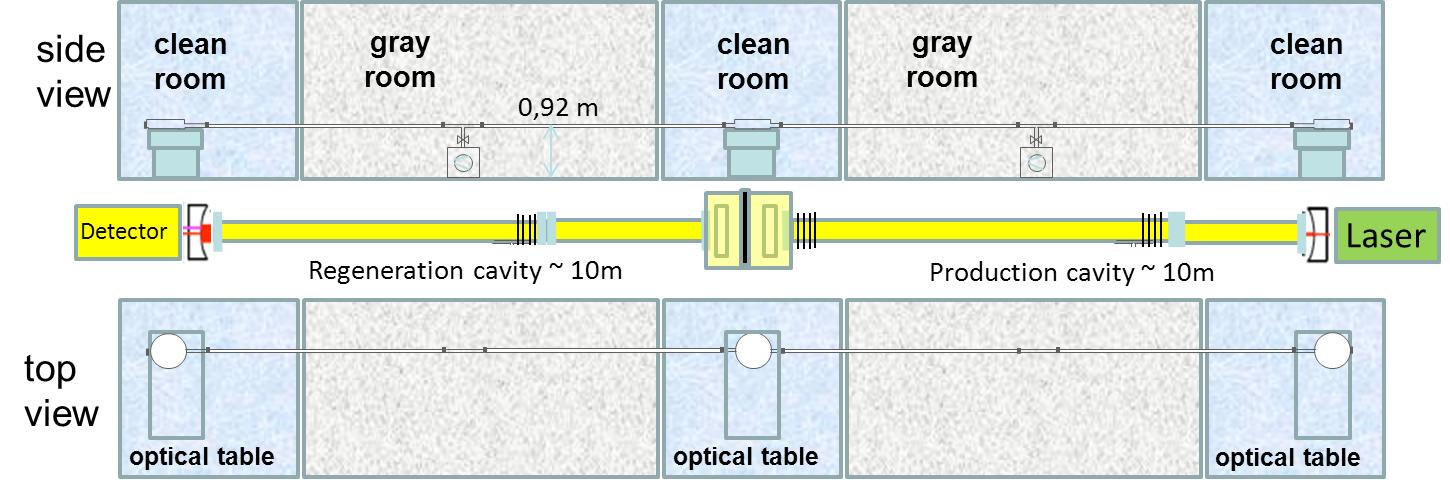}
\caption{Experimental setup for ALPS-IIa in building 50 room 607.}
\label{fig:setupa}
\end{figure}

The laboratory space for ALPS-IIa including three cleanrooms has been set up in 2011 already.
The layout of the cleanrooms for the ALPS-IIb and ALPS-IIc experimental setups is based on 
the approach used for ALPS-IIa in the laser laboratory housed presently at
DESY in building 50 (see \Figref{fig:setupa}). 
There are three cleanrooms containing:
\begin{enumerate}
 \item  the laser and the vacuum chamber with the central breadboard, 
 \item  the vacuum chamber with the second end mirror of the production 
 cavity and the first end mirror of the regeneration cavity and
 \item the vacuum chamber with the second end mirror of the regeneration cavity.   
\end{enumerate}

\noindent
Adjacent to the cleanrooms are gray-rooms which serve as airlock to the outside.
Cleanroom conditions will be ensured applying the following measures:
\begin{enumerate}
 \item The floor inside the cleanrooms and the gray-rooms is regularly cleaned.
 \item In each gray-room entrance area adhesive carpeting is used to avoid propagating dirt.    
 \item Persons use cleanroom suitable clothing and foot wear. 
 To enter a gray-room an additional cleaning lock space has to be passed, to reach the required cleanliness.
 \item 
 A flow of 
 200 ${\rm m}^3/{\rm h}$
 of fresh air is provided, sufficient for up to 4 persons. In order
 to refresh breathing air, filtered air (H12) from the outside is blown into each gray-room. 
 The resulting turbulent flow of 
 mixed air also reduces the gray-room dust particle concentration compared to the outside. 
 \item To reduce dust particle concentration in the cleanroom, clean filtered air (H14) from the adjacent gray-room is 
 blown in turbulently. The air exits the cleanroom back into the gray-room.
 Every two minutes the air volume in the cleanroom is mixed with an equal volume of filtered air. 
 In that way the dust particle concentration in 
 the gray-room is also reduced and the breathing air in the cleanroom is continuously refreshed.  
 \item The enclosure of the laser table in the cleanroom has a roof with vertical transparent blinds. 
 Within the enclosure a separate filter unit produces a clean horizontal laminar displacement
 flow across the laser table, which is always directed against the person reaching through the blinds
 and performing manual work. Dust particles from the person's moving hand or arm will not move down
 towards the optical components on the laser table, but instead leave the enclosure through the blinds.
\end{enumerate}

 Dust particle concentrations are 
 expected to be the same as in the ALPS-IIa laboratory (down to
 0.3 $\mu {\rm m}$ size):   
 \begin{itemize}
  \item 100000 particles/${\rm ft}^3$ outside, 
  \item 5000 particles/${\rm ft}^3$ in the gray-room,  
  \item 200 particles/${\rm ft}^3$ in the cleanroom and
  \item 0 - 1  particles/${\rm ft}^3$  within the laser table enclosure in front of a working person's hand reaching in.
 \end{itemize}
 
These numbers match the requirements.
About two hours after the last person has left a cleanroom of ALPS-IIa, no dust particles 
 are measured anymore by a particle counter located in the cleanroom outside the enclosure.
For more effective dust decontamination of clothing, the use of an ionization gun will
be tested in the ALPS-IIa laboratory.

At present it is planned to support the optical tables in the tunnel cleanrooms from 
the top of the HERA tunnel, as there is little space on the floor due to big pipes for 
cooling water and Helium gas (see \Figref{fig:magnets14}). 
This concept will be tested at the setup of ALPS-IIb. The cleanroom between the two
magnet strings will be located in a HERA hall and the optical table will be supported 
from below. 

Like in the ALPS-IIa setup the breadboards containing the optical elements will
be supported from the optical tables, decoupled mechanically from the
vacuum vessels, see \Figref{fig:vacuum1}, which are supported from the floor of the cleanrooms.

For ALPS-IIc the HERA kicker-bypass will be installed allowing the magnet current and the cryogenics to bypass 
the optical setup in the middle of the experiment. To compensate the force
of about 80~kN by the atmospheric 
pressure on the end flanges of the kicker-bypass, the endflanges are connected by 3 tension rods. To allow access to 
the optical elements in the vacuum vessel in the cleanroom, the tension rods have to be replaced by a steel girder construction.
The detailed design for the construction of the clean/gray rooms remains to be done. It is assumed that the construction 
will be done in a similar way as for ALPS-IIa applying a drywall installation.

\subsubsection*{Temperature Stability}
 As for the setup of ALPS-IIa, the air blown into the cleanroom will also
 be used for cleanroom temperature stabilization by mixing the input flow with the
 output from an air conditioner unit. The temperature sensor of a feedback loop will 
 be placed above the laser table. The same feedback system was used in the ALPS-I laser 
 laboratory where a long term stability of better than
 $\pm0.1$~$^\circ$C
 was reached. The three
 air conditioning units will use cold water from the
 8~$^\circ$C
 DESY cooling water system.   
 
 \subsubsection*{Laser Safety}
 
As for the ALPS-IIa laboratory, a safe laser environment will be set up in accordance
with the required safety rules (DIN/EN 60825).
An interlock system based on programmable logic controllers is foreseen
which closes the exit shutter of the laser if any door connecting
a cleanroom to its adjacent gray-room, and the gray-room exit door are open at the same time.

\subsection{Detection system}
\label{chap:tdr:detector}
The detection of the possible regenerated photons at ALPS is very
challenging, mainly due to their expected low energy and low rate. The
wavelength of the photons at ALPS-II being 1064~nm reduces
the detection options and makes silicon-based detectors less efficient due
to the proximity of the energy of the Si band gap. Caused by the low rates,
the most important requirement of a detector system for ALPS-II is extremely low
dark count rate (at least less than $10^{-3}$~s$^{-1}$), which should be coupled with a
high detection efficiency. 

For ALPS-II two detection options are pursued: 
\begin{itemize}
  \item a PIXIS 1024B CCD camera and 
  \item a Transition-Edge Sensor (TES).
\end{itemize}

Whereas the CCD camera has already been used for ALPS-I and only needs to be calibrated for the ALPS-II requirements,
an ALPS-II-specific TES set-up is currently being newly developed by the ALPS collaboration.

\subsubsection{Technical challenges}
\label{chap:tdr:detector:technical_challenges}

The main requirements for the detection of the low rates of single photons with a wavelength of 1064~nm at ALPS-II are (in the order of their importance): 
\begin{enumerate}
 \item low dark count and background rate,
  \item high efficiency,
  \item long-term stability,
  \item good energy resolution and
  \item good time resolution.  
\end{enumerate}

The CCD camera is proven to be a viable option and has the advantage of being
ready-to-use. It satisfies the first three criteria, but does not excel in
them.  The dark current of the CCD is of the order of $10^{-3}$~electrons per
pixel per second and its efficiency at the ALPS-II wavelength is about
1.2~\% (further details see \protect\Sectref{chap:tdr:detector:expected_performance}).

Transition-edge sensors~\cite{Enss:2005md} exploit the rapid change of the
resistance at the superconducting phase transition. This enables them to
reach high sensitivity~\cite{ISI:000184336600067} and achieve highest
quantum efficiencies~\cite{ ISI:000254121300021}. In addition, good energy
and timing resolutions are possible. In comparison to a Si-CCD, the quantum
efficiency of a TES at 1064~nm is larger by nearly two
orders of magnitude and additionally the dark count rate is lowered by a
considerable factor. The additional benefit of a single-photon spectroscopy
opens up attractive avenues to additionally reduce the background through
cuts on the pulse-height. The timing allows for precise definition of
good-time-intervals when the cavity is locked. Hence, an unstable
operation of the cavity poses no problem for the TES detector.
However, building and operating a TES
setup is more challenging than to use a CCD camera. Points of major
importance for the success are the control of the environment  (e.g., magnetic field etc.)
at mK temperature, and the sufficient suppression of different
possible backgrounds. In the
field of experimental searches for rare events (e.g., direct Dark Matter
experiments), TES detectors have been operated stable over the course of
months (see e.g., CRESST II \cite{2009APh....31..270A}).

\subsubsection{Conceptual design}
\label{chap:tdr:detector:conceptual_design}

\subsubsection*{CCD camera}

The design of the CCD detector is based upon the experience from the ALPS-I
experiment. The beam of regenerated photons is focused with
a lens onto the
sensor area of a CCD camera. To achieve the best performance, the size of
the focal spot should be minimal (in the ideal case one pixel).

The CCD camera in ALPS-II is the same that was used in the ALPS-I
experiment, a PIXIS 1024B by Princeton Instruments. The chip is a
back-illuminated e2v CCD47-10. It has $1024\times1024$
$13\times13$~$\mu$m sized pixels. The chip can be cooled to
a minimum of -70~$^{\circ}$C. The read-out electronics allows for two read-out
speeds of 2~MHz and 100~kHz with associated read-out noise of 13.73~$e$ and
4.09~$e$, respectively. It is capable of binning the pixels into
\emph{logical} pixels, thus reducing the impact of the read-out noise if
the signal cannot be focused on one pixel.

Neighboring pixels can be used to veto cosmic rays and ambient
radioactivity.

To optimize the focal spot, the shutter of the central breadboard will be
opened so that infrared light will enter the regeneration cavity. This light
will follow exactly the path of regenerated photons. Hence, its image can be
used to optimize the focal spot of the regenerated photons. Because of the
high intensity of this light (1.7~mW), a dedicated filter
will be installed in the beam to protect the camera, which does not change
the position of the image. This will be verified by rotating the filter
around the beam-axis.

\subsubsection*{TES detector system} 
\begin{figure}[!tb]
\centering
\includegraphics[width=0.8\textwidth]{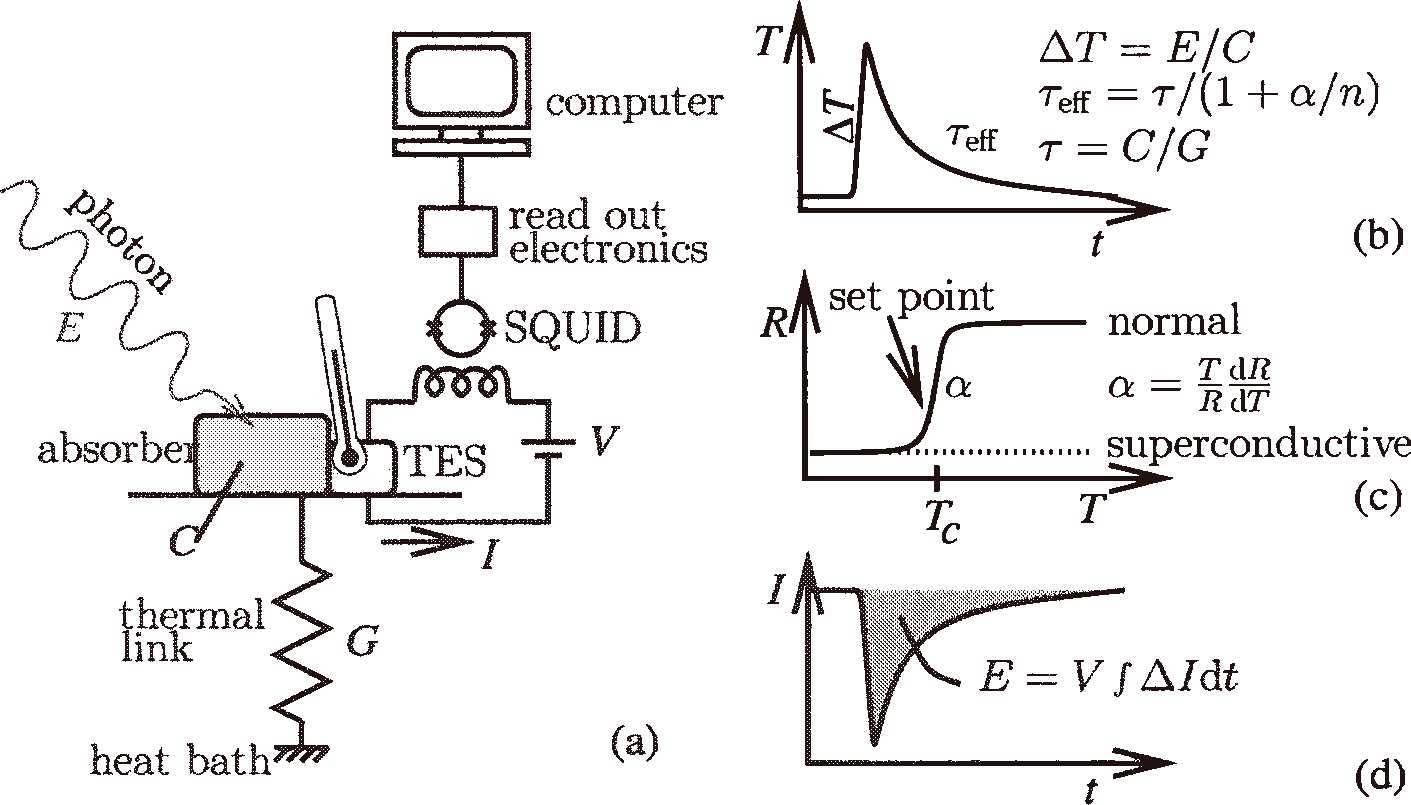}
\caption[Working principle of a TES detector.]{Working principle of a
  TES. (a) A photon sensitive
absorber with heat capacity $C$ is connected to the TES thermistor. The absorber is coupled with a
weak heat conductivity $G$ to a ``heat'' bath. The TES is voltage-biased and
the current of the bias circuit is read-out by a SQUID connected to
read-out electronics. (b) The absorption of a photon with energy $E$ causes
a rise in temperature. The TES cools
down again and is back to its set
point after the time $\tau_{\rm eff}$. (c) As the TES is operating in the superconducting transition region,
there is a change of electrical resistance $R(T)$. (d) The
current of the circuit changes, which can be readout by the SQUID, which is
magnetically coupled to the circuit. Integrating over the current
signal $I(t)$ leads to the incident photon energy. Figure from~\cite{korte2003}.}
\label{img:tes_nutshell}
\end{figure}

The working principle of a TES is briefly explained in \protect\Figref{img:tes_nutshell}.
The TES is a microcalorimeter measuring the temperature difference $\Delta$T of
the absorber material through the rapid change of the resistance at the
superconducting phase transition, which is proportional to the temperature
increase at the set point. In the case of ALPS, and in general for optical
detection, the TES detector serves as absorber and thermistor at the same time.

The research and development of TES for detection of optical/infra-red
photons is actively carried out at
 metrology
institutes like NIST (in the U.S.), AIST (in Japan), INRIM (in Italy) and PTB
(in Germany). The ALPS collaboration has established contacts to these
institutes and is collaborating
regarding the setup of the ALPS TES detector
system and the TES chips (see below). 

\begin{figure}[!tb]
\centering
	\includegraphics[width=0.95\textwidth]{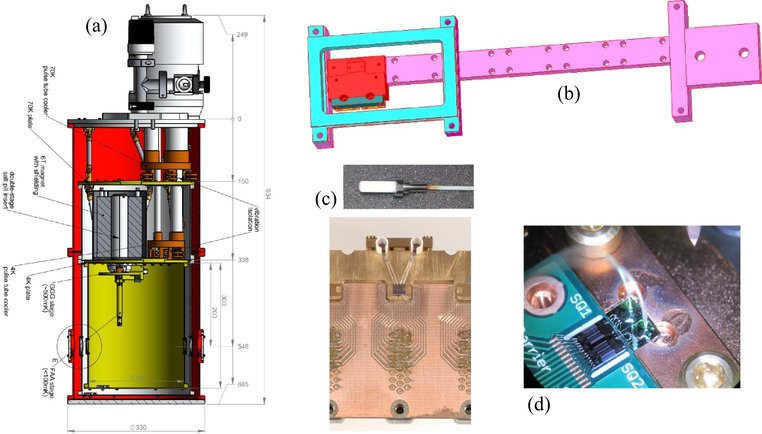}
\caption[Collection of components of TES detector.]{Collection of components of the ALPS TES detector system: (a) Sketch of the ADR: 
In the upper part there is a two-stage pulse-tube cooler connected to the
70~K and 4~K stage. The dashed gray part in the middle indicates the salt
pill with 6~T magnet for ADR cooling. The light green part is the
mK-environment with the cold finger. (b) Sketch of the detector bench attached
to the cold finger: The detector bench, made of copper, holds the
sensor module (red), where one SQUID and two TES chips are located. The sensor
module is surrounded by a brass part (light turquoise) for a magnetic field coil to compensate background magnetic fields. 
(c) Pictures of fiber end and sensor bench: In the
bottom picture a sensor module for NIST TES and PTB SQUID chips is displayed.
The two white ceramic ferrules surround the TES chips and can hold the fiber
end pictured above. 
(d) Picture of TES/SQUID design by AIST: Two fibers are glued
to TES chips, which are connected to the SQUID located on a circuit board.
[Pictures from (a) Entropy, (b) DESY, (c) NIST/PTB, (d) AIST]}
\label{img:tes_realization}
\end{figure}

In 2011, first experience was gathered through an extensive collaboration
with the Universities of Trieste and Camerino,  building up a TES detector with a dilution refrigerator.
For a TES setup essentially four components are required: 
the TES detector, a SQUID-current sensor, a cooling system to reach and maintain a sufficiently low temperature,
 and the optics to feed the light signal to the sensor. The
 realization and the status of the TES detector of ALPS-II is split into these four fields, which are 
 described in the following and summarized in \protect\Figref{img:tes_realization}. 

\paragraph*{Sensor: TES Chip}

TES detector chips are not available commercially. Research groups 
either manufacture their own detectors or obtain detectors by collaborating
with metrology institutes, most notably NIST and AIST.
NIST and AIST have developed highly efficient multilayer TES optimized for
various optical and near-infrared wavelengths.
\footnote{Both institutes have concentrated on the telecommunication
  wavelengths 1310 and 1550~nm.
}

The two following TES would be suitable for the ALPS-II wavelength of
1064~nm:
\begin{itemize}
  \item NIST TES with optical structures to couple light and metallic
    mirrors reaching 98~\% quantum efficiency for
    1064~nm~\cite{lita2010}
  \item AIST TES with optical structures to couple light and dielectric
    mirrors reaching 98~\% quantum efficiency for
    850~nm~\cite{fukuda2011}
\end{itemize}
These TES have low critical temperatures (T$_c$) to reduce the thermal
fluctuation noise and to make the device more sensitive. The NIST TES use
Tungsten (W) as sensor material and become superconductive between 100
and 200~mK, the AIST TES are made of Titanium/Gold (Ti/Au)
leading to a higher critical temperature of around
300~mK.

ALPS~is in contact with NIST and AIST and will
receive chips from both institutions in order to measure their performance for
1064~nm. It is planned to optimize the chips~\cite{tesopt}
achieving a low-noise high-efficient TES detector for single
1064~nm photons by using 
\begin{itemize}
  \item dielectric mirrors (narrowband\footnote{A narrowband detector could reduce a broadband background.}) 
instead of metallic mirrors (broadband) and 
  \item an optimized multilayer structure for coupling light with 1064~nm,
\end{itemize}
if that proves to be beneficial for the ALPS-II experiment.

\paragraph*{Read-out: SQUID-Current Sensor}

TES are mostly read-out by low-$T_c$ dc SQUID current sensors, consisting of
the SQUID itself and a read out electronic. These sensors are current to
voltage transformers (transimpedance elements): To give some numbers, a
1~eV energy input would cause a current of
$\mathcal{O}$(100~$\mu$A) in the TES circuit (see \protect\Figref{img:tes_realization}), which results in a voltage pulse in the
range of 1 to 100~mV. Low-$T_c$ SQUIDs can be operated at
mK-temperatures.

The SQUID sensors for the ALPS setup will be supplied from the
PTB\footnote{Physikalisch-Technische Bundesanstalt, Working Group 7.21:
Cryosensors} in Berlin. PTB SQUIDs are already in use with NIST and AIST
TESs,
so the PTB group is joining a collaborative effort with ALPS to find out which
SQUID-TES design is the most robust one for the ALPS-purpose and, more
generally, to test TES applications. This will allow ALPS to profit from the
great expertise of PTB regarding SQUIDs. 

ALPS has already purchased a TES/SQUID read-out electronic from Magnicon: 
Two channels are for
SQUID read-out allowing the operation of two TES sensors in parallel for cross
checks. Additionally, a compensating
magnetic field coil can be operated 
to prevent degradation
of performance of SQUID, when operated in an ADR.

\paragraph*{Optics: Fiber Feed-in}

To guide the light signal into the cryostat and finally onto the TES an optical
fiber is the method of choice. The anticipated design aims at an optimized 
optical efficiency and will have to address the following
issues:
\begin{enumerate}
  \item fiber-to-TES coupling in the cold
  \item fiber vacuum feed-through as interface between free space and cryogenic environment
  \item free-optic-to-fiber coupling 
\end{enumerate}

The fiber-to-TES coupling will be done with specific methods developed at
NIST and AIST. NIST uses a modified standard fiber connector, whereas AIST
uses UV curable resin for gluing the fiber end to the TES chip. With both
approaches low-loss couplings and a detection efficiency up to
98~\% were reached~\cite{miller2011, fukuda2011}.

For a low-loss working fiber vacuum feed-through with a continuous fiber a
standard Swagelok with a Teflon ferrule will be used,
following~\cite{abraham1998}.  DESY's workshop has already built a
single-fiber feed-through, a feed-through for two fibers is currently under
construction.

Most importantly, the incoming signal photons need to be coupled into the
fiber. The beam diameter at the coupling mirror of the regeneration cavity
(CBS2a in \Figref{fig:optical:layout}) will be 11.6~mm. This
will contain 86.5~\% of the total power (i.e., $2\sigma$ of the
Gaussian beam profile). Thus, to capture about 98.9~\% of the
total power, an optic is suitable that collimates a 17.5~mm
diameter beam into a single mode fiber with about 5~$\mu$m
diameter. First tests and experience in the ALPS collaboration and from
other groups demonstrated that with a simple collimator lens a  coupling
efficiency about 60 to 80~\% is easily reachable due to
coupling losses. 

For reducing the background there are different approaches considered to
filter out other wavelengths. The fiber itself can be used as a simple
filter by using a HI1060 fiber with a operating range of 980 to
1180~nm instead of the standard fiber SMA980 with a operating
range from 970 to 1650~nm. Other possible options are
reflecting coatings at the cold end of the fiber
and filters (including polarization) within a fiber circulator design. These
options will be further pursued, if the background level is too high.

\paragraph*{Cryogenic System: ADR}

\begin{figure}[!tb]
\centering
	\includegraphics[width=0.95\textwidth]{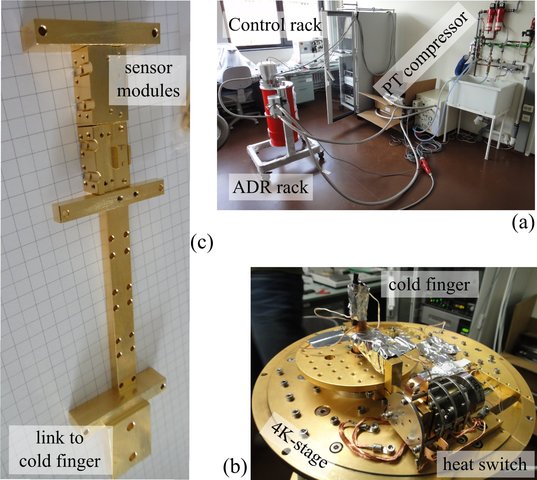}
	\caption[Status of ADR system and components.]{Status of ADR system and components: The pictures show the status at PTB, Berlin at the end of August 2012. (a) Infrastructure and components of ADR: The ADR itself (red dewar) is connected to the control rack for monitoring temperatures and pressure, operating the magnet and driving the heat switch. It is also connected to the Helium compressor of the pulse-tube cooler via 20~m pipes filled with Helium.
	(b) Open ADR: On the 4~K stage the heat switch is mounted, to connect and disconnect the 4~K bath to the salt pill. On top the cold finger is located reaching mK.
	(c) Detector bench: The copper components are ready for attaching the sensors to mK bath. One sensor module is in Japan for attaching AIST chips
}
\label{img:adr_status}
\end{figure}

An ADR was found a suitable solution
for the requirements of the TES setup in a milli-Kelvin environment.  The
University of Hamburg as collaborator of ALPS ordered in November 2011 an ADR
from the company Entropy. It was delivered in late July 2012, tested at the PTB in
Berlin (see \protect\Figref{img:adr_status}) and is now operational at DESY in Hamburg.

The ADR is a no-liquid-cryogens cryostat with a closed pre-cooling cycle:
Integrated is a two-stage pulse-tube cooler with which the cooling stages of
70~K and 4~K are reached. Then using a 6~T
magnet and a double-stage salt pill unit the milli-Kelvin environment is established
through the principle of adiabatic demagnetization.

A cryogen-free cool-down is possible within 20~h. The hold time at
100~mK is about 48~h depending on the cold mass. The
recharge time to re-establish the mK-environment is about 30~min
according to the datasheet \cite{entropy:www}.
With the ADR the aim of a compact and robust, transportable system with a quick
cool-down and easy handling is in reach.

The following components for the setup of the TES inside the ADR are currently being designed and built:
\begin{itemize}
  \item detector bench attached to the ADR cold finger for TES and
    SQUID
    chips including coils compensating for magnetic field
  \item cryo-cable for read-out from vacuum feed-through to 4~K
    stage
  \item circuit board at 4~K stage
  \item vacuum-feed-through for optical fibers
  \item thermal anchoring of cables and fibers coming from 300~K 
  \item shields against thermal and magnetic/electric radiation (currently
    several solutions being designed, the final choice will depend on the
    actually measured background levels)
\end{itemize}

After assembling the ADR and characterization of the completed TES detector the
complete detector system will be moved to Hamburg and set up at the ALPS site. The ALPS site is
fully equipped to operate the ADR with 300~V supply and purified
water for the compressor\footnote{The compressor could be a possible source of
ground vibrations. Dampening these vibrations (in a first stage by a neoprene
mat) and moving the ADR further away from the laser setup using a long optical
fiber is possible, if these steps prove necessary for the locking of the laser
cavity.} of the pulse-tube cooler. 

As a status some pictures of the ADR system at PTB are shown in \protect\Figref{img:adr_status}. The ALPS collaboration will soon reach mK-regions, 
after the adjustment and fine tuning of the heat switch. 

\subsubsection{Expected performance}
\label{chap:tdr:detector:expected_performance}

\begin{figure}[!tb]
  \centering
  \includegraphics[width=0.9\textwidth]{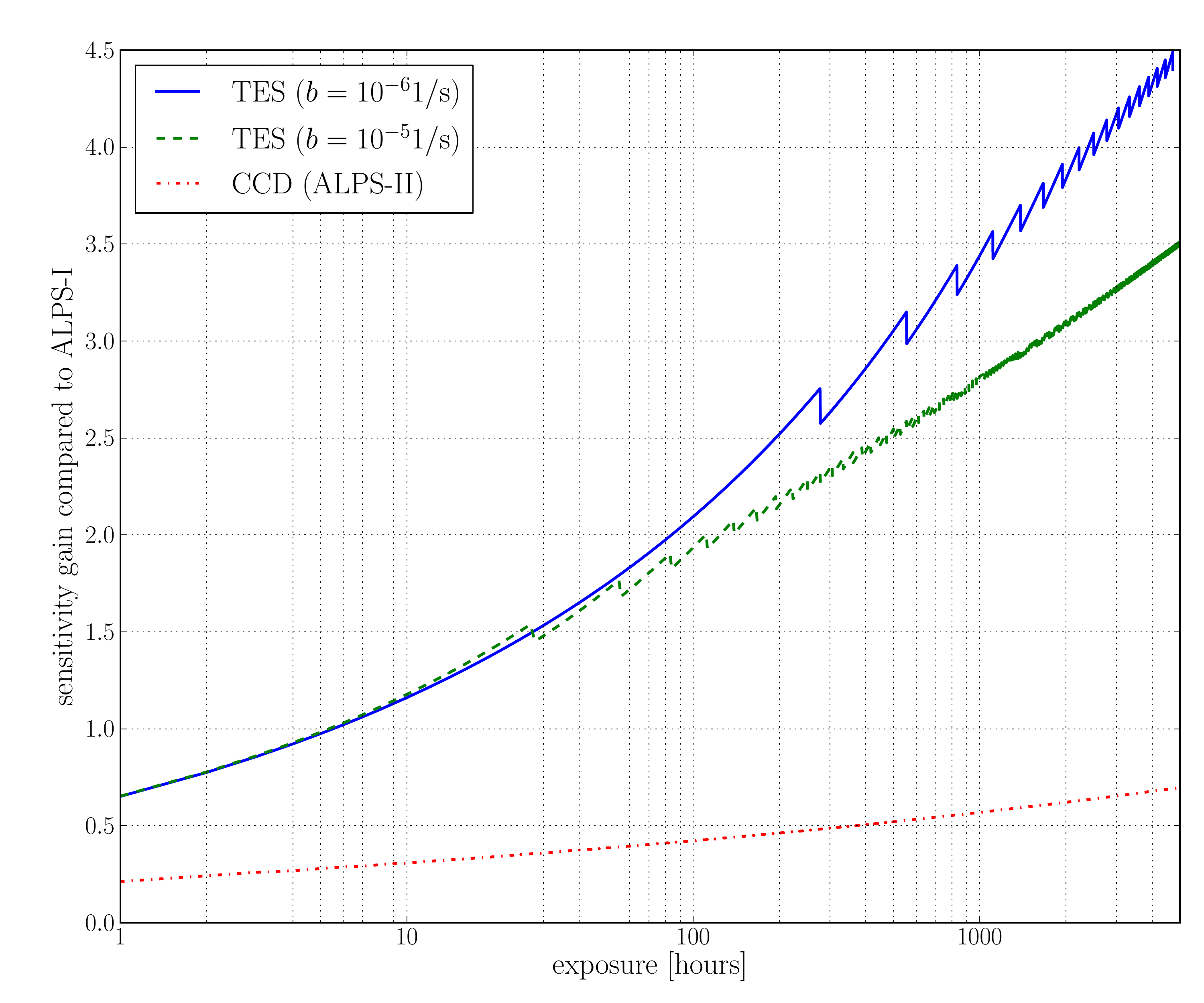}
  \caption[Expected sensitivity gain]{
    Expected sensitivity gain of the different detector systems for
    ALPS-II compared to ALPS-I:
    The dependency of the sensitivity gain on the exposure time is shown for
    the TES with a detection efficiency of 75~\% and
    background rates of $10^{-6}$~s$^{-1}$ and $10^{-5}$~s$^{-1}$,
    respectively, and for the CCD for the parameters given in
    \protect\Tabref{tab:qfactor}. The number of observed background events was
    estimated conservatively by taking the smallest integer larger than the
    expected value for the respective measurement time (i.e., always $\geq1$).
    The kinks in the TES lines are due to accumulation of additional
    background events. The comparison is made against the ALPS-I
    results for an exposure time of 27~h
    (cf.~\cite{Ehret:2010mh}).
  }
  \label{fig:sensitivity-gain}
\end{figure}

\Figref{fig:sensitivity-gain} shows the gain in sensitivity expected for
ALPS-II compared to ALPS-I. The sensitivity for the CCD
improves approximately as $t_\mathrm{meas}^{1/8}$ because it is dominated by
the accumulated dark current. The TES is dominated by the statistics of
the signal, hence the improvement is approximately $\propto
t_\mathrm{meas}^{1/4}$.

A summary of the expected detector performances and the resulting gain in
sensitivity is given in \protect\Tabref{tab:qfactor} for a measurement campaign of
two weeks corresponding to 336~h of data for the CCD and
168~h for the TES%
\footnote{This assumes that the cryogenic
  environment for the TES can be maintained 
  50~\% of the time.
}.

\begin{table}
\begin{tabular}{l|c|c|c|c}
Parameter & Impact & ALPS-I CCD & ALPS-II CCD & TES
\\ 
\hline
Efficiency $QE$ & $g_{a\gamma} \propto QE^{-1/4}$ & 0.9 & 0.012 & 0.75 \\
Detector noise $DC$ & $g_{a\gamma} \propto DC^{1/8}$ & 0.0018~s$^{-1}$ & 0.0012~s$^{-1}$ & 0.000001~s$^{-1}$
\\[4pt] 
\hline
\hline
Sensitivity gain & & 1 & 0.5 & 2.4
\end{tabular}
\caption[Parameters of the ALPS-II detector setups]{Parameters of the ALPS-II detector setups and their impact on the
  sensitivity of the experiment for a wave length of 1064~nm
  compared to the ALPS-I setup for 532~nm and 27~h
  of data. For the ALPS-II sensitivities two weeks of measurement are
  assumed. A sensitivity factor larger than 1 means an improvement of the
  sensitivity. Note that the detector noise for the TES remains to be
  demonstrated.
} 
\label{tab:qfactor}
\end{table}

In the following, details are given for both the CCD and the TES setup.

\subsubsection*{CCD:} 

\begin{figure}[!tb]
  \centering
    \includegraphics[width=0.45\textwidth]{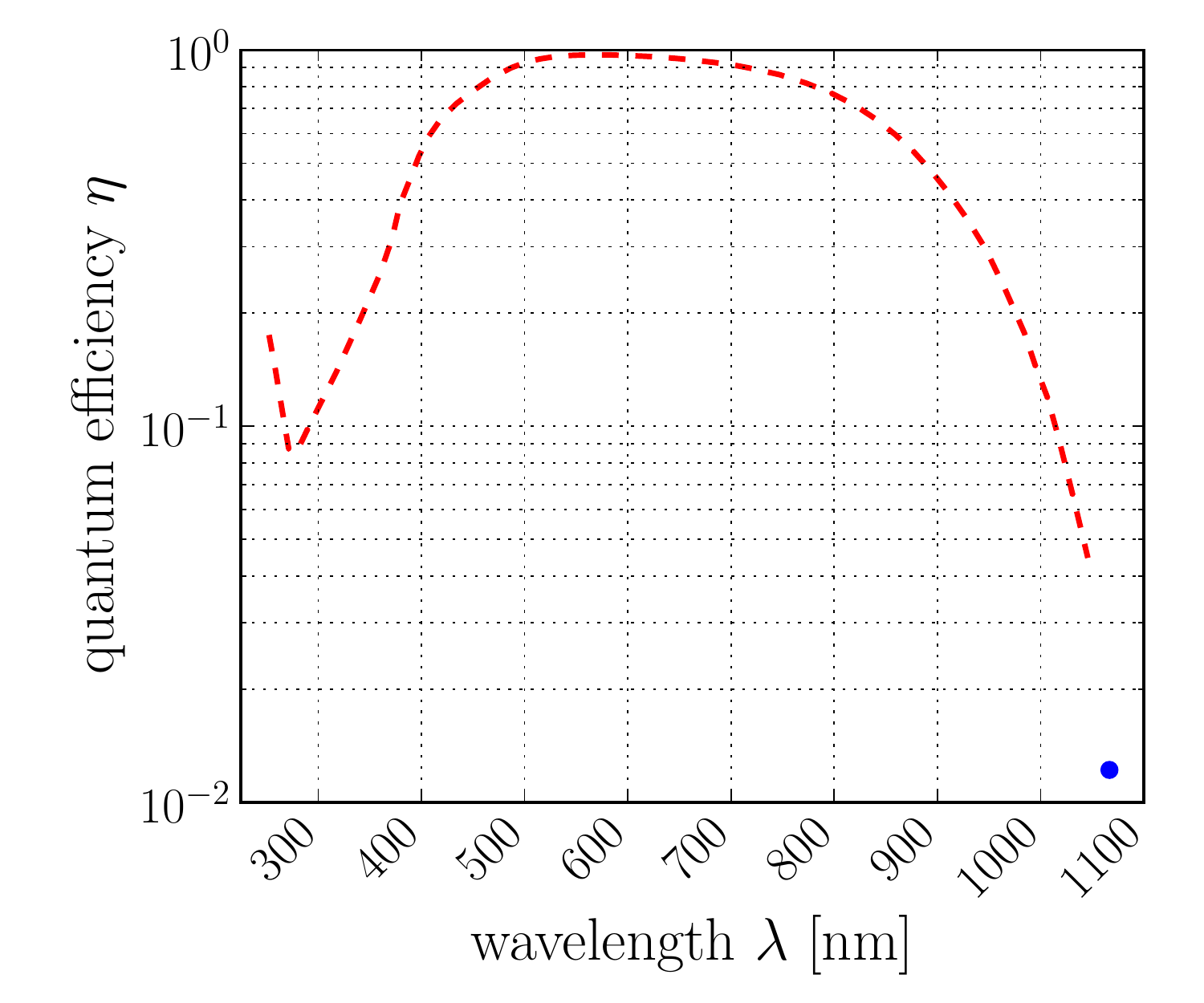}
  \caption[PIXIS 1024B quantum efficiency]{Measurement of the quantum
    efficiency of the PIXIS 1024B: The dashed line is the quantum efficiency
    measured at a temperature of T=25~$^{\circ}$C taken from the
    data sheet~\cite{pixis:datasheet}. The single bullet is the measurement
    of the quantum efficiency at T=-70~$^{\circ}$C.
  }
  \label{img:pixis_quantum_efficiency}
\end{figure}

The dark current of the CCD is known from the data sheet and the experience
in ALPS-I to be 0.0008~$e$~s$^{-1}$ per pixel. The
read-out noise has been determined
\footnote{This value has been established in measurements and differs
  slightly from the value given by the manufacturer
  (3.85~$e$).
}
to be 4.09~$e$
at 100~kHz read-out speed and a quantum efficiency
1.2~\% was measured for a wavelength of 1066.7~nm
and a chip temperature of -70~$^{\circ}$C.
\protect\Figref{img:pixis_quantum_efficiency} shows the quantum efficiency as stated
by the data sheet~\cite{pixis:datasheet} together with the measured value.

Tests showed that the CCD can be operated reliably with exposure times up to
6000~s. For longer observation times, the probability of
contaminating the frame with charged particle background (cosmics and decay
products) leads to a loss of frames.

In ALPS-I the end-mirrors of the production cavity had a limited lifetime
and, thus, had to be replaced between measurements. Therefore, the position
of the signal region changed between data runs and its shape had to be
optimized with a time-consuming procedure resulting in a minimal achieved
size of $3\times 3$ pixel. During all phases of ALPS-II, the position
of the signal region will be fixed. Hence, we expect to achieve a smaller
signal region of $1\times 1$ pixel.

\subsubsection*{TES:} 
For the setup using a Transition-Edge Sensor an intrinsic background level
of less than $10^{-4}$ counts per second can be achieved. Only upper limits
have been set by previous studies~\cite{ISI:000184336600067, schwarzk} and a
reduction up to rates as low as  $10^{-7}$ counts per second might be
possible as estimated for the ALPS setup based
on~\cite{schwarzk}\footnote{For the calculation of the sensitivity gain
  compared to the detector setup in ALPS-I in \Tabref{tab:qfactor}, a background level of   $10^{-6}$ counts per second is assumed.}.

The quantum efficiency of the TES sensor was shown to be as high as 98~\% and of a whole detection setup of 95~\%, 
if optimized~\cite{ ISI:000254121300021}. Within the ALPS environment and especially the different fiber couplings described above, 
a detection efficiency of 60 to 80~\% is expected.

The energy resolution will be about $\frac{E}{\Delta E} = $ 5 to 6, which could improve the background reduction in principle.
A good timing resolution of about 1~$\mu$s easily enables selecting events based on, e.g., shutter conditions.

\clearpage

\subsection{Magnets and cryogenics}
\label{sec:tdr:magnets}

\subsubsection {Introduction} 
To increase the sensitivity for the detection of axion-like particles, the ALPS-II collaboration plans to set up optical cavities both 
on the production and the regeneration side \cite{Hoogeveen:1990vq,Sikivie:2007qm} of the experiment with a power buildup of 5000 and 40000, 
respectively, and magnet strings of superconducting HERA dipoles as long as possible, as the sensitivity for the detection of axion-like particles
scales with the product of magnetic field strength $B$ and magnetic length $L$ \cite{Sikivie:1983ip,Sikivie:1985yu,Raffelt:1987im,Arias:2010bh},
see also \Eqref{eq:ALPs_osci_prob}.

\begin{figure}
\centering
\includegraphics[trim=7cm 8cm 7cm 8cm, clip=true,width=0.75\textwidth]{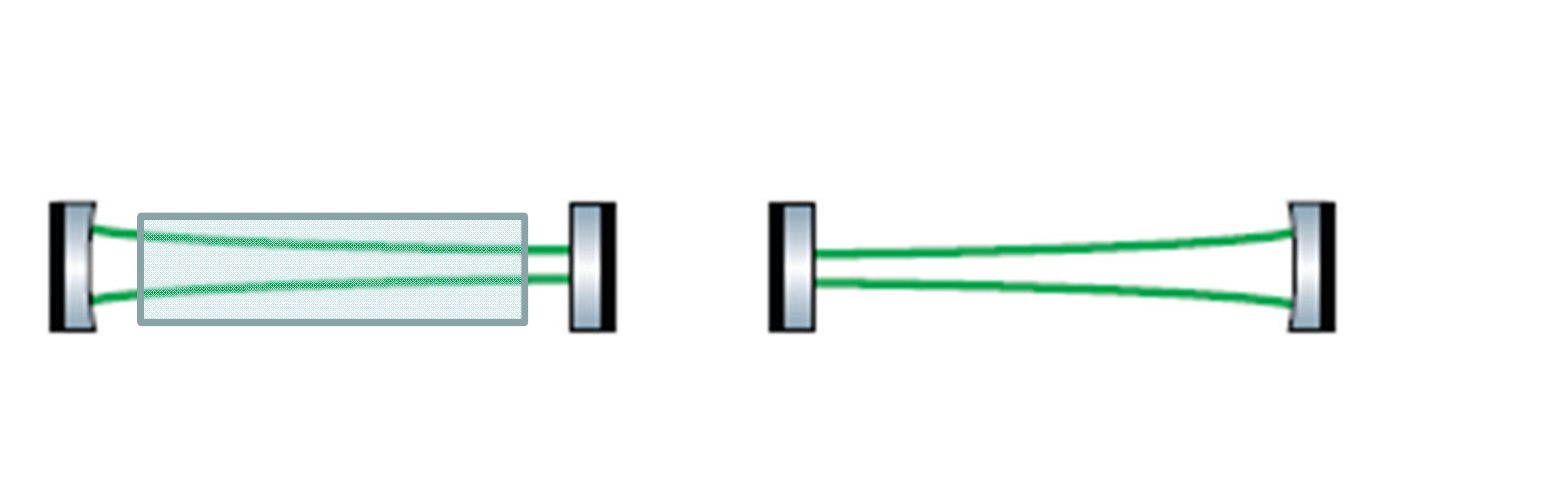}
\caption{Shape of the photon beam in the optical cavity within the surrounding vacuum pipe. 
The vacuum pipe is only shown on one side of the setup (left).}
\label{fig:magnets15}
\end{figure}

The achievable power buildup of an optical cavity with a certain length inside a string of dipoles depends 
on the aperture of the vacuum pipe in the dipoles \cite{3}. The diameter of the circulating photon beam near the 
focusing mirror will be larger for larger lengths of the optical resonator (see \Figref{fig:magnets15}). 
Therefore the clipping losses in the resonator will increase with length, limiting the number of dipoles per string for a given aperture.

 \Tabref{tab:magnets1} shows the maximum number of HERA dipoles, allowing for an optical cavity with a power buildup of 40000 at a 
 wavelength of 
 1064nm for different apertures. The free aperture is taken to be 
 10mm
 smaller, to account for alignment tolerances, the waviness of the vacuum pipe, and fluctuations of the laser position. 
 For comparison the apertures and the values for $B\cdot L$ for an experimental setup using LHC dipoles \cite{Wyss:1999cw} 
 as proposed in \cite{Sikivie:2007qm} or Tevatron dipoles \cite{Edwards:1986fn}, as considered for REAPR \cite{Mueller:2009wt,42,38} are included.

The inner diameter of the vacuum pipe in the superconducting HERA dipole is 
55mm. Due to the curvature of the dipoles 
built for HERA\footnote{See for example \cite{Kaiser:1986ye}. Key people for the development of the superconducting HERA magnets were 
H.~Kaiser, K.H.~Mess, P.~Schm\"user and S.~Wolff.}, the free horizontal aperture is reduced to $\approx$  
35mm. 
However, as will be described below, HERA dipoles can be straightened by a modification of the cold mass. 
This modification yields the full aperture of
 55mm
and allows for the best possible setup for ALPS-II of $2\cdot 12$ dipoles.
However, the total costs of this operation, although not known at present, are expected to be considerable.

Therefore a cheaper way to increase the aperture in the dipoles has been investigated (see below). By a ``brute force'' 
deformation of the magnet yoke an aperture of
 50mm 
(almost straight) can be obtained, allowing for a setup of $2\cdot 10$  dipole with a power buildup of 40000. 
This is the base line layout for the ALPS-IIc proposal.

\begin{table}
\begin{center}
\small
\begin{tabular}{|c|c||c|c|c|} 
\hline
\multicolumn{2}{|c||}{Dipole aperture} & Number of dipoles & $B\cdot L \: [{\rm Tm}]$ &
Length of single string
\\
mm & experiment &  &  &  plus 5m optical setup  \\
\hline  \hline
35 & HERA\footnotemark & $2\cdot 4$& 187 & 44\\ \hline
40 & LHC\footnotemark~\cite{Sikivie:2007qm} &$2\cdot 4$ & 515 & 72 \\ \hline
48 & REAPR\footnotemark~\cite{38} & $2\cdot 6$ & 184  &44 \\ \hline
50 & HERA& $2\cdot 10$ & 468 & 103\\[-1.4ex]
   & {\tiny almost straight} & & & \\
\hline
55 & HERA& $2\cdot 12$ & 562 & 122 \\[-1.4ex] 
   &  {\tiny straight}& & & \\
\hline

\end{tabular}
\caption{Comparison of $B\cdot L$ for HERA dipoles with different apertures and proposals using LHC or Tevatron dipoles.}
\label{tab:magnets1}
\end{center}
\end{table}
\addtocounter{footnote}{-2}
 \footnotetext{HERA: magnetic length 8.83~m, field 5.3~T.}
\addtocounter{footnote}{1}
 \footnotetext{LHC magnetic length 14.3~m, field 9~T.}
\addtocounter{footnote}{1}
 \footnotetext{REAPR magnetic length 6.12~m, field 5~T.}

It should be noted that already a setup for ALPS-IIc with $2\cdot 4$ standard HERA dipoles would result in a
larger sensitivity with respect to $B\cdot L$, compared to laboratory experiments performed up to now\footnote{As detailed
in \Sectref{chap:tdr:goals},  $2\cdot 10$ magnets at the envisaged power build-up suffice to improve current helioscope limits and tackle parameter
regions favored by theory and astrophysical hints.}
(see \Tabsref{tab:magnets1} and \ref{tab:magnets2}). With straight HERA dipoles and also with the 
almost straight dipoles ALPS-IIc will be clearly competitive to the other experimental setups proposed with respect to $B\cdot L$.

\begin{table}
\begin{center}
\begin{tabular}{|c|c|c|c|c|} 
\hline
\small{Experiment} & \small{Year of} & \small{Magnetic length} & \small{Magnetic field} &  \\
& \small{publication} &$L$[m] &$B$ [T] & $B\cdot L$ [Tm]\\ \hline \hline
BFRT~\cite{Cameron:1993mr}&1993 & 4.4 & 3.7 & 16.3\\ \hline
BMV~\cite{Robilliard:2007bq}& 2007& 0.25& 11& 2.8\\ \hline
LIPPS~\cite{Afanasev:2008jt}& 2008& 1& 1.7& 1.7\\ \hline
PVLAS~\cite{Zavattini:2007ee}& 2008& 1& 5& 5\\ \hline
GammeV~\cite{Chou:2007zzc}& 2008& 3&5 &15.0 \\ \hline
ALPS-I~\cite{Ehret:2010mh}& 2010& 4.42& 5.0& 22.1\\ \hline
OSQAR~\cite{Schott:2011fm}& 2011& 14.3& 9& 128.7\\ \hline
\end{tabular}
\caption{Values of $B\cdot L$ for previous laboratory experiments.}
\label{tab:magnets2}
\end{center}
\end{table}

The infrastructure needed for the operation of a superconducting dipole string is specific for each type of dipole, i.e., the LHC, HERA, 
or Tevatron dipole. This refers for example to the cryogenics,
the cryogenic boxes, the power supply, or the quench protection system. Although it is possible to set up a string of the 
three types of superconducting dipoles, considered here at any major laboratory, 
the effort needed would be substantially larger compared to the ``home'' laboratory of the magnet. Therefore we did not consider the 
use of LHC or Tevatron dipoles on the DESY site.

 \subsubsection{Straight HERA dipoles}
Originally the iron yoke, the clamps, the coils, and the vacuum pipe of the HERA dipole were fabricated 
straight; only the outer vacuum vessel was formed as a polygon. 
The welding of the half cylinders of the Helium vessel around the iron yoke was performed in a big tool which forced the cold mass 
to a given curvature.
The beam pipe was forced to follow the curvature by spacers glued to the outside of the pipe.

\begin{figure}
\centering
\includegraphics[trim=3.cm 1.5cm 3.5cm 1.5cm, clip=true, width=0.95\textwidth]{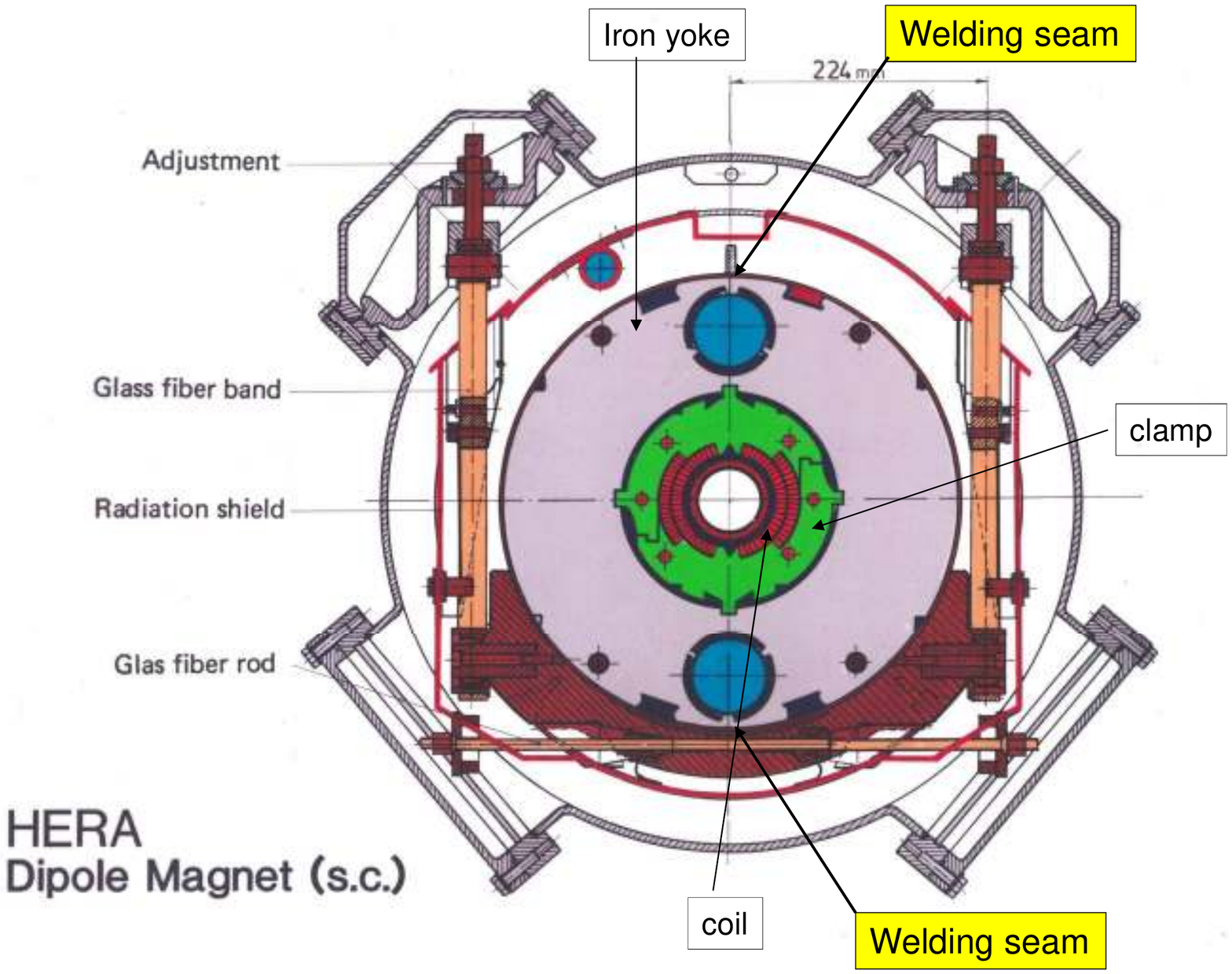}
\caption{Cross section of the HERA dipole cryostat. The welding of the half cylinders of the Helium vessel around the iron yoke was performed in a
big tool which forced the cold mass to a given curvature. The beam pipe was forced to follow the curvature by spacers glued to 
the outside of the pipe.}
\label{fig:magnets1}
\end{figure}

Therefore by cutting the welding seam of the Helium vessel (see \Figref{fig:magnets1}), straightening the yoke and 
welding two straight half cylinders around the yoke, 
it should be possible to obtain a straight dipole magnet. However, this procedure requires the complete disassembly of the magnet 
cryostat and the rebuilding of part of the tooling used originally. 
The know-how for this procedure still exists in one company which originally assembled half of the total number of dipoles. 
The cost for this method of straightening HERA dipoles however, 
although not known at present, is expected to be considerable.

We therefore looked for a simpler and more importantly
cheaper way of straightening the dipole.  Engineering studies showed that a straightening 
of the yoke and thus the beam pipe should be possible by a brute force deformation with
 $\approx$40~kN
from the outer vacuum vessel at the 3 planes of support of the dipole\footnote{Concept by R\"udiger Bandelmann; 
development by Gerhard Meyer. We learned later from~\cite{Sikivie:2007qm} that a similar method had been considered at CERN by 
P.~Pugnat for LHC dipoles.}. 
    
\begin{figure}
\centering
\includegraphics[width=0.5\textwidth]{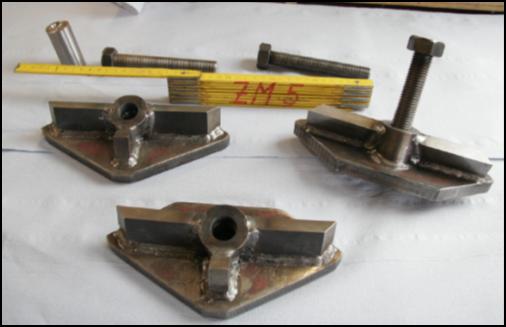}
\includegraphics[width=0.4\textwidth]{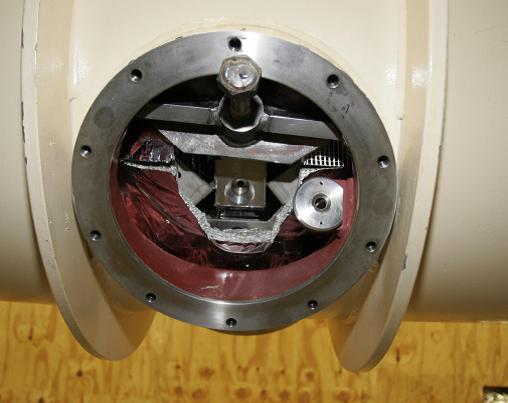}

\caption{Tools for straightening the cold mass (left) and the inserted tools (right).}
\label{fig:magnets2}
\end{figure}

\begin{figure}
\centering
\includegraphics[width=0.75\textwidth]{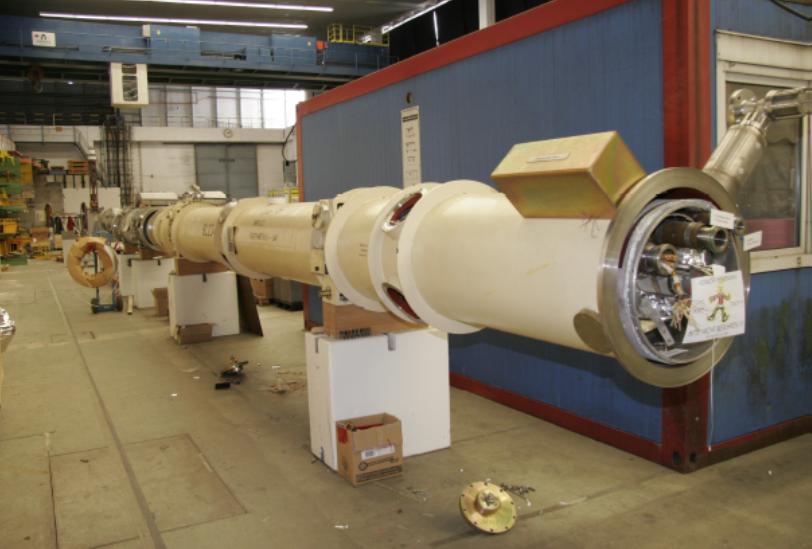}
\caption{Setup of a HERA dipole to test the deformation method. The dipole was used as an exhibit 
for many years and was not foreseen as a spare for HERA.}
\label{fig:magnets4}
\end{figure}

In May 2011 the deformation method was first tried on a HERA dipole serving as an exhibit before. The tools to bend the magnet yoke 
(see \Figref{fig:magnets2}) were inserted into the lower flanges of the outer vacuum vessel at three positions, near the ends 
of the dipole on one side of the vessel and in the middle on the other side (see \Figsref{fig:magnets2} and \ref{fig:magnets4}).

\begin{figure}
\centering
\includegraphics[width=0.9\textwidth]{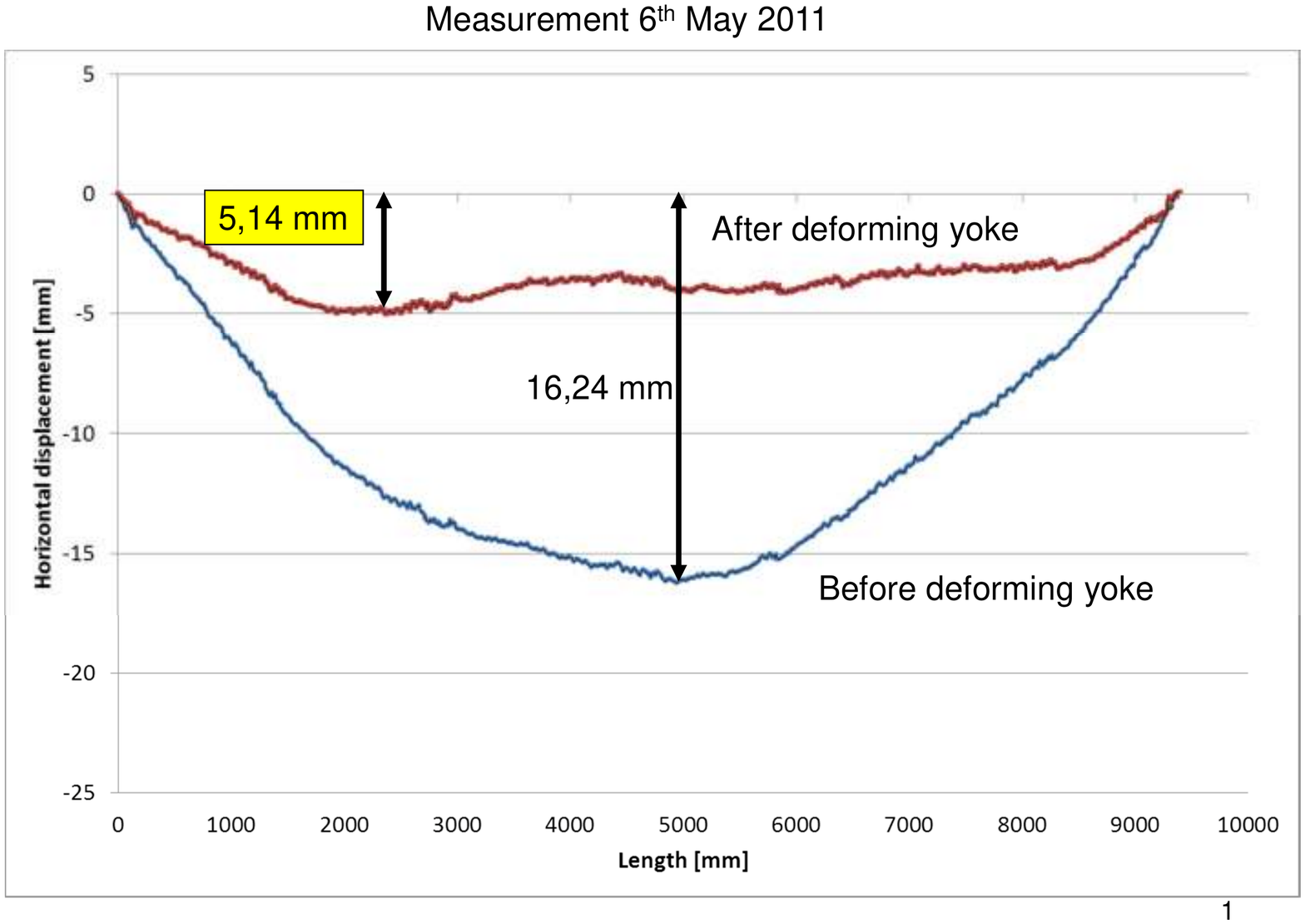}
\caption{Position of the beam pipe axis along the magnet.}
\label{fig:magnets5}
\end{figure}

\Figref{fig:magnets5} shows the position of the center of the beam pipe before and after 
the application of the deforming forces. The deviation from a straight line was 
reduced\footnote{Measurements of the beam pipe were performed by the DESY survey group MEA.} from 
 16.24~mm to 5.24~mm, yielding about 90~\% of the maximum horizontal aperture of 55~mm.

This result proves the applicability of the ``brute force'' method to substantially enlarge the 
horizontal aperture of the vacuum pipe of the superconducting HERA
dipoles.                                                                               
\begin{figure}
\centering
\includegraphics[width=0.9\textwidth]{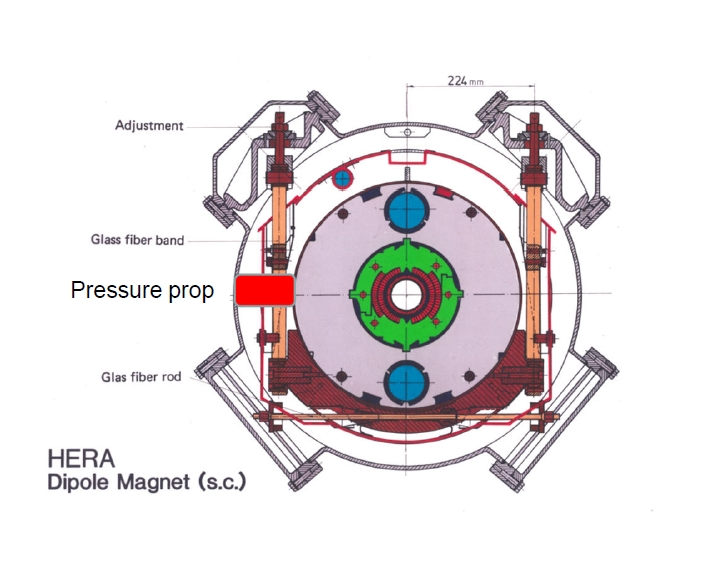}
\caption{Position of a pressure prop in the dipole cryostat.}
\label{fig:magnets6}
\end{figure}

As the deformation of the yoke is elastic, the deforming force has to be maintained during operation at cryogenic temperatures.
Therefore pressure props were designed, 
which keep the thermal flux from the vacuum vessel at room temperature to the yoke at liquid Helium temperature within acceptable limits. 
The pressure props replace the deformation tools after the straightening (see \Figref{fig:magnets6}). 

\begin{figure}
\centering
\includegraphics[width=0.45\textwidth]{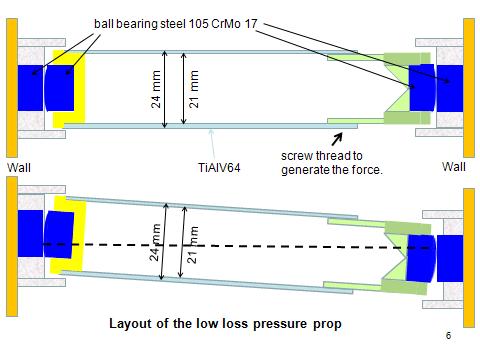}
\includegraphics[width=0.45\textwidth]{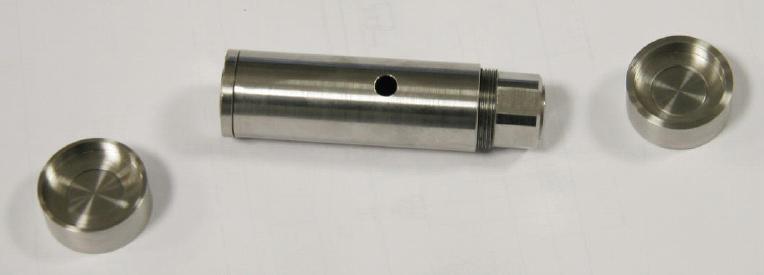}
\caption{Principal layout of a pressure prop with low thermal conductivity and the first prototype.}
\label{fig:magnets7}
\end{figure}

The props near the ends of the dipole must allow for the length change of the yoke during cool-down 
and warm-up and yet maintain the deforming pressure.  A simple solution to this problem is to use a section 
of a sphere which rolls with the motion of the cold mass\footnote{Concept and realization by Gerhard Meyer.}. 
This way the distance between the vacuum vessel at room temperature and the yoke does not change during the 
thermal motion except for the small thermal shrinkage of the sphere. The forces are always perpendicular to the 
surfaces thus avoiding a momentum on the prop and the danger of its tipping over. The thermal heat flow from room temperature to the cold mass at 
 4~K
through the thin walled titanium tube of the pressure prop, see \Figref{fig:magnets7}, amounts to about
 1~Watt
per prop.

A prototype pressure prop was subjected to functional tests in vacuum and at liquid Nitrogen temperature to make sure about the proper 
choice of materials and the validity of the concept.
The test validated the concept and the choice of materials.

\begin{figure}
\centering
\includegraphics[width=0.5\textwidth]{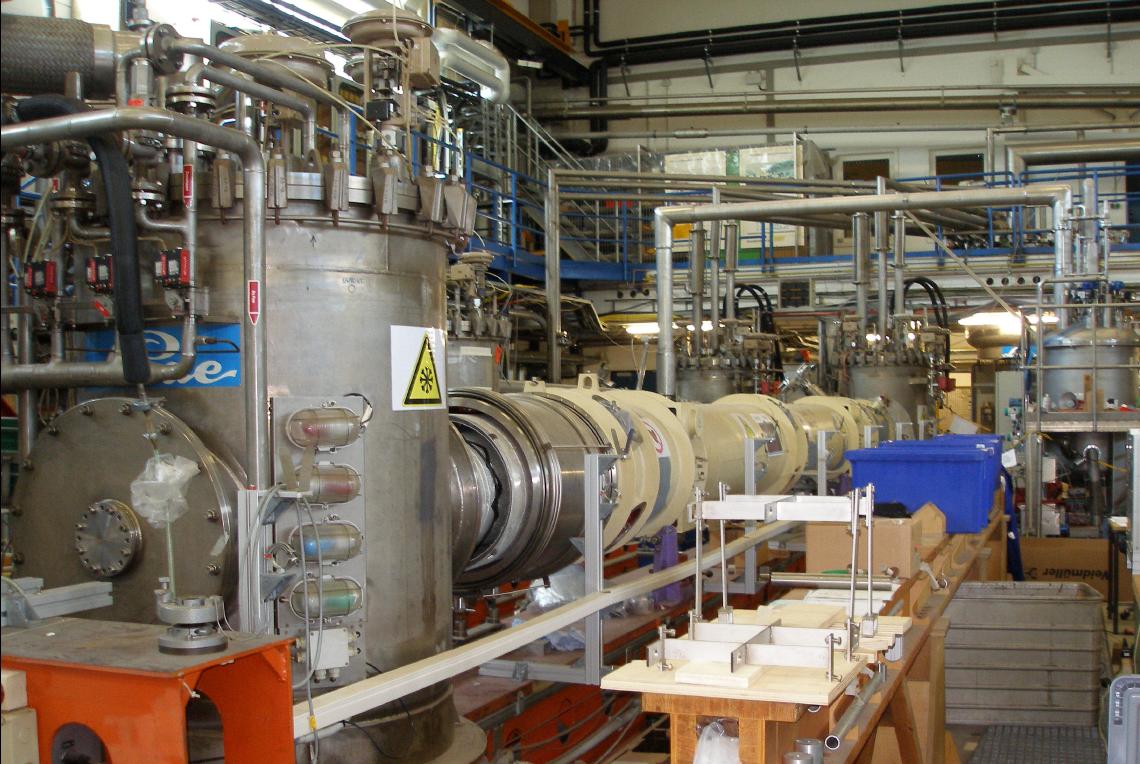}
\caption{Test bench with the HERA dipole used for the ALPS-I experiment. It is located in hall 55 and will be used to
test all dipoles foreseen for ALPS-IIc.}
\label{fig:magnets8}
\end{figure}

\begin{figure}
\centering
\includegraphics[width=0.55\textwidth]{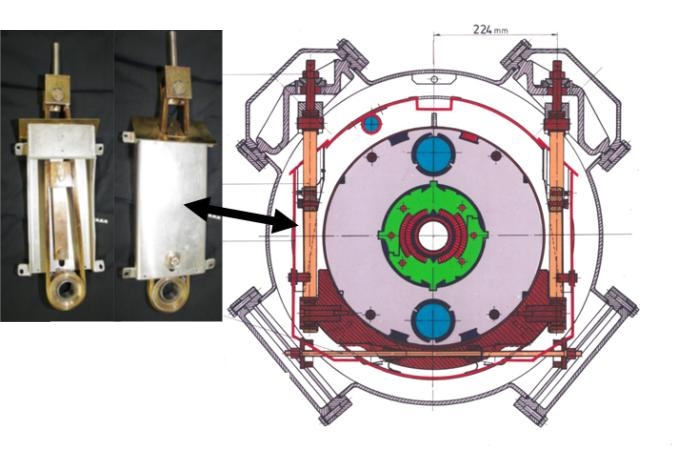}
\includegraphics[width=0.35\textwidth]{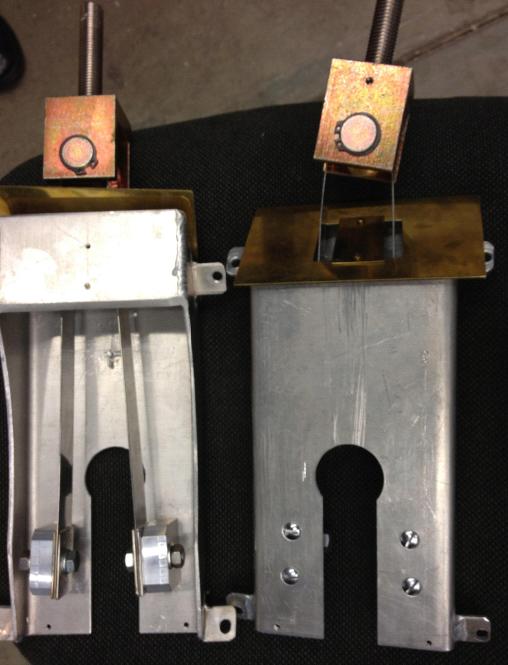}
\caption{Suspension of radiation shield and the modification to allow the insertion of a pressure prop.}
\label{fig:magnets9}
\end{figure}

For the insertion of the deforming tools, 3 of the total 6 suspensions of the cold mass and the radiation shield have to be removed. 
The installed pressure props interfere with a re-installation of the suspensions. While the cold mass is still well supported by the
three remaining suspensions, 
the radiation shield, being a rather soft structure, needs to be supported. By replacing the G10 (glas fiber enforced epoxy) 
loops of the suspensions by 
steel strips (see \Figref{fig:magnets9}), 
connected to the shield part of the suspension off center, and by machining a slit into the shield, 
to give room for the pressure props during installation, 
the suspension of the radiation shield can be reestablished in 
essentially the same way as before.

Recently (August 2012) the HERA dipole at the magnet test bench in hall 55, which was used for the ALPS-I experiment (see \Figref{fig:magnets8}), 
has been straightened on the test bench, with the Helium pipes of the cryogenic boxes remaining connected to the dipole. 

The straightened dipole was operated in September 2012 at cryogenic temperatures reaching a quench current of 
6050 amperes\footnote{The measurements were performed by members of the DESY group MKS.}  . 
This current is slightly higher than the quench current measured on the unmodified dipole  
(5920 amperes) in August 2011. The straightened dipole was operated for 30 hours continuously at the 
design current for ALPS-IIc of 
5700 amperes. The cryogenic losses of the straightened dipole were as for the unmodified dipole.

The straightening procedure 
will be reviewed based on the experience gained during the first straightening experiments.  
Better assembly tools will have to be developed, as the correct positioning 
of the pressure props turned out to be difficult. After the testing of the new tools on a magnet,
pressure props will be fabricated to start the straightening of the magnets foreseen for the ALPS-IIc strings. 
Measurements of the position of the vacuum chambers along the dipoles (see below) will allow for a selection of their location 
in the magnet strings, which yields the largest overall aperture for the optical resonators.

\subsubsection{The ALPS-IIc magnet string \label{sec:alps_magnet_string}} 
A string of $2\cdot 10$ HERA dipoles, straightened with the described deformation method, 
will supply the necessary horizontal aperture for the optical cavities with sufficiently 
low clipping losses. Twenty-four spare HERA dipoles are available.   

\begin{figure}
\centering
\includegraphics[width=0.85\textwidth]{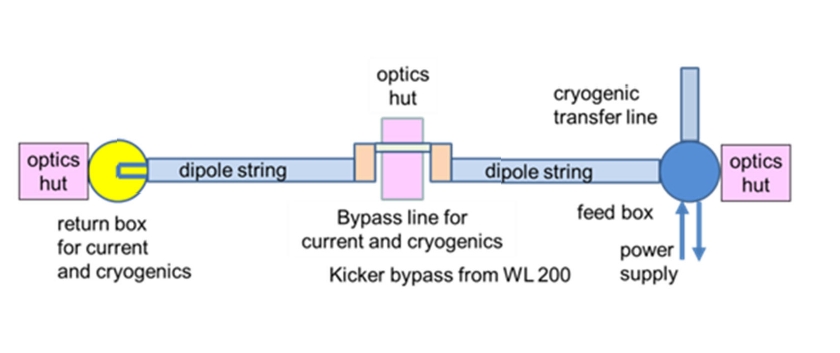}
\caption{Schematic layout of the experimental setup for ALPS-IIc. Cold Helium and the magnet current are fed into a cryogenic box on one side 
of the experiment and run through both dipole strings to a cryogenic box at the other end of the experiment setup, 
where the flow is directed back through the 
dipole strings. Helium and electrical current are led around the optical setup in the middle by a bypass line.
Please note that
all optical elements are placed outside the cold part of the experiment and thus
are not expected to be affected by the cool-down 
of the magnets.}
\label{fig:magnets10}
\end{figure}
 
\begin{figure}
\centering
\includegraphics[width=0.5\textwidth]{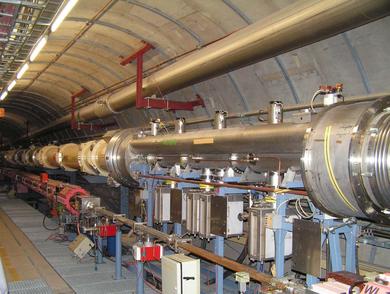}
 \caption{Kicker bypass in the HERA tunnel (West left 200~m). 
The bypass will be installed in the middle of the experiment connecting the two dipole strings of ALPS-IIc 
and pass the magnet current and the cryogenics around the optical setup.}
\label{fig:magnets11}
\end{figure}

\Figref{fig:magnets10} shows the setup of the experiment with the two magnet strings and the optics huts schematically. 
The cryogenic boxes at the ends of the magnet strings are taken from the straight section of HERA selected for the setup of the experiment 
(see \Sectref{sec:tdr:site}
for site considerations). 
To pass the magnet current and cryogenics  around the optical setup in the middle, the ``Kicker Bypass'' from the HERA section 
WL 200 will be used (see \Figref{fig:magnets11}). Essential systems like the quench gas collection line, other warm Helium 
pipes or the dump resistors are not shown.

\subsubsection{Alignment}
\label{sec:tdr:alignment}

\begin{figure}
\centering
\includegraphics[width=0.5\textwidth]{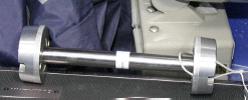}
\includegraphics[width=0.4\textwidth]{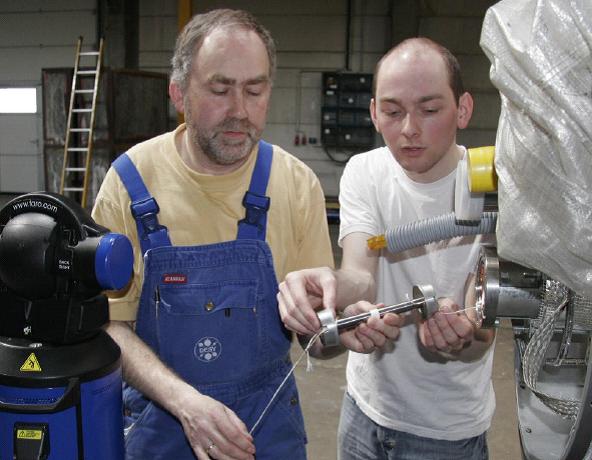}
\caption{The ``mouse'' survey tool, used to measure the position of the middle of the beam pipe along the magnet.}
\label{fig:magnets20}
\end{figure}

The position of the beam pipe along the magnet will be measured for each dipole. This will be done using a simple 
tool called ``mouse''\footnote{Measurements of the beam pipe were performed by the DESY survey group MEA.}, 
which is pulled through the pipe with a string (see \Figref{fig:magnets20}). From the data, the positioning of each dipole can be evaluated yielding 
the largest horizontal aperture in the string of magnets. 
This information will be transferred to existing survey marks on the outer vacuum vessel to allow 
for the proper positioning of the magnet in the HERA tunnel.

\begin{figure}
\centering
\includegraphics[width=0.4\textwidth]{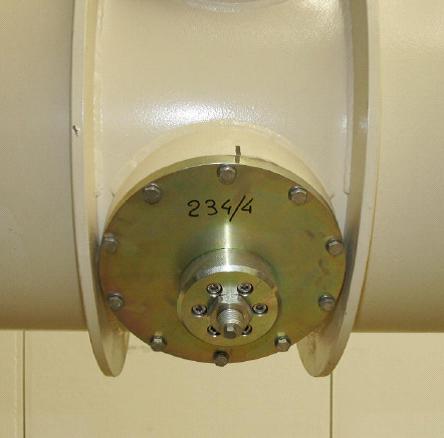}
\caption{Transportation fixture.}
\label{fig:magnets21}
\end{figure}

The dipoles foreseen for ALPS-IIc are stored outside the HERA tunnel in a hall on the DESY site.
For the transportation of the dipoles, the magnet yoke has 
to be fixed to the outer vacuum vessel by special fixtures (see \Figref{fig:magnets21}) to avoid damage to parts connecting the cold mass with the 
vacuum vessel, like the magnet and shield suspensions, or pipes containing the cables for the quench detection. 
This is especially important for the transport into the HERA tunnel, as the magnet has to be rotated out of the horizontal 
to fit through the access shaft.  
It remains to be measured, whether the insertion of the fixtures and the transportation of the magnet change the position of the beam pipe.

The design of the supports of the Helium vessel guarantees that the position of the middle of 
the magnet with respect to the outer vacuum is the same at room 
temperature and at the operating temperature of
 4~K.
However, it remains to be measured whether this also holds for the position of the beam pipe along the magnet. 
This measurement can be performed at the magnet test bench, where the achievable magnetic field will be measured for every dipole 
before the installation into the
HERA tunnel. It is conceivable, to use the ``mouse'' on the test stand by pulling the string by stepping motors 
in the insulating vacuum. 
However, the method suited best for this measurement remains to be determined.

It should be noted that the dipoles can be moved within the completed setup horizontally and vertically by a few millimeters, 
even when the string is operated at 
liquid Helium temperature.

 \subsubsection*{Power Supply for the Magnet String of ALPS-IIc \label{sec:power_supply}}
To operate the dipole strings at a magnetic field of
 5.3~T
a power supply has to deliver a current of
 5690~A
at a voltage given by the resistance of the cable connections 
between the power supply  in the hall and the magnet string in the tunnel. 
For the existing connection of the HERA superconducting magnet ring in hall West 
with a resistance of 
4~m$\Omega$
the 
voltage amounts to
 $\approx$23 Volt
at full current.

The power supply has also to supply additional voltage during the ramp up of the 
current due to the inductive voltage $L\cdot \mathrm{d}I/\mathrm{d}t$ 
at the magnet string.  A ramp rate of 
 2.6 A/s,
corresponding to a ramp time of 36.5 minutes, would require an additional voltage of
 $\approx$3 Volts
at the 
power supply for a string of $2\cdot 10$ dipoles.

An adequate power supply is available in hall West, the original power supply for the superconducting magnets of the HERA proton ring.  
In principle this power supply can be moved to any of the three other halls. 
However, operation of the power supply at the required values, at HERA halls other than hall West, would need investments for additional 
transformers and cabling 
\cite{6}.

 This expenditure can be reduced using the power supply in hall North, originally used for 
 the operation of the superconducting solenoid of the H1-experiment. 
 By reducing the cable resistance between the power supply and the magnet strings to below 
 3~m$\Omega$ the capacity of the power supply of 
 6000~A
 at
  20 Volts
 is sufficient to ramp the current and operate a string of $2\cdot 10$ dipoles. There is an excellent proposal \cite{6} to achieve this 
 reduced resistance, by using the main bus bar system of the 
 HERA electron dipoles in parallel to the cables from the proton ring.

\subsubsection*{Quench Protection}

\begin{figure}[tb]
\centering
\includegraphics[width=0.75\textwidth]{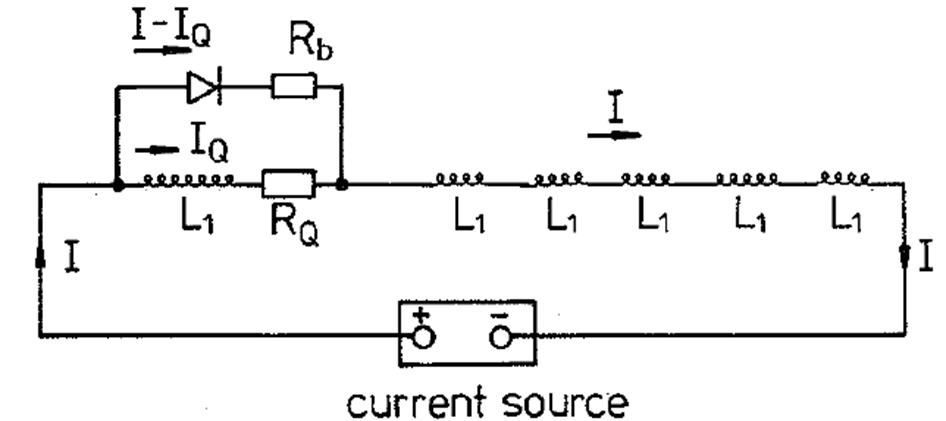}
\caption{Schematics of the HERA quench protection.}
\label{fig:magnets12}
\end{figure}

If a quench happens in one of the dipoles, the current in the coil must be reduced to zero in a short time
 ($<$1~sec)
to avoid overheating and possibly a 
destruction of the normal conducting part of the coil \cite{Mess:1996zu}. 
The solution used at HERA is to bypass each magnet in the chain with a diode; if a magnet quenches, 
the current in the chain is reduced slowly but is guided 
around the quenched coil (see \Figref{fig:magnets12}) by means 
of the diode \cite{Mess:1996zu}.

If a quench is detected, the power supply is switched off, and switches to dump resistors (stainless steel tubes of
 57~mm
diameter and
 2.9~mm
thick walls) are opened  to dissipate the stored energy \cite{Niessen:1991zm}. 
The current decays with the time constant $L/R$, where $L$ is the inductance of the magnet string
 (1.16~Henry)
and $R$ is the resistance of the dump resistors plus the cable connections to the power supply. Two dump switches 
increase the safety in case one switch fails and generate a symmetric voltage profile across the string. 
Requiring that the inductive voltage per individual dipole during the decay of the current be equal to the operation of 
HERA
 ($\approx$13 Volt),
yields a value of 
24 m$\Omega$ for the resistance of the dump resistors, which can be realized by slight modifications to the existing dump resistors.

For the operation of the dipole strings of ALPS-IIc a new quench detection- and protection-electronics system 
is required, as the components used for the HERA system are not fully functional anymore and spares for the outdated
electronics and controls are not available. The work on this system was started in August 2011. 
It is planned to install and test a prototype of the new system at the magnet test bench setup in hall 55 at DESY in 2014. 
The complete quench protection system has to be available after the magnet string installation at the end of 2016. 
This leaves sufficient time for testing and fabrication of the complete system ($\approx$1 year). 

\subsubsection*{Magnet Strings with alternating polarity}
The current in the HERA dipoles can flow in only one direction due to the diodes. 
A string of dipoles with the standard interface connection between magnets can therefore supply only a single direction of the magnetic field. 
A segmentation of the magnetic field into regions of alternating polarity to increase the mass range of the experiment, 
as proposed in reference \cite{VanBibber:1987rq}, is not possible without major efforts like special connection 
boxes between individual dipoles. However, a comparable increase in sensitivity can be achieved  without much effort 
by changing the refractive index in the optical resonators
(see \Sectref{sec:tdr:vacuum}, vacuum system).

\subsection*{Cryogenics} 
Supplying cold Helium to the ALPS-IIc setup is possible in principle to any of the HERA halls from the cryogenics plant on the DESY 
site \cite{Horlitz:1985vr,Clausen:1987hf}.  The effort and the cost of operation depend on the straight section chosen for the setup.

The purchase of a new smaller stand-alone refrigerator was considered, but has been discarded due to the high cost compared to 
the use of the DESY plant. 

There are many advantages in supplying the ALPS-IIc magnet strings with cold Helium from the DESY cryogenics plant like a 
trained and experienced operations crew, or the availability of large storage tanks for the Helium gas, 
when the strings have to be warmed up to room temperature (18 tanks with
 367 $\mathrm{m}^3$
at
 20~bar
are available). 
The storage of
 $\approx$ 170$\mathrm{m}^3$
required for the warm up of a $2\cdot 10$ dipole string is therefore easily possible and --it should be noted-- 
without any limitation to the warm up of the XFEL or FLASH 
accelerator modules with superconducting cavities. 

\begin{figure}[tb]
\centering
\includegraphics[width=0.7\textwidth]{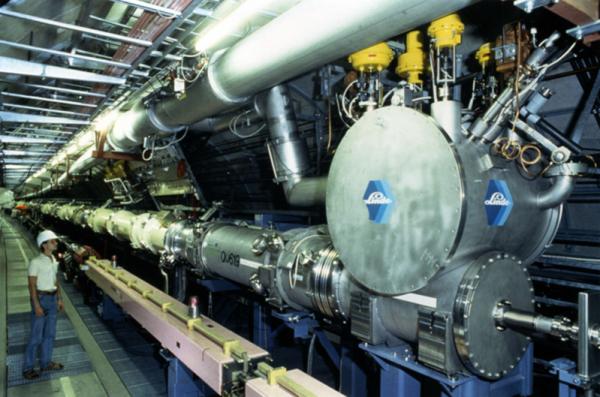}

\caption{Cryogenic feed box for an octant of superconducting magnets at HERA.}
\label{fig:magnets16}
\end{figure}

For ALPS-IIc cold Helium will be transported from the cryogenics plant 
into the tunnel to a box (see \Figref{fig:magnets10}) by the available 
Helium transfer line \cite{Eschricht:1988wg}, which feeds the Helium into the strings at 
one side and returns the Helium to the plant after its passage through the strings. 
At the other end of the strings a similar box will be connected to the magnets, which returns the Helium and 
the current through the strings. It is 
intended to use the cryogenic boxes presently installed in the HERA tunnel at the end of the straight 
section chosen for the setup of ALPS-IIc. However, the boxes, 
as installed for HERA (see \Figref{fig:magnets16}), have to be moved to the other side of the straight 
section to match the connection pattern of the dipoles. They must be disconnected from the Helium transfer line and other cryogenics 
equipment like pre-cooler or valve stations and be reconnected at their new position. 

\begin{figure}[tb]
\centering
\includegraphics[width=0.8\textwidth]{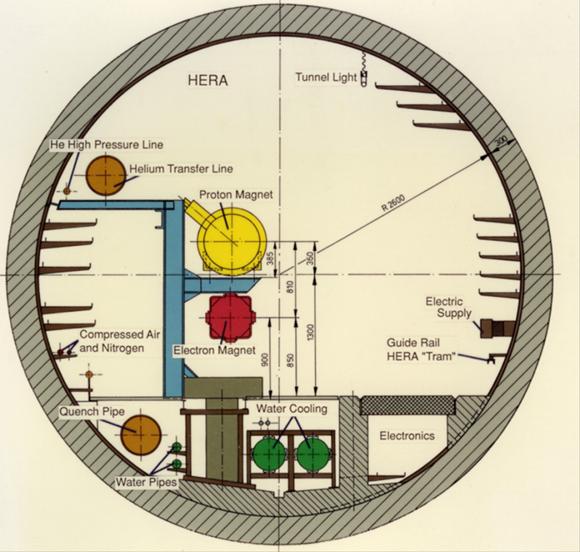}
\caption{Cross section of the HERA tunnel in the arcs. 
The ALPS-IIc setup in the straight sections of the HERA tunnel will look very similar, 
the main difference being that the components of the HERA electron ring will have been dismantled.}
\label{fig:magnets14}
\end{figure}

\begin{figure}[tb]
\centering
\includegraphics[width=0.45\textwidth]{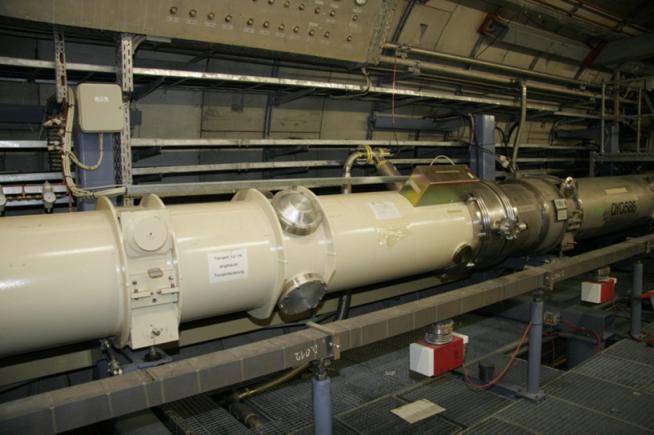}
\includegraphics[width=0.45\textwidth]{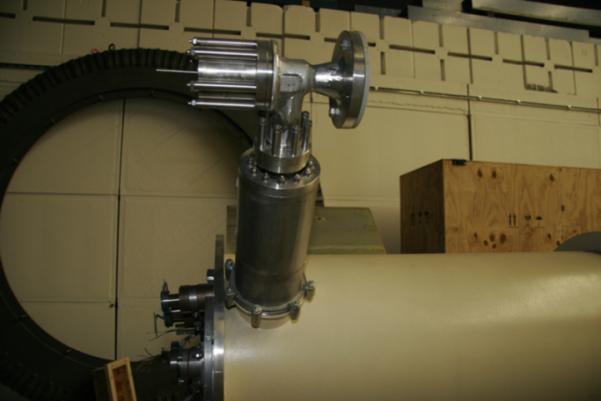}
\caption{``Kautzky''-relief valve connected to the quench gas collection line (left)  and ``Kautzky''-relief valve (right).}
\label{fig:magnets18}
\end{figure}

In case of a quench in the dipoles or a failure of the insulation vacuum, the Helium pressure 
in the magnets will increase and eventually has to be relieved before it reaches a level
($\approx 20$ bar)
which might damage the magnets. 
The Helium is released via an automatic valve (see right side of \Figref{fig:magnets16}),
named after its inventor Kautzky from Fermilab, into the quench gas collection tube
(see \Figsref{fig:magnets14} and \ref{fig:magnets18}),
which guides the gas back to the cryogenics plant. In the valve a reference pressure and 
the pressure in the magnets are compared, thus defining a threshold for the pressure relief.
The reference pressure is supplied from a high pressure line (see \Figsref{fig:magnets14} 
and \ref{fig:magnets18}) and will be set to
 14~bar
like at HERA.
The existing quench gas collection tubes and the high pressure lines can be used for 
ALPS-IIc by adding new connections to the Kautzky valves of the magnets in the strings.

To operate the magnet strings, pressure code regulations have to be met. 
The necessary pressure tests will be performed as for HERA.

The total heat load at the
 4 K
level of the straightened magnet strings for ALPS-IIc amounts to about
 140 Watt,
substantially less than the heat load of about
 400 Watt
for an octant, which is the smallest cryogenic unit of HERA's superconducting magnet ring.   

As the clipping losses from the stored laser beam power of
 150~kWatt
on the vacuum pipe aperture are well below 
 1~Watt
in the entire magnet string,
 it can be concluded that the cryogenic operation of the dipole strings will not pose any problems. 

\subsection{Vacuum system}
\label{sec:tdr:vacuum}

The vacuum system for the optical resonators of ALPS-IIb and
ALPS-IIc and the insulating vacuum system for the superconducting
magnet strings will be described. As we will outline below, here, we profit
very much from the experience gained at HERA. In addition the concepts for
changing the refractive index in the cold pipe within the optical resonators
of ALPS-IIc will be described. 

\subsubsection{The vacuum system for the optical resonators of ALPS-IIb}

In the straight section HERA West two optical cavities of about
100m length can be set up with very little effort needed for the
vacuum system. The still existing vacuum pipe of the HERA proton ring can be
used for ALPS-IIb as it is straight over about $2\cdot 160$m in
contrast to the other straight sections of HERA, where vertical bending
magnets limit the available straight length in the vacuum pipe to about
$2\cdot 60$m. The aperture of the proton beam pipe is large
(cf.~\Figref{fig:aperature_west}) -- except for one location which can be
easily modified -- and allows the operation of optical cavities at very low
clipping losses.

Locations  in hall West and in the tunnel on both sides of the hall have
been identified, where little effort is required to remove the existing
accelerator installation to free space for the cleanrooms at an adequate
distance from the hall. The vacuum vessels for the optical components will
be identical to the ones used at ALPS-IIc (see below).

The existing vacuum system is equipped with a large number of ion getter
pumps and titanium sublimation pumps which will allow to obtain very low
vacuum pressures in the optical resonators. At HERA a pressure of
$10^{-10}$~mbar
was achieved. For the regeneration cavity the operation of
ion getter pumps is problematic as they are a potential source of light from
glow discharges in the pumps. Therefore they will be switched off during the
measurements searching for hidden photons. However, the pressure in the pipe
can be easily maintained at about
$10^{-6}$~mbar
by turbo-molecular pump-stations.

\subsubsection{The vacuum system for the optical resonators of ALPS-IIc} 

\begin{figure}[tb]
\centering
\includegraphics[width=0.5\textwidth]{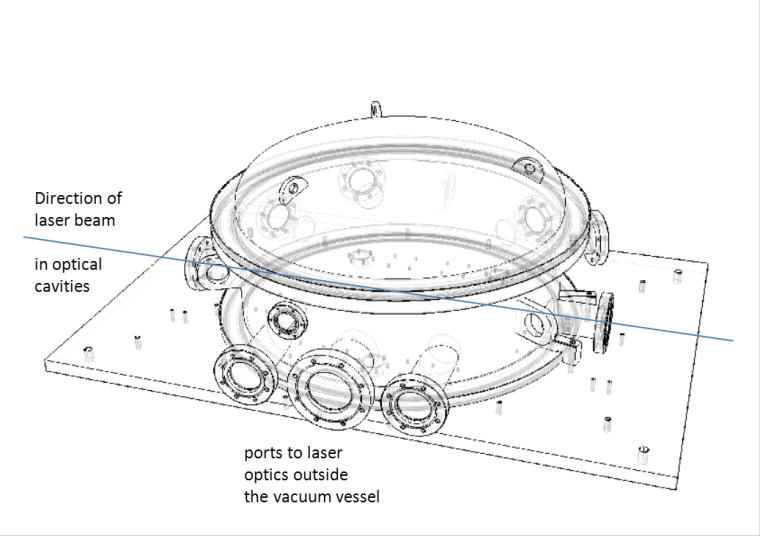}
\includegraphics[width=0.35\textwidth]{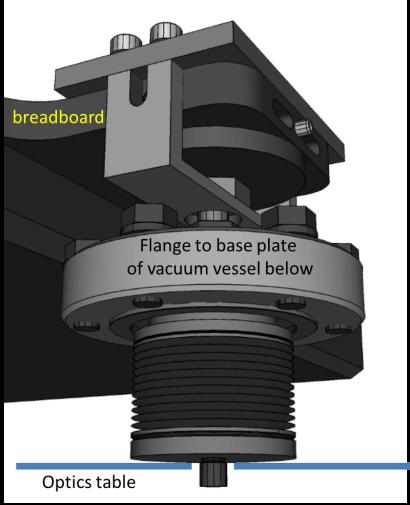}
\caption{Vacuum vessel for the optical elements in the middle of the experimental setup and one support of the optical breadboard.}
\label{fig:vacuum1}
\end{figure}

Three vacuum vessels at room temperature contain the optical elements of the
experiment at the ends of the magnet strings and in the middle between the
strings (see \Figref{fig:vacuum1}). 
In each vessel the breadboard for the optical elements is supported from the
optical table at three points. The supports are mechanically decoupled from
the vacuum vessels by soft bellows (see \Figref{fig:vacuum1}). 
The vessels will be supported from the floor of the cleanrooms. This concept
will be tested at ALPS-IIa.

The vessels will be separable from the vacuum in the magnet strings by all-metal gate valves, to allow venting of the vessels for work on 
the optical components, while the magnet strings are at liquid Helium temperature. 
If the work on the optical components in the vented vessel requires transmission of the laser light through both production and regeneration 
cavity, i.e., an opening of the gate valves, a complete warm up of the magnet strings
will be necessary. Flanges on the vessels with electrical feed-throughs will allow the connection of electro-optical elements and sensors. 
A flange in the central vessel will allow the operation of the ``light tight wall'' shutter.

The large flanges at the top and at the bottom of the vessels will be sealed by O-rings to ease the access to the optical elements in the 
vessels. To reduce the permeation through the O-rings there will be two rings per flange 
spaced a few mm apart. The space between the O-rings will be pumped to about
1 mbar
by a small membrane pump. This concept allows to obtain pressures below 
$10^{-8}$~mbar
in the vessels.

The vessels will be pumped by oil-free turbo molecular pump stations and NEG pumps. 
The use of NEG pumps which pump only chemically active gases, will allow the insertion of Helium gas into the 
vacuum system (see below), while maintaining a low partial pressure of active gases.

\begin{figure}[tb]
\centering
\includegraphics[width=0.4\textwidth]{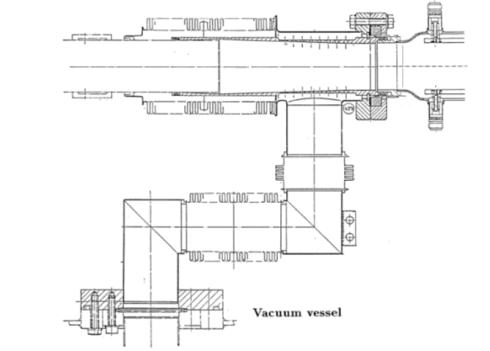}
\includegraphics[width=0.45\textwidth]{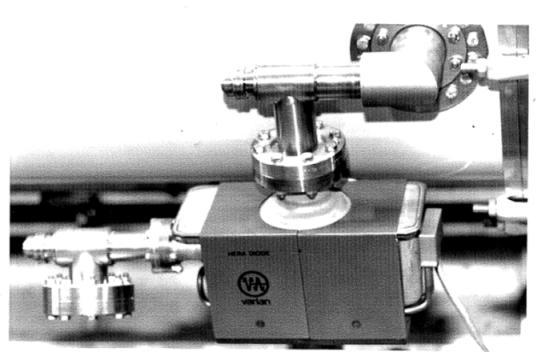}
\caption{Pumping port (left) and pumping port with Ion Getter Pump (right).}
\label{fig:vacuum2}
\end{figure}

\begin{figure}[tb]
\centering
\includegraphics[width=0.35\textwidth]{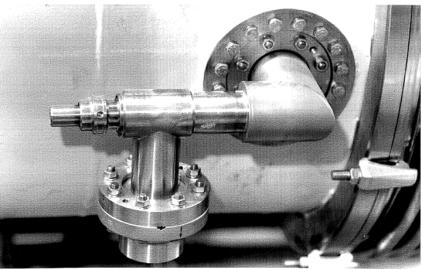}
\includegraphics[width=0.3\textwidth]{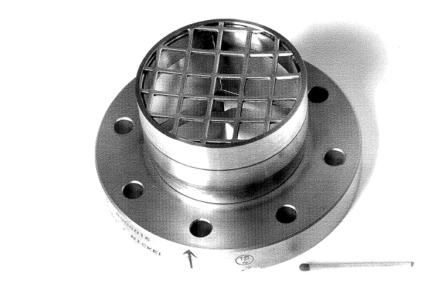}
\caption{Pumping port with rupture disk.}
\label{fig:vacuum4}
\end{figure}

\begin{figure}[tb]
\centering
\includegraphics[width=0.6\textwidth]{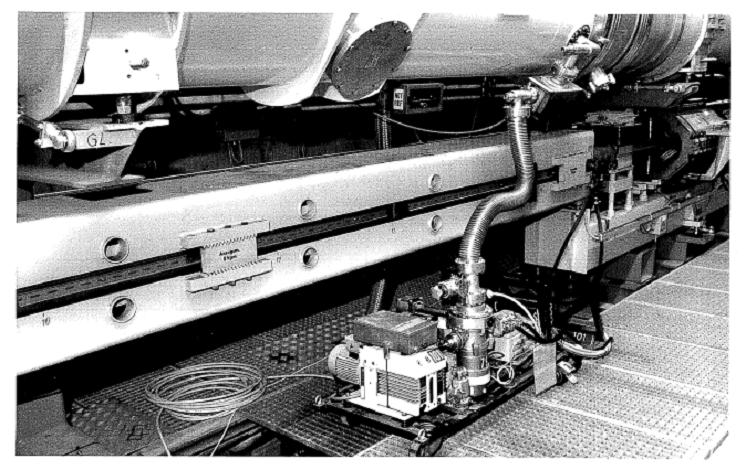}
\caption{Turbo-molecular pump station.}
\label{fig:vacuum5}
\end{figure}

Each string of 10 superconducting HERA dipoles will be containing 2 dipoles
with a pumping port (see \Figref{fig:vacuum2}), one being connected via a
manual valve to an Ion getter pump of 60 liters per second
\cite{Bohnert:1992ra}, the other to a rupture disk, which serves as an
overpressure valve (see \Figref{fig:vacuum4}). For pump down from atmospheric
pressure to 
$<10^{-5}$mbar,
a pump station with a turbo molecular
pump will be connected (see \Figref{fig:vacuum5}). Once the beam tubes in the
magnet strings approach liquid Helium temperature external pumps are not needed
anymore as the wall of the tubes acts as a cryo-pump, and the external pumps
will be disconnected by the manual valves. The pressure in the cold sections
will be well below
$10^{-10}$~mbar.

The procedures for assembly and leak checks will be the same as for HERA~\cite{Bohnert:1991jv}. 
     
\subsubsection{The insulation vacuum system}

At HERA the insulation vacuum for the chain of superconducting magnets was
segmented into sections by vacuum barriers in the quadrupole cryostats
\cite{Bohnert:1991jv}. At ALPS-IIc there are no vacuum barriers, so the two
100~m
long strings of cryostats, connected by the kicker bypass line (see
Sect.~\ref{sec:alps_magnet_string}), have a common insulating vacuum. This
will not pose any problem, as demonstrated at HERA, where during operation
the vacuum barriers were bridged by vacuum bypass lines, forming one
$\approx$600~m
long insulation vacuum section.  The insulation vacuum of the
magnet strings will be continuously pumped by a few turbo molecular pump
stations as at HERA to a pressure of about
$10^{-6}$~mbar.

The procedures for assembly and leak checks will be the same as for HERA~\cite{Bohnert:1992ra}. 

\subsubsection{Increasing the refractive index in the optical resonators of ALPS-IIc }
To improve the range of ALPS-IIc concerning the coupling constant and the mass of the searched for ALPs, the momentum transfer $q$ 
between the laser photon and the axion-like particle has to be varied, cf. Sect.~\ref{sec:alps_osci}. 
This can be achieved  by a change of the refractive index $n$ in the beam pipe for both the production and the regeneration side of the experiment. 
The momentum transfer is given by:
\begin{equation}
 q=n\cdot \omega-\sqrt{\omega^2-m^2} 
\label{eq:refraction}
\end{equation} 
where $\omega$ is the photon energy and $m$ is the mass of the axion-like particle. 

In ALPS-I~\cite{Ehret:2010mh}, the refractive index was changed by injecting Argon gas into the vacuum pipe. 
This procedure is applicable also to ALPS-IIb.
However, as the wall of the vacuum pipe at ALPS-IIc is at liquid Helium temperature, in contrast 
to the experimental setup of ALPS-I , gases will condense on the cold surface with too low a vapor pressure to be of any use.

\begin{figure}[tb]
\centering
\includegraphics[width=0.70\textwidth]{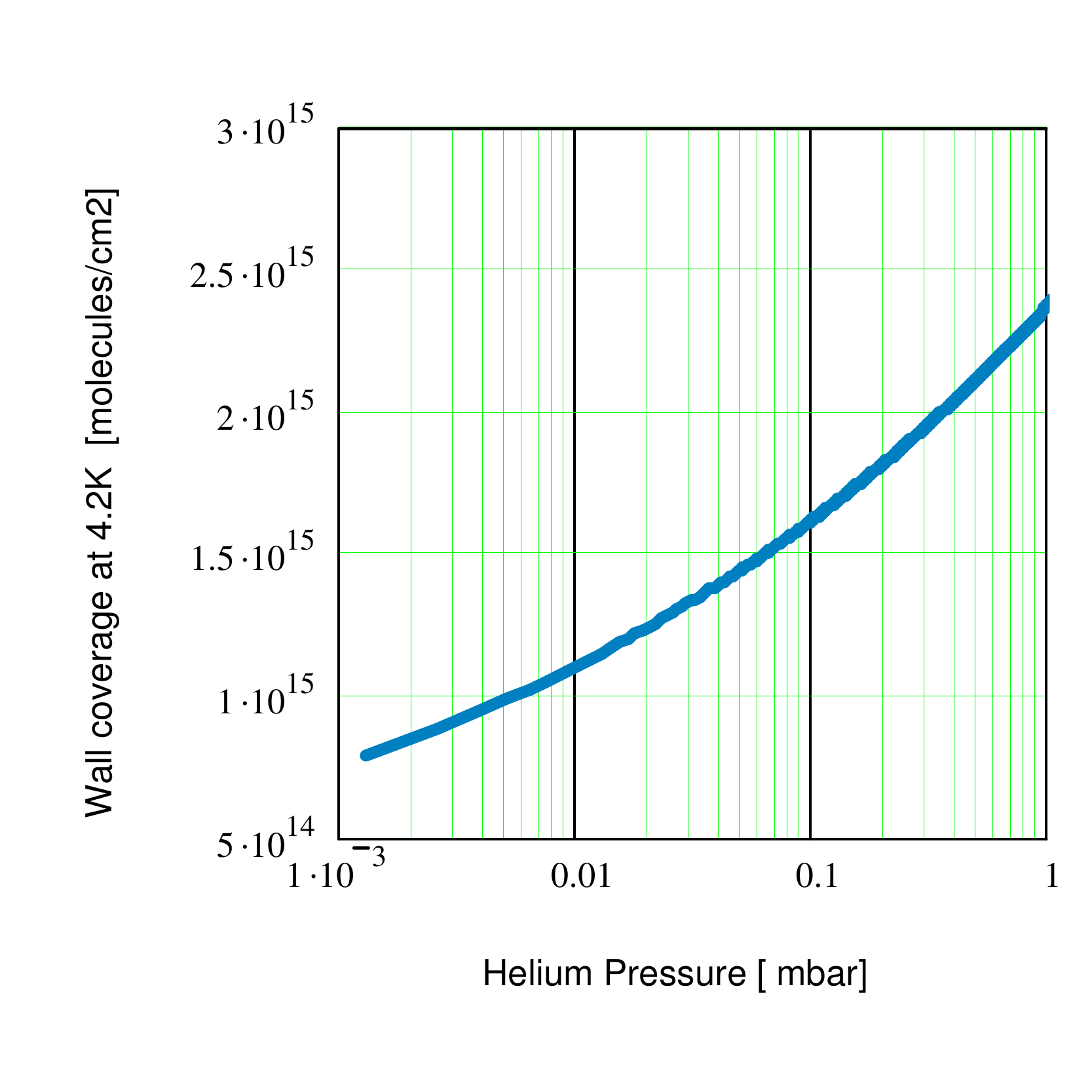}
\caption{Wall coverage in $\frac{\rm{molecules}}{\rm{cm}^2}$ as a function of pressure. 
The scaling formula is taken from E. Wallen~\cite{ISI:A1997YG05800019}.}
\label{fig:vacuum6}
\end{figure}

Only the insertion of Helium gas can lead to a pressure increase in the pipe, yielding adequate values for the refractive index. 
Helium atoms will also condensate on the wall of the pipe, but as soon as the coverage exceeds a 
monolayer ($\approx10^{15}\frac{\rm{molecules}}{\rm{cm}^2}$) the gas pressure in the pipe at liquid Helium 
temperature will reach values (see \Figref{fig:vacuum6}), 
which lead to a sufficiently large increase of the refractive index and thus the momentum transfer $q$. 


\begin{figure}[tb]
\centering
\includegraphics[width=0.80\textwidth]{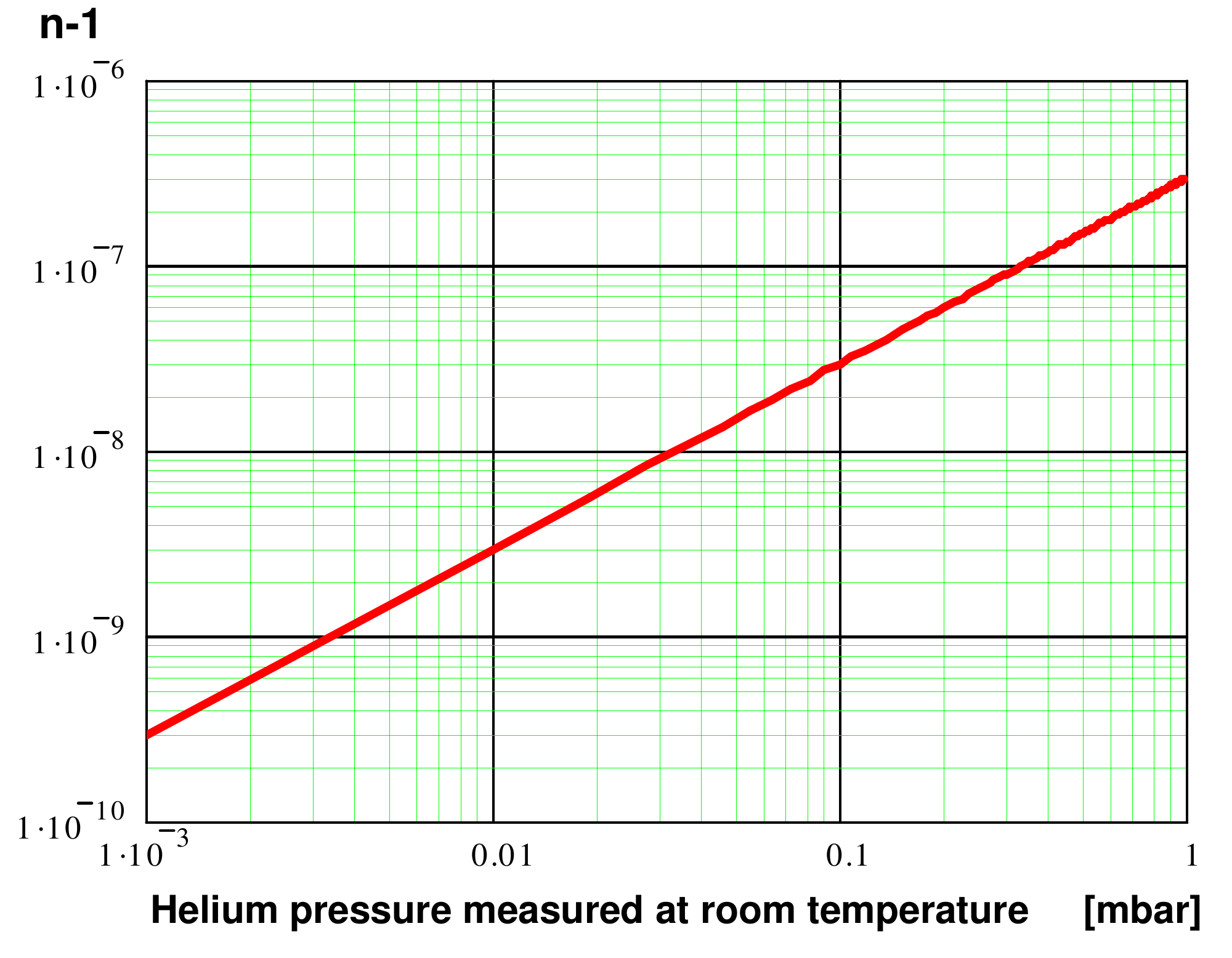}
\caption{Relation between Helium pressure measured at room temperature and deviation from vacuum refractive index $n-1$.}
\label{fig:vacuum7}
\end{figure}

The pressure shown in \Figref{fig:vacuum6} is measured at room temperature, for example at the pumping port shown in \Figref{fig:vacuum2}. 
The pressure $p_{\rm w}$ at room temperature ($T_{\rm w}$) and the pressure $p_{\rm c}$ in the cold pipe ($T_{\rm c}$) are connected by the Knudsen relation
\begin{equation}
  \frac{p_{\rm c}}{p_{\rm w}}=\sqrt{ \frac{T_{{\rm c}}}{T_{{\rm w}}} } \ .
\label{eq:knudsen}
\end{equation} 
 Using this relation and the Lorentz-Lorenz law, one can determine the refractive index \cite{Ruoso:1992nx} in the cold pipe from 
 the pressure $p_{\rm w}$ measured at room temperature (see \Figref{fig:vacuum7}), through
 \begin{equation}
  (n-1)= \sqrt{\frac{T_{\rm w}}{T_{\rm c}}} (n_{\rm w} -1) \frac{p_{\rm w}}{1000 \ {\rm mbar}} \ ,
\label{eq:pressure_vac}
\end{equation} 
where 
$T_{{\rm w}}=300$K,
$T_{{\rm c}}=4$K
and $n_{\rm w}$ the refractive index at room temperature and 
1000~mbar.

\begin{figure}[tb]
\centering
\includegraphics[width=0.90\textwidth]{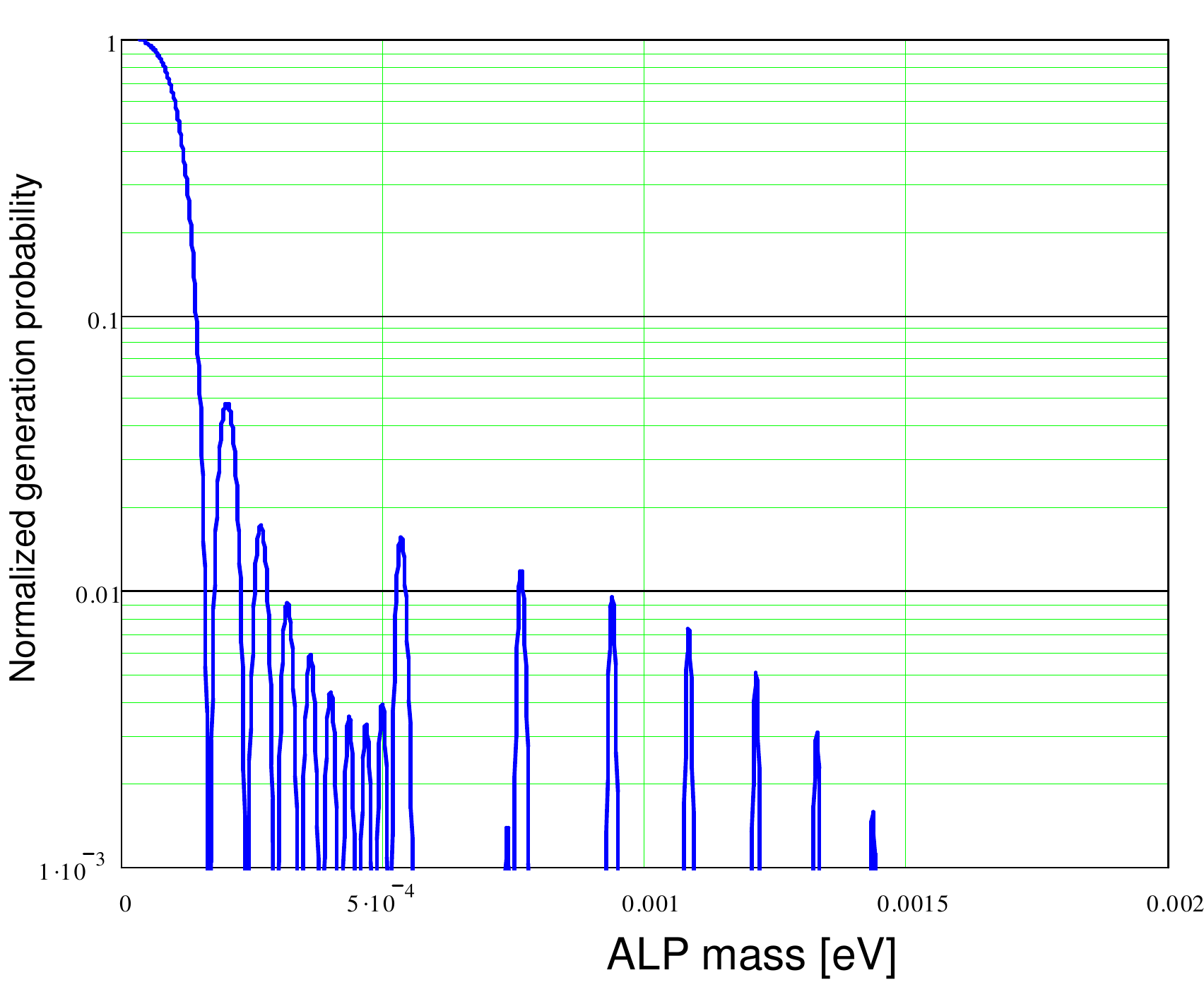}
 \caption{Normalized production probability at $10^{-3}$~mbar as a function of  ALP mass.}
\label{fig:vacuum8}
\end{figure}
The production probability for axion-like particles \cite{Arias:2010bh} (see \Eqref{eq:ALPs_osci_prob}) does essentially not change up 
to a pressure (at room temperature) of
$\approx 0.001$~mbar
for the chosen photon wavelength of
1064~nm.
\Figref{fig:vacuum8} shows the spiky structure of the probability for a pressure of
$10^{-3}$~mbar. 
Varying the pressure to higher values (up to 
 $\approx$0.5~mbar)
will shift the position of the spikes and thus leading to a smooth coverage
of the ALP's mass range.

A pressure stability of
0.001~mbar
will keep the conversion probability in the production and regeneration 
regions stable at a level better than
99\%. 
This will be feasible by measuring the pressure 
with capacity manometers, as the partial pressure of
Helium is dominant in the range of interest, compared to other 
gases in the warm part of the vacuum system. 

To establish a certain Helium pressure in the cold pipe, Helium gas will be injected into the vessels at 
room temperature at the ends of the magnet strings. 
Helium atoms will condensate first on the wall of the cold pipe close to the 
point of injection. Then the wall coverage will slowly propagate along the pipe, eventually 
reaching the far end of the pipe \cite{ISI:A1997YG05800019}. 
In the pressure range of interest for ALPS-IIc it will take a few 
hours to reach a homogeneous and stable coverage, indicated by equal Helium pressure in the vacuum vessels at room temperature.

For the variation of the momentum transfer $q$ the Helium pressure will be changed upwards from
0.001 mbar
to 
0.5 mbar,
as the removal of Helium from 
the walls of the cold pipe will require a warm up of the magnet strings well above 
4.2K.

The round trip losses in the optical cavities due to Rayleigh scattering off Helium atoms will be very small ($2\cdot 10^{-8}$ at 
0.5~mbar), 
thus not leading to a reduction of the power buildup of the cavities. The additional thermal load on the magnet strings due the scattered photons is 
negligible
 ($\approx$0.01~Watt).

\subsection{Data acquisition}


The data acquisition system (in the following DAQ) will provide the
front-end to the detector read-out and various subsystems (including the
slow-control).  The front-end will  include control over various subsystems
and will record monitoring informations during the operation of the
experiment in data-taking and calibration modes. The various data-streams
from the detector and the monitoring systems will be collected and stored
for later analyses. 

\paragraph{Technical challenges}

The DAQ has to be scalable to the different stages of the ALPS-II
experiment and to be flexible for changes in the photon detector during the
different setups.  The amount of data to be read-out and handled is modest
and within the performance of available technology.  We  are expecting event
rates well below kHz in triggered mode (mostly during calibration runs) for
the TES and CCD read-out rates probably well below a Hz
for calibration and even smaller for data-taking.

\paragraph{Conceptual design}

\begin{figure}
 \includegraphics[width=\linewidth]{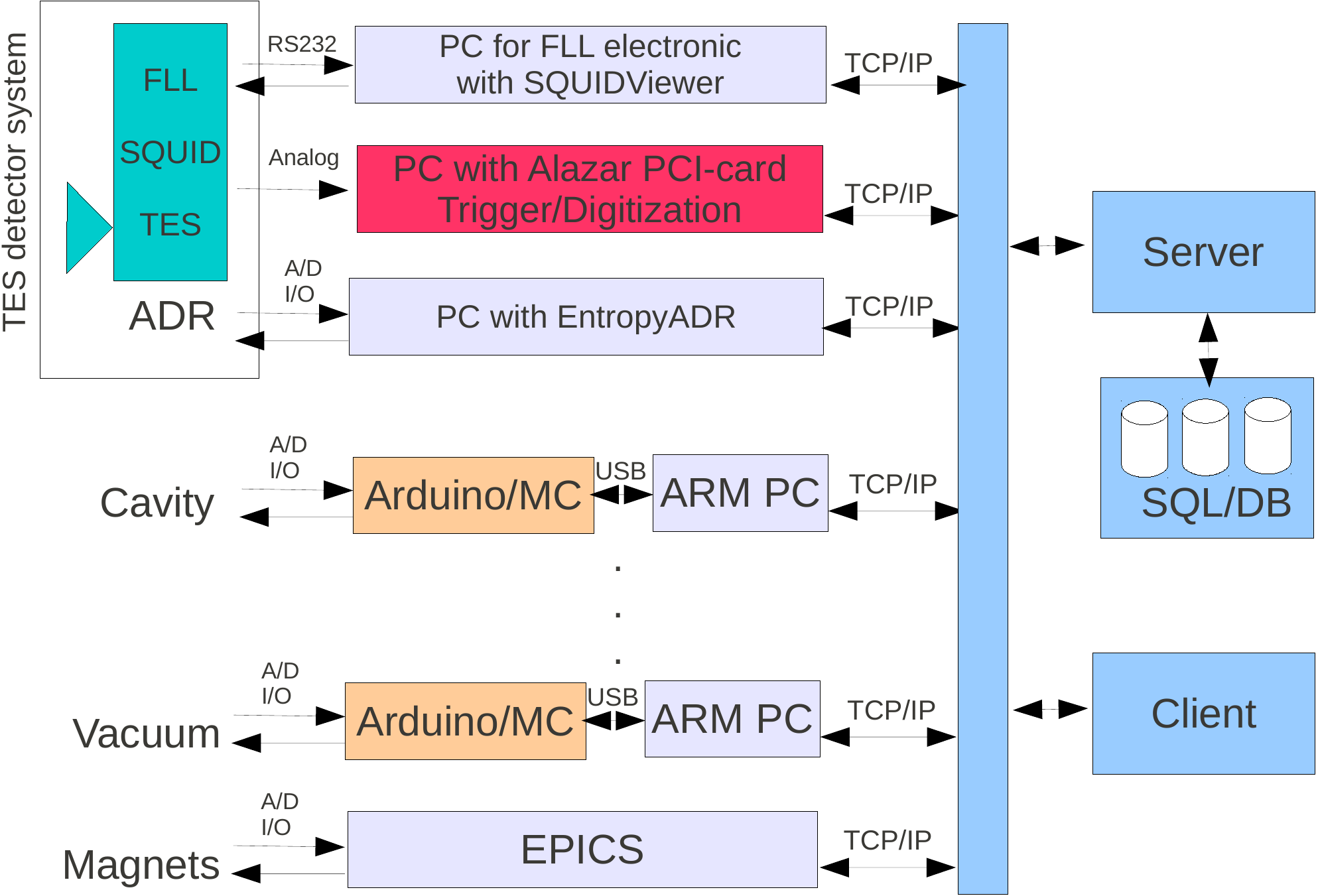}
 \caption{ \label{fig:daq}  Scheme of the distributed data acquisition and control system.}
\end{figure}

The suggested design of the TES-readout is based on a commercially
developed digitizing system (ATS9462 from AlazarTech 
\cite{ATS9462}) 
for the
actual data taking.  The digitizing board allows to sample and digitize up
to 512 MSamples with a frequency of 180 MSamples/s and 16 Bit dynamic range.
This PCIe based board has been in use for the digitization of TES
signals at the group 7.21 at the Physikalisch Technisches Institut in
Berlin.  For the initial setup and tests an already available DRS-4 board of
the ETH Z\"urich with 4 channels in self-triggering mode can be used
\cite{Ritt:2010zz} as well. In case of a CCD-camera as detector system,
the camera can be connected directly via a USB interface to the read-out PC.
The node (detector and read-out) handles the communication and transfer of
data to a server system. There, the data is collected and stored in a
relational data-base system (preferentially MySQL). The inter-process
communication will be based upon a state-machine system which defines states
(e.g., inactive, configuring, ready) and allowed transitions between the
states. The unified state-machine approach will provide well-defined access
and control over the various sub-systems. The actual protocol for the
inter-process communication and control is based upon XML (e.g., XML-RPC
\cite{xmlrpc}) 
(see \Figref{fig:daq} for a scheme of the DAQ system).

In a similar way, the interface to additional 
slow-control systems will be incorporated including 
\begin{enumerate}
 \item{Magnet control:} The superconducting magnets will be controlled by
the cryogroup at DESY. The control environment is based on the
EPICS (Experimental Physics and Industrial Control System), which 
is in wide use at accelerator facilities. The ALPS-I experiment used 
a thin client which would communicate via the http protocol to the actual
user. A similar scheme will be used, extending the HTML syntax to XML. 
The parameters passed to the DAQ will include the electrical current, the
temperature, as well as error messages in order to diagnose a quench. 
\item{Safety loop for the laser:} The safety loop will monitor the 
status of the doors which connect the grey room from the outside, the emergency
switch-off buttons, as well as the integrity of the laser protection covers 
(personal interlock). The technical interlock will interrupt the operation
if the room temperature exceeds a defined limit,  
the pressure in the pipes is
too large (vacuum system), or the superconducting magnet quenches.
Additional warnings will be provided if any other system is in a non-ready
state. 
\end{enumerate}
For safety reasons, any breaches of the interlock system will have to 
lead to an immediate shut-down of
the laser system. 

The temperature switch can be implemented by using the 
Uniflex CI45 module from PMA Prozess-
und Maschinen-Automation 
\cite{pma}. 
These modules need a 24~V power supply and 
are mounted on an industrial standard (DIN) rail (Hutschiene). The measured
temperature can be read out directly by the module. 
The modules can be programmed so
that a relay switches at an upper or lower limit. This relay can
be connected directly with the safety loop. 
The combination of ARM based computer nodes
(e.g., Plug-PC or RASPBERRY PI) 
which act as interfaces to micro-controller  units
(e.g., Arduino boards: these boards provide 14 D/A ports  and 6 A/D ports, 
access through a USB interface). These ready-to-use
micro-controller boards are 
a convenient solution based upon commercially available hardware that is low-priced and very flexible. 

Again, the individual sub-systems will be logically accessed as a
state-machine from the server. Data-streams will be stored in a data-base.
The actual number of monitoring systems is either limited by the address
space of the TCP/IP stack (unlikely to be relevant) or the required
bandwidth that can be handled by the server which is used for communication.
For the initial setup, tens of monitoring channels distributed  over five
nodes will be used. 

The front-end to the user will be available through a client program that
can be run on any computer within the same sub-domain of address space. At a
later stage, remote-operation through secure network protocols will be
implemented (at a lower priority).  The client will have control over the
type of data-taking performed (regular data taking and calibration data
taking), will be able to set parameters of the data-taking (e.g., exposure
times of the CCD, triggering threshold for the TES) and access and
control all sub-systems. The parameters will be stored in the data-base in
order to recover the state of the entire system at any time.

The actual data retrieval for analysis will be handled through access of the
MySQL data-base. A number of simple client scripts allow to select and
retrieve data and calibration information preferentially in FITS-type files. 

\paragraph{Expected performance}

The required performance of the system is well-matched with the proposed
system. In case of a TES detector system, the most challenging
component is the read-out of the data during calibration runs. In this
particular case, it may be necessary to digitize with a high sampling
frequency in the ideal case continuously the signal for a fraction of a
second. During regular data-taking, the expected background rate of the
TES will be small ($10^{-4}$~Hz) and will be of no concern. The
proposed PCIe-based digitizer system will allow both continuous data-taking
for a few seconds to monitor the base-line accurately as well as triggered
operation with additional on-board digital pulse-shaping (if needed).  

The other systems will have modest read-out rates below kHz.
Thus, the traffic of tens of nodes can be handled easily with regular
network bandwidth.

\subsection{Site considerations for ALPS-IIb and IIc}
\label{sec:tdr:site}

\subsubsection*{ALPS-IIb}

\begin{figure}[b!]
 \includegraphics[width=0.5\textwidth]{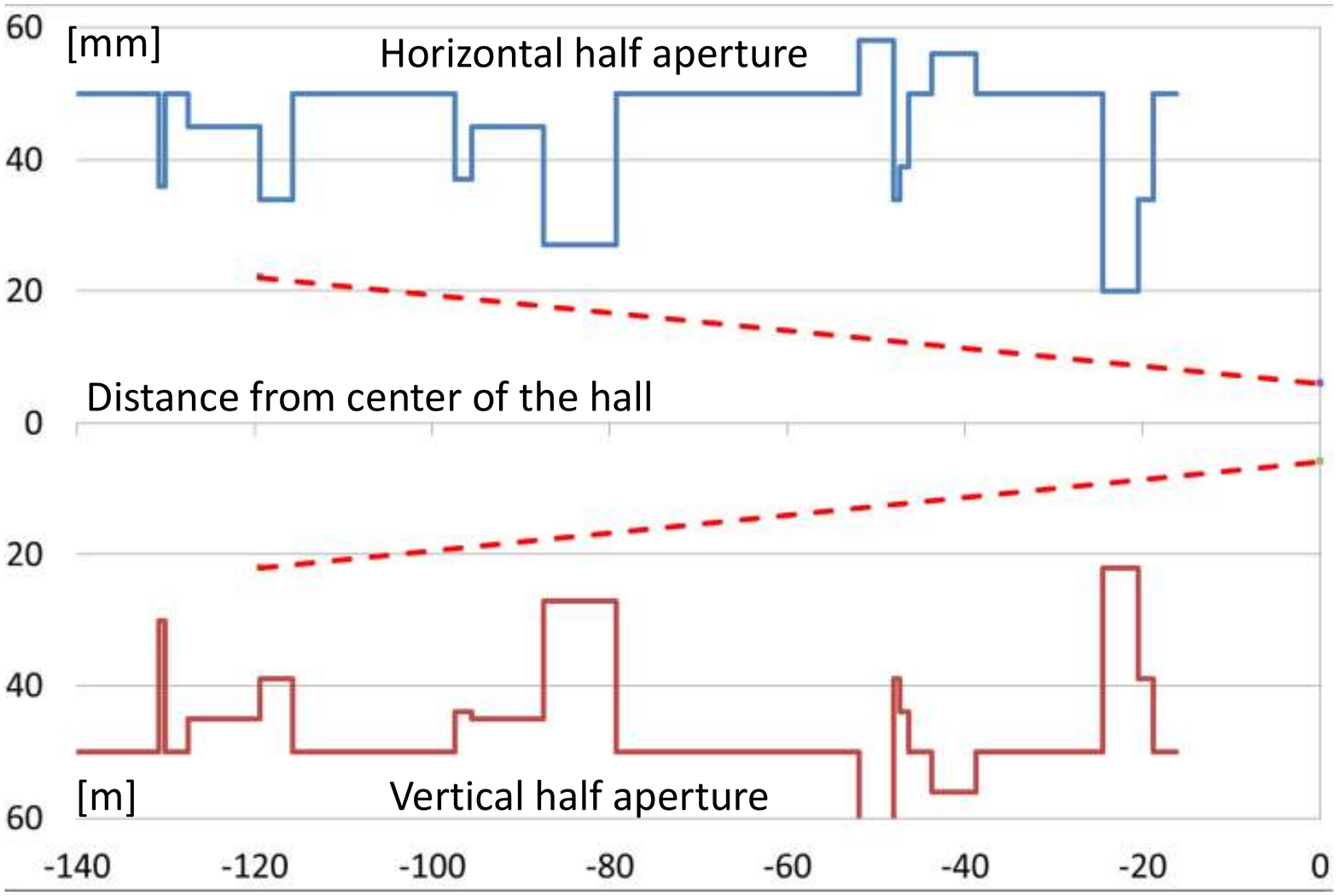}
  \includegraphics[width=0.5\textwidth]{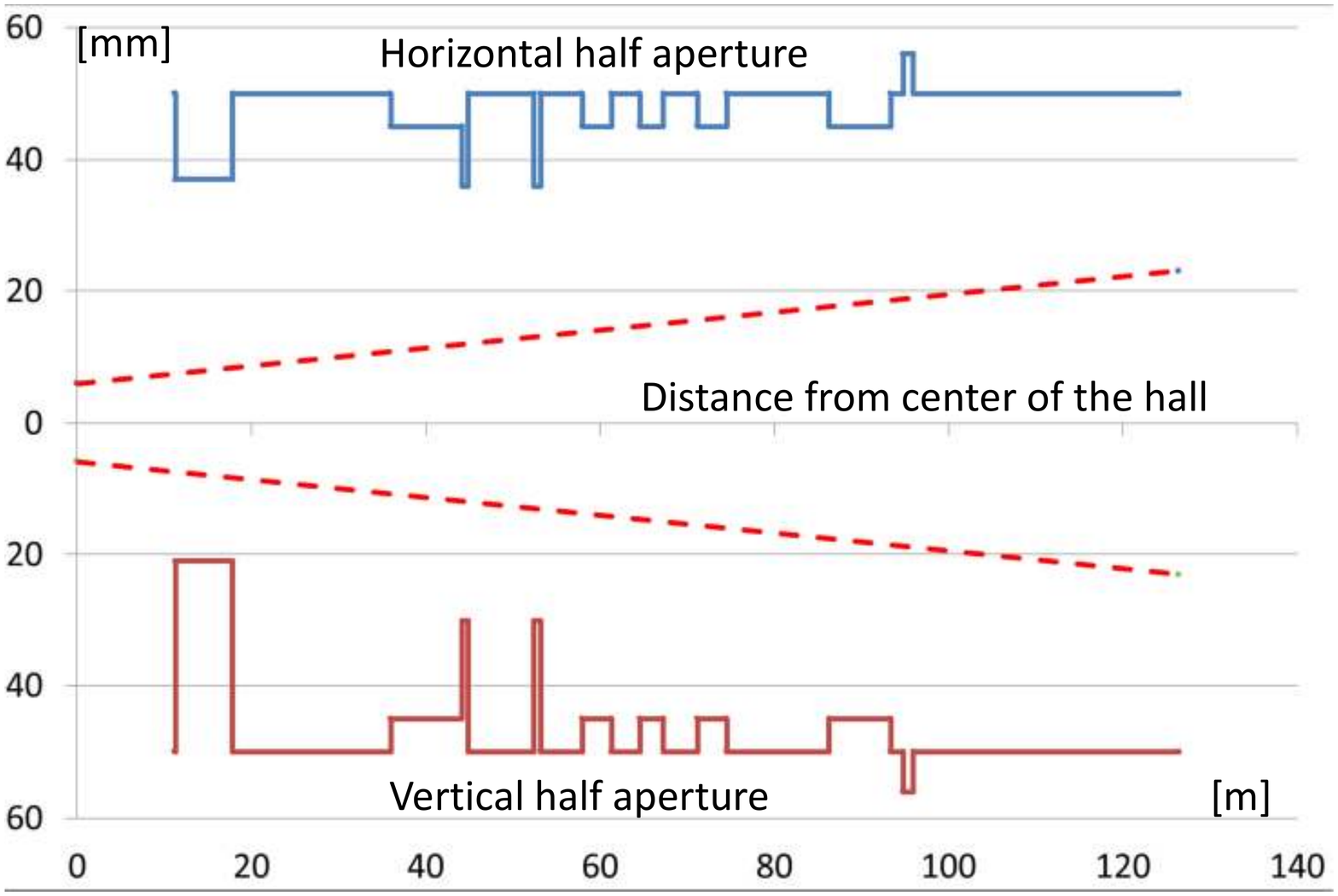}
\caption{The figure shows the horizontal and vertical apertures of the proton beam pipe to both 
sides of HERA hall west.
The Gaussian laser beam with a size required for a power buildup of
40000
is sketched as a red dotted line. 
As can be seen there is ample space to accommodate the laser beam on both sides. }
\label{fig:aperature_west}
\end{figure}

For ALPS-IIb  two optical cavities of about 100 m length will be set up with very little infrastructure 
effort in the straight 
section HERA West. The still existing vacuum pipe of the HERA proton ring can be used for ALPS-IIb as it is 
straight over about 
$2\cdot 160$~m 
in contrast to the other straight sections of HERA, where vertical 
bending magnets limit the available straight length in the vacuum pipe to
about
$2\cdot 60$~m.
The aperture of the proton beam pipe (see \Figref{fig:aperature_west}) is large -- except for one 
location which can be easily modified -- and allows
the operation of optical cavities as described before. 

\begin{figure}[b!]
\begin{center}
 \includegraphics[width=0.7\textwidth]{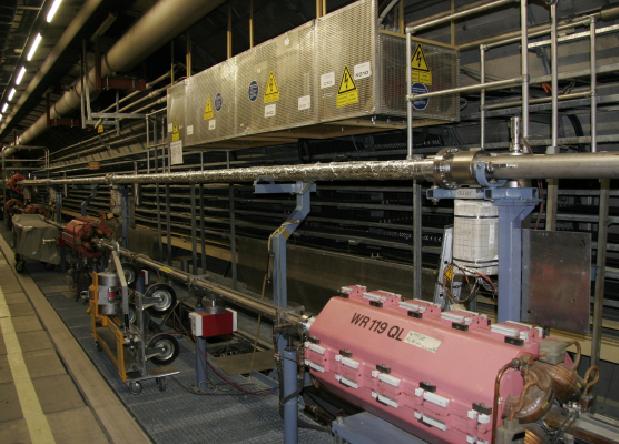}
\end{center}
\caption{This picture shows a part of the HERA West right section. Here,
  there are about
  25~m space
for an optical setup requiring little effort for disassembly.}
\label{fig:west_righ_99}
\end{figure}

Locations  
in hall West and in the tunnel on both sides of the hall have been identified (see \Figref{fig:west_righ_99}), 
where little effort is required to remove the existing accelerator installation to free space for the clean 
rooms at an adequate distance from the hall.

\subsubsection*{ALPS-IIc}

For ALPS-IIc the best possible setup -- allowing for a power buildup of
40000 in the  optical cavity on 
the regeneration side of the 
experiment- consists of $2\cdot 12$ straight HERA dipoles. With the overall dipole length of 
9.766~m such a setup requires a total length of 
about 
250~m including the space for the cleanrooms and laser huts.
The base line design for ALPS-IIc with $2\cdot 10$ almost straight dipoles requires a total length 
of about 
210~m. 

The natural choice for the setup of dipole strings with these lengths is a straight section of 
the HERA tunnel due to the principal availability of 
infrastructure like cryogenics. The available length in the long straight sections East and
West amounts to 
320~m, long enough for a setup of $2\cdot 12$ HERA dipoles for ALPS-IIc. 
The short straight sections South and North with a length of
220~m allows the setup 
of $2\cdot 10$ HERA dipoles. 

An installation outside of the HERA tunnel on the DESY site has also been considered. 
There are possible locations with sufficient length along the eastern borderline of the DESY site. 
Due to the additional costs of such an installation for buildings, cryogenics, and infrastructure, 
which have been estimated to about 12 million Euros, this possibility has
not been followed any further. 

The installation of the dipole strings connected properly to cryogenic boxes requires the 
removal of the existing accelerator installation in any straight section of HERA chosen for this 
purpose. This is mainly due to the necessary exchange of the cryogenic boxes from one end of 
the straight section to the other, to match the connection pattern of the dipoles. 
The effort in this respect is largest in the straight section West due to the large number of 
special and complicated accelerator systems like the proton injection line, the superconducting
cavities of the electron ring, the RF cavities of the proton ring, and the proton beam dump.
The beam dump would have to be disposed of \cite{43}, which would cost a few
million Euros. 

The setup of HERA dipole strings in the straight section East would require the 
highest operation cost for the cryogenics of all straight sections.  
In addition the power supply for the superconducting magnet chain of HERA would have to be moved 
from hall West to East.
However, the straight section East remains the location for a setup of  $2\cdot 12$ 
dipoles string with lower cost as compared to West, due to the disposal cost for the beam dump.

For the base line design of the ALPS-IIc experiment with $2\cdot 10$ dipoles the total cost is 
lowest in the straight section HERA North, which therefore has been selected as best 
suited for the setup of ALPS-IIc.

\subsection{Setup and installation}
\label{sec:tdr:setup}

\subsection*{ALPS-IIb}

In the straight section HERA West the infrastructure for two optical
cavities of about 
100m length will be 
set up with very little infrastructure effort. The existing vacuum pipe of the HERA proton ring can be used for 
ALPS-IIb as it is straight over about 
$2\cdot 160$m in contrast to the other straight sections of HERA, where 
vertical bending magnets limit the available straight length in the vacuum
pipe to about 
$2\cdot 60$m.

Locations  in hall West and in the tunnel on both sides of the hall have been identified, where only minor 
activities are required to remove the existing accelerator installation to free space for the clean rooms at an 
adequate distance from the hall.

\subsubsection*{2013}

The accelerator installation at the locations for the cleanrooms can be dismantled by the DESY group MEA, 
whenever manpower is available throughout the whole year, creating space for the installation of the cleanrooms.  
The construction of the laser huts will be designed and orders for the construction as well as for the clean
room equipment will be placed. 

\subsubsection*{2014}

The cleanrooms will be installed within the first half of the year.  
With the completion of the ALPS-IIa experiment in the middle of 2014 (see \Sectref{sec:tdr:laser})  
the optical components and the optical tables become available  and will be installed in the cleanrooms.

\subsubsection*{2015}

The 
 $\sim100$m long production and regeneration cavities  will be setup and commissioned, 
allowing physics runs in the second half of the year.

\subsection*{ALPS-IIc}

\subsubsection*{2015}

In the beginning of 2015, with the completion of the XFEL project, manpower will be available at DESY to 
install the ALPS-IIc setup and also cryogenics capacity, allowing the cryogenic tests of the HERA dipoles for 
ALPS-IIc on the magnet test bench.

 In parallel to the magnet tests, the accelerator installations in the straight section North will be disassembled.
 For the disassembly of the normal conducting magnets and cavities in
 the straight section the ``HERA tram'' (see left side of \Figref{fig:magnets22}) will be used, which has been transfered 
 from hall West after the removal of the Kicker bypass from the HERA tunnel.
 
 After the deinstallation of all elements in the straight section, the cryogenic boxes at the ends of the straight section will be
 disconnected and moved to the opposite sides.

Then activities will follow, which are easier without the magnet strings in place, like the modification of the 
quench gas collection pipe, modification of dump resistors, connections for cooling water of the laser huts, or 
work on the main power cables.

\subsubsection*{2016}

New supports for the dipoles will be installed. For the installation of the dipoles the ``HERA tram'' will be used. 
In general the installation of the 
dipole strings will follow the procedures
established for the installation of the superconducting magnets at HERA \cite{Borchardt:1991dn}.

After the installation of the superconducting dipoles, the HERA kicker-bypass will be installed between the two strings. 
To compensate the force of about 
80 kN by the atmospheric pressure on the end flanges of 
the kicker-bypass, the endflanges are connected by 3 tension rods, which have to be replaced by a steel girder construction to allow 
later after the installation 
of the cleanroom access to the optical elements in the vacuum vessel.

\begin{figure}[tb]
\centering
\includegraphics[width=0.4\textwidth]{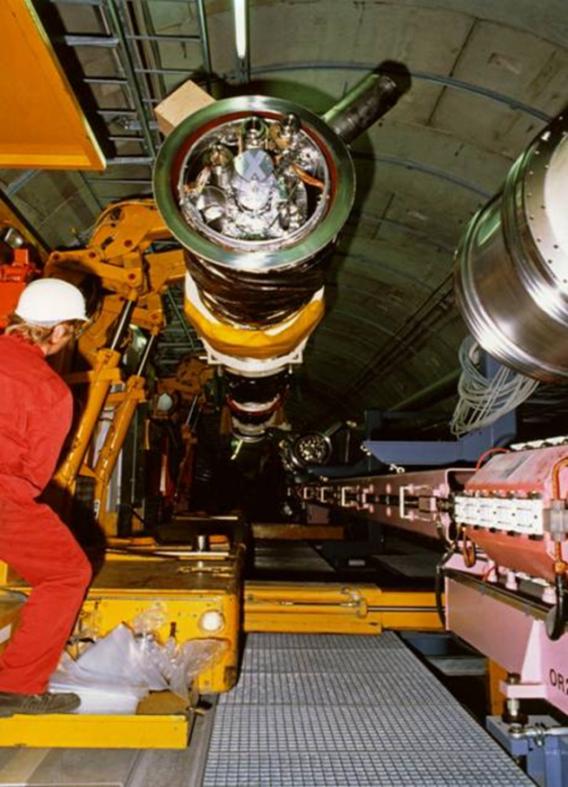}
\includegraphics[width=0.5\textwidth]{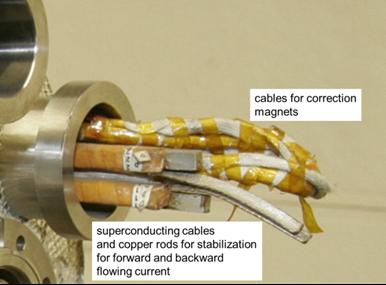}
\caption{HERA tram (left) and superconducting cables in 
4K Helium tube (right).}
\label{fig:magnets22}
\end{figure}

The proper electrical connection of the superconducting cables for forward
and return current between adjacent dipoles (see right side of \Figref{fig:magnets22}) 
is of uttermost importance for the operation of the magnet strings.
The heat generated by the resistivity of the connection must be small enough to be absorbed by the surrounding liquid Helium, 
to keep the temperature of the 
connection stable. For HERA a special tool was used to braze the connections
properly by well-trained people \cite{Borchardt:1993pb}. Most of those people are in retirement by now, but a few are still at 
DESY, who will be capable of 
connecting the cables after some retraining. 
Video material\footnote{Video material by Otto Peters retired from the DESY group MKS.} is available which shows the procedures. 
The video material also shows 
how to insulate and to fix the cables after brazing to avoid an electrical discharge.  
Once the electrical connections are made and tested, the Helium pipes of adjacent dipoles will be joined, 
using special 
welding tools from HERA, shown in the available video material. These tools are still available and also knowledgeable people to weld the connections.

The electronics for the quench detection and protection will be placed in the tunnel close to the magnets after 
the completion of the magnet installation.

 The laser huts 
 ($2.25\cdot {\rm m^2}$) 
 with the optical tables at the end of the strings  and in the middle will be installed, 
 using the available material from the completed ALPS-IIb experiment for the setup of the optical 
 resonators and the detector. The cooldown of the magnet strings will be done once the commissioning of the optics will be completed, as venting of 
 the vacuum pipe in the magnets for the commissioning of the optics can only be done when the
 magnets are at room temperature.

\subsubsection*{2017}

After the commissioning of the optical resonators, the detector  and commissioning of the magnet string physics runs will start.

\subsection{Operation and run procedure}
\label{sec:tdr:operation}

\begin{figure}[!b]
\tikzset{>=stealth'}
\tikzstyle{mode}=[rectangle,%
  rounded corners,%
  draw=black, very thick,%
  text width=10em,%
  minimum height=2em,%
  text centered,%
  join]
\tikzstyle{box}=[rectangle,
    draw=gray,
    thick,
    dashed,
    inner sep=1em]
\tikzstyle{every join}=[->]
\resizebox{\textwidth}{!}{
  \begin{tikzpicture}[
    start chain=1 going below,
    start chain=2 going below,
    start chain=3 going below,
    node distance = 1.5em
    ]
    \begin{scope}[every node/.style={mode,on chain=1}]
      \node (startPrep) {optics};
      \node[join] {detector alignment};
      \node[join] (stopPrep) {detector calibration};
    \end{scope}

    \begin{scope}[every node/.style={mode,on chain=2}]
      \node[right=5em of startPrep] (startData) {optics};
      \node[join] {detector dark};
      \node[join] {optics};
      \node[join] {detector signal};
      \node[join] {\ldots};
      \node[join] {optics};
      \node[join] (stopData) {detector dark};
    \end{scope}

    \begin{scope}[every node/.style={mode,on chain=3}]
      \node[right=5em of startData] (startConcl) {optics};
      \node[join] {detector calibration};
      \node[join] (stopConcl) {detector alignment};
    \end{scope}

  \node[box,fit=(startPrep) (stopPrep)] (prepBox) {};
  \node[above] at (prepBox.north) {run preparation};

  \node[box,fit=(startData) (stopData)] (dataBox) {};
  \node[above] at (dataBox.north) {data taking};

  \node[box,fit=(startConcl) (stopConcl)] (conclBox) {};
  \node[above] at (conclBox.north) {run conclusion};

  \pgfmathsetmacro{\MyOffset}{1.5}

  \draw[->,thick] ($(prepBox.north east) - (0, \MyOffset)$) -- ($(dataBox.north west) - (0, \MyOffset)$);
  \draw[->,thick] ($(dataBox.north east) - (0, \MyOffset)$) -- ($(conclBox.north west) - (0, \MyOffset)$);

\end{tikzpicture}
}
\caption{A typical run of \ac{ALPS}-II in routine operation.}
\label{fig:run:typical}
\end{figure}
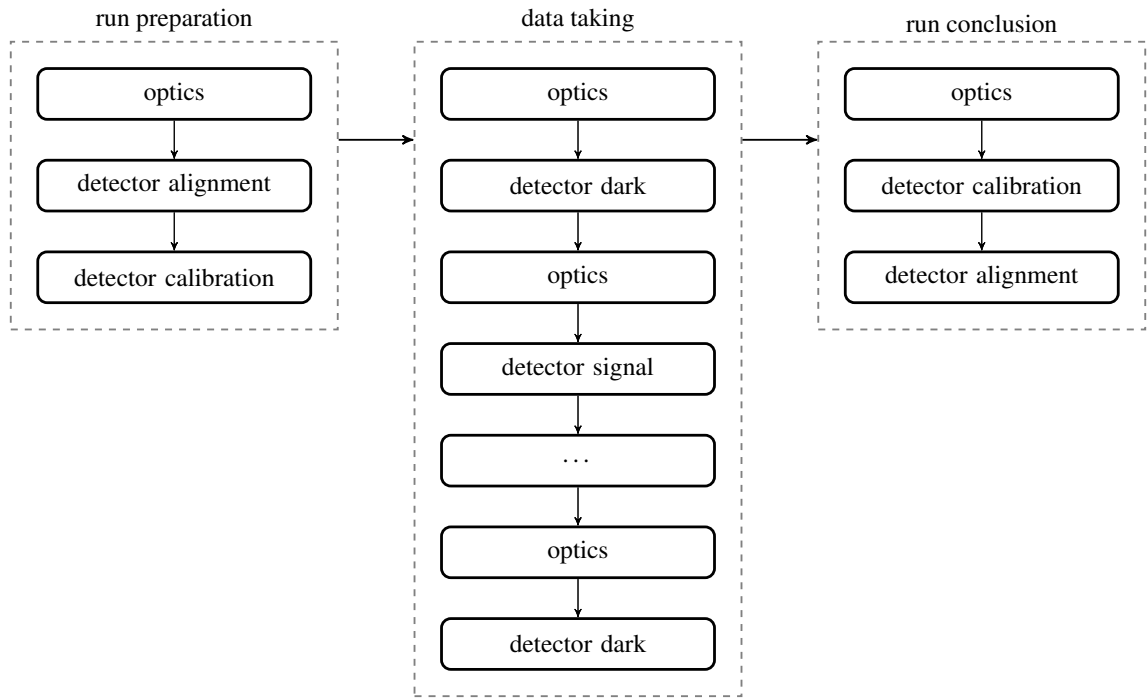

As discussed in \Sectref{chap:tdr:goals}, some parameters will be varied
over the course of the experiment in order to interpret a potential signal
and distinguish the different types of \acp{WISP}, and to fill the
sensitivity gaps (cf.~\Figsref{fig:ALPsvac} and~\ref{fig:hpgas}). These
parameters include the magnetic field strength, the orientation of the laser
polarization w.r.t.~the magnetic field, and the pressure of the rest gas.
The data taken with one set of these parameters will result in a
\emph{run}. During each run, the experiment will operate in different
\emph{modes}, which are dedicated either to assert the proper condition of
the experiment or to take data. These are described below. Depending on the
actual performance of the subsystems of the experiment and during the commissioning phase, additional modes may
be necessary. 
A typical run is shown in \Figref{fig:run:typical}.

\subsubsection*{Condition Assertion}

\paragraph{\emph{Optics} Mode}

During this mode, the optics system is brought up. This includes checking
the alignment of the production and regeneration cavities, locking the two
cavities and setting the resonance condition of the regeneration cavity for
$1064~{\rm nm}$
as necessary for the targeted data type (see below).
The resonance condition for
$1064~{\rm nm}$
can be checked by
monitoring QPD7 and opening the shutter on the central breadboard
(\Figref{fig:optical:layout}). If the shutter is open, infrared light enters
the (resonant) regeneration cavity and is recorded by QPD7.

Because the shutter may be open in this mode,
$1.7~{\rm mW}$
of
$1064~{\rm nm}$
($10^{16}$
photons per second) will leave the
regeneration cavity and potentially reach the detector. Hence, the detector
has to be protected by, e.g., a flip mirror, which redirects the light to a
beam dump.

\paragraph{\emph{Detector Alignment} Mode}

In this mode the alignment of the detector to the beam of reconverted
photons is verified. To this end, the shutter on the central breadboard will
be open. Infrared light will enter the regeneration cavity, following
exactly the path of potentially regenerated photons. As mentioned above,
$1.7~{\rm mW}$
of
$1064~{\rm nm}$
radiation will hit the
detector. Hence, a dedicated filter will be necessary to protect the
detector from any damage due to this intensive radiation as discussed in
\Sectref{chap:tdr:detector:conceptual_design}.

\subsubsection*{Data Taking}

It is foreseen to routinely operate the \ac{CCD} and the \ac{TES} in three distinct
modes: \emph{calibration}, \emph{signal}, and \emph{dark}. Signal and
dark mode will only happen during stable operation of the cavities. They
differ only in the resonance condition of the regeneration cavity. In signal
mode the regeneration cavity will be resonant for infrared light while in
dark mode it will be resonant only for
$532~{\rm nm}$.
This will be
achieved with the \ac{AOM} on the central breadboard
(\Figref{fig:optical:layout}). Thus, the regeneration probability is
inhibited by a factor of the power build-up of the regeneration cavity for
infrared light
($\sim40000$).
This makes it possible to record all experimental background
(e.g., stray light, fluorescence effects).

\paragraph{CCD Operation}

The calibration of the \ac{CCD} will include the read-out of a bias-frame, a
dark-frame, and a flat-field, which will be used to assure a stable
operation of the \ac{CCD}. The expected stability during routine operation will
reduce the need to take calibration data very frequently. However,
complete calibration data will be taken before and after operation in signal
or dark mode.

Regular data-taking in signal or dark mode will consist of a series of
frames. The exposure time will be chosen such that the read-out
noise and dark current will balance, which results in roughly one hour
(cf.~\Sectref{chap:tdr:detector:expected_performance}).

\paragraph{TES Operation}

The specific calibration and data-taking routine for the \ac{TES} based
read-out is at this point less-well defined as for the \ac{CCD}.
Generically, a calibration mode will be required to calibrate and
characterize the detector, the \ac{SQUID} amplification, and the digitization.
All components are potentially a source of background and require
independent calibration to characterize the baseline and its fluctuations.
Currently, we envisage to use a triggered digitization with up to 180
MSample/s frequency.  A continuous read-out of up to 512 MSamples can be
stored on-board, for longer data-taking, direct memory access is possible to
store longer data-streams. This will however be only an option in
exceptional cases. In the case of the DRS4-board, the maximum bandwidth of
the read-out limits the read-out rate to a few hundred
Hz
(based upon the
experience with the DRS4 board, the maximum rate is
$530~{\rm Hz}$).
With a random
trigger of the read-out, the baseline can be sampled  in an unbiased way.
However, any irregularity at frequencies smaller than kHz will not be
visible. This is in principle not an issue given that the signals are much
faster.

Dedicated calibration runs will be used to read-out pulses registered with
the \ac{TES} illuminated by light. The pulses will be analyzed to accumulate
a photon number spectrum to calibrate the single-photon response and to set
the trigger threshold.

During regular data-taking, the triggered signals are recorded (without any
on-line pulse-shaping) in a time window of
$100~{\rm ms}$
in order
to cover the pulse as well as the base-line before and after the pulse. The
trigger time is accurate to
$8~{\rm ns}$
and will be stored for each
triggered event. 


\subsection{Data analysis}
\newcommand\formcomma{\:,}
\newcommand\formstop{\:.}
\newcommand\vc{\ensuremath{v_\mathrm{c}}\xspace}
\newcommand\srodc{\ensuremath{\sigma_{\mathrm{ro}+\mathrm{dc}}}\xspace}
\newcommand\pois{\ensuremath{\mathrm{Pois}}\xspace}
\newcommand\gaus{\ensuremath{\mathrm{Gauss}}\xspace}
\newcommand\offset{\ensuremath{\mathit{Offset}}\xspace}

The analysis of the photon detector data will result in the number of
regenerated photons, which has to be combined with other experimental
parameters to determine the coupling strength to photons and possibly also
the mass of a \acp{WISP}. More accurately, a confidence interval will be
estimated, yielding either an upper limit or an interval for the number of
regenerated photons. Because the \ac{TES} and the \ac{CCD} deliver different
types of measurement (the \ac{TES} is a true single-photon detector while
the \ac{CCD} integrates the incoming photon flux during exposure), the
procedure to estimate the number of regenerated photons has to by tailored
for each detector. 

The method to estimate the rate of reconverted photons for the two detectors
is described below. The interpretation of a signal of re-converted photons
is sketched in \Sectref{subsec:tdr:overview-modes}.


The signal recorded by any of the two detectors will be the sum of different
components: the flux of reconverted photons, irreducible background, noise
of the read-out, and a potential offset. The value of the signal, $v$, can
be described by the following statistical model,
\begin{equation}
  v \sim \pois(\lambda_\mathrm{sig})
  \oplus \pois(\lambda_\mathrm{bgd})
  \oplus \gaus(\delta, \sigma_\mathrm{ro})
  \oplus \offset
  \formcomma
  \label{eq:statistical:model}
\end{equation}
where $\pois(\lambda)$ are Poisson-distributed components due to reconverted
photons (sig) and background photons (bgd), $\gaus(\mu, \sigma)$ the
component due to noise of the read-out (ro), which is assumed to be
Gaussian, and \offset a constant. The measured value, $v$, is distributed
like the convolution of the distributions of the components, which is
represented by $\oplus$.

\paragraph{CCD}

The \ac{CCD} was already used in \ac{ALPS}-I and later tested in detail to
determine its performance for
$1064 {\rm nm}$
photons. The analysis
procedure is based on these experiences and, naturally, more matured than
the approach for the \ac{TES}, where the data analysis will depend on the
results of the ongoing R\&D.

Interpreting the components of \Eqref{eq:statistical:model} in the
case of the \ac{CCD} is straight forward. The Poissonian background is due
to the accumulation of thermally produced dark charge; the read-out noise is
caused by the amplification and digitization during the read-out of the
camera; and an offset will be present due to biasing the sensor chip. Pixels
outside of the signal region on the camera can be described by the same
model by setting $\lambda_\mathrm{sig}$ to zero.

It was found that the bias level of the camera shows approximately linear,
spatial variations similar for each frame (fixed pattern) and varies by a
global offset on a per-frame basis. To correct this per-frame variation of
the offset, the value of the signal pixel will be corrected by subtracting
the average of a large number of pixels in the vicinity of the signal pixel.
Because the variation of the bias is approximately linear, the spatial
variation of the bias is averaged and can be dropped if these pixels are
distributed symmetrically around the signal pixel. Thus, the corrected
signal pixel value, \vc, is described by
\begin{equation*}
  \vc \sim \pois(\lambda_\mathrm{sig})
  \oplus \gaus(\delta, \srodc)
  \formcomma
\end{equation*}
where the Gaussian now contains the components of the read-out noise and
dark current of the signal pixel as well as of the averaged pixels and
incorporates a possible systematic offset, $\delta$, between these.

Any signal is searched for by comparing so-called data frames (data, where
re-converted photons from \acp{WISP} can be expected) with dark frames
without any re-converted photons. Corresponding details are sketched in
\Sectref{subsec:tdr:overview-modes}. To estimate $\lambda_\mathrm{sig}$ from
a set of $N$ signal and $M$ dark frames, we will use the difference of the
averages of the corrected pixel values, $t$,
\begin{equation*}
  t = \frac{1}{N}\sum_{i=1}^N \vc^\mathrm{(sig)}(i) -
  \frac{1}{M}\sum_{j=1}^M \vc^\mathrm{(dark)}(j) 
  \formcomma
\end{equation*}
which is described statistically by
\begin{equation*}
  t \sim \frac{1}{N} \bigoplus_{i=1}^N \pois(\lambda_\mathrm{sig})
  \oplus \gaus(0, \sigma_t)
  \formcomma
\end{equation*}
where the first term represents the averaging of $N$ Poisson-distributed
observables and $\sigma_t$ depends on the sizes of the signal and dark data
sets, $N$ and $M$, and on the noise from the read-out and the dark current,
\srodc,
\begin{equation*}
  \sigma_t^2 = \left(\frac{1}{N} + \frac{1}{M}\right) \srodc^2
  \formstop
\end{equation*}

An algorithm based on \cite{Feldman:1997qc} has already been implemented to
estimate $\lambda_\mathrm{sig}$ for an observed value of $t$, where \srodc
is estimated from the dark frames. Monte Carlo simulations showed that for
$N>10$ and $M\geq10\cdot N$ this gives precise coverage. Systematic tests of
this method will include using different groups of pixels for the offset
correction and applying the method on non-signal pixels.

\paragraph{TES}

The \ac{TES} provides single-photon informations which can be extracted from
the pulse shape (e.g., rise-time, height, pulse-width). The pulse has a
typical risetime of
$300 {\rm ns}$. With the proposed read-out
system that samples at a frequency of 180\:MSamples/s, the rise-time of the
pulse will be very well resolved with $\approx$ 50\:samples.

In the simplest scheme, a trigger threshold will be set to suppress the
base-line noise and will trigger the read-out of the digitized pulse and the
adjacent base-line\footnote{A more refined triggering would be based upon
the image shape and could be implemented in the FPGA of the digitizer}.

The trigger threshold will be regulated to suppress unwanted noise of the
base-line while maintaining a high photon detection efficiency. The
resulting background will be given by real photons (mostly thermal)
($\lambda_\mathrm{bgd}$) as well as read-out noise ($\gaus(\delta,\sigma)$).
The analysis of the events can be used to suppress background photons which
are not in the right wavelength band (pulse-height) or do not show the
characteristic pulse-profile. The former will reduce the value of
$\lambda_\mathrm{bgd}$ while the latter will help to reduce the impact of
base-line/readout. A possible statistical analysis of the \ac{TES} data
could in principle be carried out in a similar fashion as for the \ac{CCD}.
However, it may be more sensitive to develop an analysis which will benefit
from the individual pulse measurements. Such an analysis would calculate a
probability for each observed pulse to be produced from a re-converted
photon, a background photon, or to originate from noise. In such an
analysis, hypothesis testing and parameter estimates would be based upon a
likelihood method.




\section{Summary \& concluding thoughts}
\label{chap:tdr:summary}

Low-energy particle physics experiments enable us to explore fundamental physics in a complementary
way to accelerator-based searches
by looking for new light particles with tiny couplings.
Such new particles arise naturally in many 
extensions of the Standard Model and might also explain observations that are not accounted for
within the particle physics known today (important examples being the absence of CP violation in 
the strong interactions
and the nature of Dark Matter).

We have presented here the technical design for the proposed ALPS-II experiment at DESY which could 
contribute strongly to 
exploring the widely unchartered territory of Weakly Interacting Slim Particles (WISPs).
ALPS-II would improve the best present-day laboratory sensitivity achieved by ALPS-I by more than three orders of 
magnitude and even surpass indirect limits on WISP properties from astrophysics and solar observations.
Thus, parameter regions motivated by astroparticle physics phenomena and from predictions of
string theory are in reach of ALPS-II.

ALPS-II necessitates a collaboration of research fields which 
are usually not in close contact: the know-how of large 
particle physics accelerator projects, optical expertise from gravitational 
wave interferometers and background-free single photon counting
similar to the requirements in quantum optics and communication technology. 
Based on the experience with ALPS-I we are confident to realize
ALPS-II with data taking completed in 2017 given the resources and the time schedule presented in this TDR.

ALPS-II could be implemented at DESY in a very cost effective manner by re-using
 available spare HERA dipole magnets and the HERA infrastructure.
For investments and operation about 2\,MEuro are required in total within the next five years.
To significantly go beyond the ALPS-II reach, R\&D activities to tackle technological challenges have been 
identified:
Examples are a further increase of the laser power in the production cavity to 
several MW and operating with green light
allowing for much longer installations due to a smaller laser beam divergence.
Corresponding research has started already.
Future ALPS-II-like experiments could strongly benefit from magnet development ongoing for a possible 
energy upgrade of LHC.
However, such large and probably costly installations are far from being ready for decision.
Besides the technology issues one has to await new results from astrophysics, laboratory WISP searches and theory.

To conclude: Although challenges remain, the rewards of exploring the low-energy frontier of particle 
physics with ALPS-II at DESY could be enormous,
yielding fundamental insights into long-standing astro and particle physics puzzles.

\acknowledgments

\label{cha:ack}
\vspace{0.3cm}
The ALPS collaboration gratefully acknowledges support by the LEXI Hamburg and DFG-SFB 676.
Also, the ALPS collaboration cordially acknowledges the advice and active support of a number of people who have contributed to the R\&D of the proposed experiment or the preparation of this TDR.\\

For advice and helpful discussions in theory and phenomenology we thank
P.~Arias (Universidad Cat\'{o}lica, Santiago, Chile), C.~Burrage (University of Nottingham),
M.~Cicoli (ICTP Trieste), M.~Goodsell (CPhT, Paris),  J.~Jaeckel (
University of Heidelberg), F.~Karbstein (Helmholtz Institute Jena) and J.~Redondo (MPI Munich ASC Munich).

For advice or active support on optics, opto-mechanics and control we thank
M.~Frede (neoLASE GmbH), T.~Meier, P.~Schauzu and A.~Weidner (AEI).

Regarding the development of the detection system we thank
F.~Borges, K.~Ehret, O.~Hellmig, 
M.~Kowalski,  H.~Maser, A.~Zuber (DESY) and G.~Wiedemann (University of Hamburg),
as well as our Italian TES collaborators
G.~Di Giuseppe, M.~Karuza, M.~Lucamarini, R.~Natali, D.~Vitali and his local crew from the University of Camerino and G.~Cantatore with his team from INFN and University of Trieste.
We also thank  H.~Barthelmess (Magnicon GmbH), U.~\"Ochsner (Sch\"after + Kirchhoff GmbH), K.~Phelan, and D.~Wernicke (ENTROPY GmbH).
For introducing us to the TES community and much very valuable support we thank J.~Beyer, I.~Novikov, M.~Schmidt and the local crew from PTB Berlin as well as
D.~Fukuda (AIST Japan), S.~W.~Nam (NIST USA), L.~Lolli (INRIM Torino), M.~Guistina, A.~Mech,  R.~Ursin (University of Vienna) and A.~J.~Miller (Albion College USA).

For advice and support regarding the development of the magnet, vacuum and cryogenics systems we thank the DESY staff, in particular
C.~Albrecht, S.~Baark, R.~Bandelmann, W.~ Benecke, M.~Berg, M.~B\"ohnert, H.~Br\"uck, F.~Czempik, H.J.~Eckoldt,  U.~Eggerts, K.~Escherich, J.~Eschke,P.D.~Gall, C.~Hagedorn, B.~Hager, U.~Hahn, K.~Harries,
S.~Holm, D.~Hoppe, K.~Jensch, A.~Jung, S.~Karstensen, L.~Klein, A.~Kock,
M.~K\"orfer, T.~Kurps, U.~Laatzen, N.~Lass, A.~Leuschner, H.~Lewin, H.P.~Lohrmann, K.~Ludwig, G.~Meyer, M.~Noack, F.~Obier, 
U.~Packheiser, C.~Peitzmann, O.~Peters, B.~Petersen, K.~Petersen,  J.~Prenting, B.~Racky, U.~Reinhold, H.~Remde, S.~Rettig-Labusga, J.~Schaffran, B.~Schmidt, T.~Schnautz, C.~Schulz, M.~Schwalger,  
C.~Schwalm, D.~Sellman, M.~Staack, L.~Steffen, M.~Stolper, J.~Teichmeier, N.~Tesch, K.~Tiedtke, P.~Toedten, W.~von Schr\"oder, H.P.~Wedekind, G.~Weichert, G.~Wolgast and J.~Zajac.

From DESY we thank also
W.~Bialowons, C.~B\"uhrig, B.~Conrad, U.~Djunda, M.~Ebert, B.D.~Elie, M.~K\"opke, B.~Lange, A.~Matheisen, N.~Meyners, H.~Ruge, A.~Schleiermacher, M.~Schl\"osser, B.~Schmidt, S.~Schrader,  
B.~Sparr, F.R.~Ullrich, S.~Warratz and O.~Wrobel for their support and advice concerning the cleanroom construction.\\

For uncountable and fascinating discussions of WISP physics as well as 
for the fun of organizing workshops, we thank A.~Afanasev 
(Jefferson Laboratory and George Washington University),
V.~Anastassopoulos (University of Patras),
O.~K.~Baker (Yale University),
L.~Baudis (Zurich University),
J.~Jaeckel 
(University of Heidelberg),
J.~Redondo (MPI Munich),
M.~Schumann (University of Bern)
D.~Tanner (University of Florida),
W.~Wester (FNAL),
K.~Zioutas (University of Patras and CERN)
as well as all the attendees and contributors to the PATRAS worshop series.

We are grateful to PIER at Hamburg for supporting the WISPy lecture day and 
the ALPS seminar series.\\

Last but not least: For a careful proofreading of the manuscript, we 
thank U.~Schneekloth as well as P.~Hendrikman Verstegen. 
Any remaining mistakes are to be blamed to the authors of this TDR.

\bibliography{ALPS-II}
\end{document}